\documentclass[12pt]{iopart}
\pdfoutput=1

\usepackage{iopams}
\usepackage{amsmath}
\usepackage{bm}
\usepackage[mathcal]{euscript}
\usepackage{graphicx}
\usepackage{subfigure}
\usepackage{cite}
\usepackage{ifthen}

\eqnobysec

\graphicspath{{figs/}{figs_lo/}}

\newcommand{\ie}{i.e.}

\newcommand{\mathnotation}[2]{\newcommand{#1}{\ensuremath{#2}}}

\makeatletter
\def\Left#1#2\Right{\begingroup%
\def\ts@r{\nulldelimiterspace=0pt \mathsurround=0pt}%
\let\@hat=#1%
\def\sht@im{#2}%
\def\@t{{\mathchoice{\def\@fen{\displaystyle}\k@fel}%
{\def\@fen{\textstyle}\k@fel}%
{\def\@fen{\scriptstyle}\k@fel}%
{\def\@fen{\scriptscriptstyle}\k@fel}}}%
\def\g@rin{\ts@r\left\@hat\vphantom{\sht@im}\right.}%
\def\k@fel{\setbox0=\hbox{$\@fen\g@rin$}\hbox{%
$\@fen \kern.3875\wd0 \copy0 \kern-.3875\wd0%
\llap{\copy0}\kern.3875\wd0$}}%
\def\pt@h{\mathopen\@t}\pt@h\sht@im%
\Right}%
\def\Right#1{\let\@hat=#1%
\def\st@m{\mathclose\@t}%
\st@m\endgroup}
\makeatother

\makeatletter
\def\Leftt#1#2\Rightt{\begingroup%
\def\ts@r{\nulldelimiterspace=0pt \mathsurround=0pt}%
\let\@hat=#1%
\def\sht@im{#2}%
\def\@t{{\mathchoice{\def\@fen{\displaystyle}\k@fel}%
{\def\@fen{\textstyle}\k@fel}%
{\def\@fen{\scriptstyle}\k@fel}%
{\def\@fen{\scriptscriptstyle}\k@fel}}}%
\def\g@rin{\ts@r\@hat\vphantom{\sht@im}}%
\def\k@fel{\setbox0=\hbox{$\@fen\g@rin$}\hbox{%
$\@fen \kern.3875\wd0 \copy0 \kern-.3875\wd0%
\llap{\copy0}\kern.3875\wd0$}}%
\def\pt@h{\mathopen\@t}\pt@h\sht@im%
\Rightt}%
\def\Rightt#1{\let\@hat=#1%
\def\st@m{\mathclose\@t}%
\st@m\endgroup}
\makeatother

\newcommand{\norm}[1]{\l\lVert#1\r\rVert}    
\newcommand{\normt}[1]{\lVert#1\rVert}       
\newcommand{\savg}[1]{\l\langle #1\r\rangle} 
\newcommand{\tavg}[1]{\overline{#1}}         
\newcommand{\stavg}[1]{\Left<#1\Right>}      
\newcommand{\stavgt}[1]{\Leftt\langle#1\Rightt\rangle}
\newcommand{\pair}[2]{\savg{#1\,,\,#2}}
\newcommand{\Ltnorm}[1]{\norm{#1}}
\newcommand{\Ltnormt}[1]{\normt{#1}}
\newcommand{\tavgnormt}[1]{\overline{\normt{#1}\,\,}\!\!}
\newcommand{\tavgnormtwo}[1]{\overline{\norm{#1}^2}}

\newcommand{\brak}[2]{\l[#1\,,\,#2\r]}       

\newcommand{\Order}[1]{\mathrm{O}(#1)}

\renewcommand{\l}{\left}            
\renewcommand{\r}{\right}           
\renewcommand{\time}{t}             
\mathnotation{\pd}{\partial}        
\mathnotation{\grad}{\nabla}        
\renewcommand{\div}{\nabla\cdot}    
\mathnotation{\lapl}{\Delta}        
\mathnotation{\mlapl}{(-\Delta)}    
\mathnotation{\imi}{\mathit{i}}     
\mathnotation{\ldef}{\mathrel{\raisebox{.069ex}{:}\!\!=}}
\mathnotation{\rdef}{\mathrel{=\!\!\raisebox{.069ex}{:}}}
\mathnotation{\dint}{\,{\mathrm{d}}}

\mathnotation{\xc}{x}               
\mathnotation{\xv}{\bm{\xc}}        
\mathnotation{\yc}{y}               
\mathnotation{\kc}{k}               
\mathnotation{\kv}{{\bm{\kc}}}      
\mathnotation{\km}{\kc}             
\mathnotation{\uc}{u}               
\mathnotation{\uv}{\bm{\uc}}        
\mathnotation{\Uc}{U}               
\mathnotation{\Lsc}{L}              
\mathnotation{\Vol}{\Omega}         
\mathnotation{\dVol}{\dint\Vol}     
\mathnotation{\mVol}{\lvert\Vol\rvert}
\mathnotation{\sdim}{d}             

\mathnotation{\Eff}{\mathcal E}      
\mathnotation{\LL}{{\mathcal L}}    
\mathnotation{\Var}{\text{Var}}     
\mathnotation{\src}{s}              
\mathnotation{\srck}{{\hat \src}_\kv}
\mathnotation{\Pe}{\mathrm{Pe}}     
\mathnotation{\qq}{q}               
\mathnotation{\pexp}{\qq}           
\mathnotation{\fw}{f}               
\mathnotation{\cw}{c}               
\mathnotation{\gt}{g}               
\mathnotation{\T}{T}                
\mathnotation{\Asigma}{\mathcal{A}} 
\mathnotation{\Ltwo}{L^2}           
\mathnotation{\Linf}{L^\infty}      
\mathnotation{\Sobo}{H}             
\mathnotation{\hSobo}{\dot\Sobo}    

\mathnotation{\mixnxint}{x'}
\mathnotation{\mixnd}{d}
\mathnotation{\mixnp}{x}
\mathnotation{\mixns}{w}
\mathnotation{\mixnPhi}{\Phi}

\mathnotation{\Cu}{C}               
\mathnotation{\NK}{K}               

\mathnotation{\cA}{\mathcal A}      %
\mathnotation{\cAz}{{\widetilde \cA}}
\mathnotation{\LLz}{\widetilde \LL}
\mathnotation{\thetaz}{{\widetilde \theta}}
\mathnotation{\rsrc}{r}             

\mathnotation{\ac}{\alpha}
\mathnotation{\acv}{\bm{\alpha}}
\mathnotation{\nc}{n}
\mathnotation{\afunc}{W}
\mathnotation{\elm}{z}
\mathnotation{\Alm}{\eta}
\mathnotation{\timef}{\time_f}
\mathnotation{\RRc}{R}
\mathnotation{\RRv}{\bm{\RRc}}
\mathnotation{\Act}{A}

\mathnotation{\Proj}{\mathbb{P}}
\mathnotation{\ftheta}{\phi}
\mathnotation{\dunorm}{\gamma}
\mathnotation{\ev}{\Lambda}
\mathnotation{\uve}{\uv_{\mathrm{e}}}
\mathnotation{\uvp}{\uv_{\mathrm{p}}}

\mathnotation{\K}{K}                
\mathnotation{\kms}{\kc_{\src}}     

\mathnotation{\xx}{\kc}      
\mathnotation{\Oms}{\dunorm} 
\mathnotation{\Tms}{\lambda} 
\mathnotation{\entr}{\chi}   
\mathnotation{\Aa}{A}
\mathnotation{\NN}{N}        
\mathnotation{\Deff}{D}      
\mathnotation{\rexp}{\beta}  

\mathnotation{\ld}{\ell_{\mathrm{d}}}   
\mathnotation{\kd}{\kc_{\mathrm{d}}}    
\mathnotation{\lB}{\ell_{\mathrm{B}}}   
\mathnotation{\lu}{\ell_{\uc}}          
\mathnotation{\ls}{\ell_{\src}}         
\mathnotation{\xits}{\xi_{\theta,\src}} 
\mathnotation{\ci}{c_1}
\mathnotation{\cii}{c_2}
\mathnotation{\ciii}{c_3}
\mathnotation{\Peu}{\Pe_{\uc}}
\mathnotation{\luls}{\delta}

\mathnotation{\Deq}{D^{\mathrm{(eq)}}}  
\mathnotation{\Deqt}{\mathbb{D}^{\mathrm{(eq)}}}  
\mathnotation{\Src}{S}                  
\mathnotation{\ssrc}{\hat\src}          
\mathnotation{\Chomo}{C}                
\mathnotation{\xhc}{\xi}                
\mathnotation{\xhv}{\bm{\xhc}}          
\mathnotation{\timeh}{\tau}             
\mathnotation{\eye}{\mathbb{I}}         
\mathnotation{\Defft}{\mathbb{D}}       

\mathnotation{\func}{\mathcal{F}}

\begin{document}

\review[Using multiscale norms to quantify mixing and transport]
  {Using multiscale norms to quantify \\ mixing and transport}

\author{Jean-Luc Thiffeault}

\address{Department of Mathematics, University of Wisconsin --
  Madison, \\ Madison, WI, USA}
\ead{jeanluc@math.wisc.edu}

\begin{abstract}
  Mixing is relevant to many areas of science and engineering,
  including the pharmaceutical and food industries, oceanography,
  atmospheric sciences, and civil engineering.  In all these
  situations one goal is to quantify and often then to improve the
  degree of homogenisation of a substance being stirred, referred to
  as a passive scalar or tracer.  A classical measure of mixing is the
  variance of the concentration of the scalar, which can be related to
  the $\Ltwo$ norm of the concentration field.  Recently other norms
  have been used to quantify mixing, in particular the \emph{mix-norm}
  as well as negative Sobolev norms.  These norms have the advantage
  that unlike variance they decay even in the absence of diffusion,
  and their decay corresponds to the flow being mixing in the sense of
  ergodic theory.  General Sobolev norms weigh scalar gradients
  differently, and are known as \emph{multiscale norms} for mixing.
  We review the applications of such norms to mixing and transport,
  and show how they can be used to optimise the stirring and mixing of
  a decaying passive scalar.  We then review recent work on the
  less-studied case of a continuously-replenished scalar field --- the
  source-sink problem.  In that case the flows that optimally reduce
  the norms are associated with transport rather than mixing: they
  push sources onto sinks, and vice versa.
\end{abstract}



\section{Introduction}
\label{sec:intro}

One of the most vexing questions about fluid mixing is how to measure
it.  People typically know it when they see it, but specific
applications require customised measures.  For example, a measure
might be too fine-grained for some applications that don't require
thorough mixing.  Some measures, such as residence time distributions,
are designed for open-flow situations where fluid particles are only
mixed for a certain amount of time.  Others, such as the rigorous
definition of a mixing flow in ergodic theory, are better suited to an
idealised mathematical treatment.  Finally, one of the main points of
this review is that measures used to quantify mixing in the
initial-value decaying problem must be interpreted very differently
when sources and sinks are present.

One of the earliest attempts to quantify mixing was by the chemical
engineer and bomb disposal officer Peter
V.~Danckwerts~\cite{Denbigh1986}.  Danckwerts realised that scale was
an important consideration; he identified the large-scale breakup of
fluid into clumps and the subsequent homogenisation at small scales
due to diffusion as separate processes~\cite{Danckwerts1952}:
\begin{quote}
  The breaking-up and the interdiffusion are, in the case of liquids,
  largely independent processes which produce distinguishable results.
  The former reduces the size of the clumps, while the latter tends to
  obliterate differences of concentration between neighbouring regions
  of the mixture. It therefore seems desirable to use two quantities
  to describe the degree of mixing --- namely the \emph{scale of
    segregation} and the \emph{intensity of segregation}.
\end{quote}
Two other pioneers are the oceanographers Carl
Eckart~\cite{Eckart1948} and Pierre Welander~\cite{Welander1955}, who
also identified the complementary roles of mechanical stirring and
diffusion.  Following Eckart, modern parlance refers to these two
stages as \emph{stirring} and \emph{mixing}, the distinguishing
feature being that stirring is a mechanical action, whilst mixing is
diffusion-driven or the result of coarse-graining.  Welander in
particular was emphatic about the important role of stirring in
creating \emph{filaments}, which can subsequently be smoothed by
diffusion.  It was then Batchelor~\cite{Batchelor1959} who identified
the length scale of the filaments, at which stirring and diffusion
achieve a balance, now known as the Batchelor scale.

Let us pin a mathematical meaning on the ideas above.  Danckwert's
scale of segregation is a correlation length of the concentration of a
mixture.  His intensity of segregation is a normalised variance of the
concentration.  It is through the variance that the connection to
norms first appears, in this case as the~$\Ltwo$-norm of the
concentration field.  If~$\theta(\xv,\time)$ is the concentration of a
passive scalar --- such as temperature, dye, or salt --- then the
variance is
\begin{equation}
  \Var\,\theta = \Ltnorm{\theta}^2 - \savg{\theta}^2,
\end{equation}
where
\begin{equation}
  \Ltnorm{\theta}^2 = \frac{1}{\mVol}\int_\Vol\theta^2\dVol,\qquad
  \savg{\theta} = \frac{1}{\mVol}\int_\Vol\theta\dVol
  \label{eq:L2andmean}
\end{equation}
are the~$\Ltwo$-norm and mean value of~$\theta$, respectively,
with~$\Vol$ the spatial domain and~$\mVol$ its volume (or area in two
dimensions).

Why is the variance a good measure of mixing quality?  A first answer
is that it measures fluctuations from the mean, and a mixed state is
exactly one where the concentration is equal to the mean --- \ie, it
is uniform.  But there is a second, more intimate reason why variance
is important: from the classical advection-diffusion equation for an
incompressible velocity field~$\uv(\xv,\time)$ and diffusion
coefficient~$\kappa$,
\begin{equation}
  \frac{\pd\theta}{\pd\time} + \uv\cdot\grad\theta =
  \kappa\lapl\theta,
  \qquad \div\uv=0,
  \label{eq:AD}
\end{equation}
we find that the concentration variance is monotonically driven to
zero in the absence of sources (see~\sref{sec:norms}).  Indeed, after
a few integrations by parts and assuming boundary conditions that
conserve the total amount of~$\theta$ (no-flux or periodic, see
\sref{sec:AD}), we find the~$\Ltwo$-norm and mean~\eref{eq:L2andmean}
obey
\begin{equation}
  \frac{d}{d\time}\savg{\theta} = 0,
  \qquad
  \frac{d}{d\time}\Ltnorm{\theta}^2
  = -2\kappa\Ltnorm{\grad\theta}^2,
\end{equation}
so that
\begin{equation}
  \frac{d}{d\time}\Var\,\theta = -2\kappa\Ltnorm{\grad\theta}^2.
  \label{eq:vardecay}
\end{equation}
Observe that the right-hand side is negative-definite
unless~$\theta=\text{const.}$, \ie, the concentration is uniform.
Hence, the advection--diffusion equation says that the concentration
is driven to a uniform state with~$\Var\,\theta=0$, at a rate dictated
by the product~$2\kappa\lVert\grad\theta\rVert_2^2$.  Determining this
rate is an important aspect of the \emph{scalar decay} or
\emph{initial value} problem, where we have some initial concentration
field~$\theta(\xv,\time)=\theta_0(\xv)$ and want to know how fast it
is mixed by a velocity field~$\uv(\xv,\time)$.  There is a vast
literature focused on determining and estimating this decay rate.
Note that the advecting velocity field does not appear directly
in~\eref{eq:vardecay}: its role is to increase gradients of
concentration.

Thus, monitoring variance is a simple way of quantifying the
effectiveness of a mixing process.  It has, as mentioned above, the
dual advantages of being intuitive and of being mathematically sound.
In what situations, then, is the variance a less-than-ideal measure of
mixing quality?  The answer is: when the diffusivity~$\kappa$ is very
small and the stirring process is very effective.  The typical
situation in that case is that the decay
term~$2\kappa\lVert\grad\theta\rVert_2^2$ becomes \emph{independent}
of~$\kappa$.  The physical picture is that the stirring sharpens
gradients of~$\theta$ until they are large enough that diffusion
easily smooths them out.  A balance between advection and diffusion is
then reached, and the variance decays at its optimal rate.

In theory, this is fine; in practice, it is disastrous.  If we are
trying to optimise the mixing process, it means we have to keep track
of scales down to lengths of order~$\kappa^{1/2}$, which can be very
small.  Our simulations are limited by the small quantity~$\kappa$,
and yet the decay rate is independent of~$\kappa$.  However, if we try
and avoid this by setting~$\kappa=0$ in~\eref{eq:AD}, we find that
variance is exactly conserved.  The limit is singular: taking~$\kappa$
to zero is completely different from setting~$\kappa=0$.

This problem is similar to what happens for turbulence: letting the
viscosity tend to zero in the Navier--Stokes equations does not
necessarily recover solutions to Euler's equations.  However, here the
problem is less severe, since the concentration~$\theta$ does not feed
back on the flow.  Hence, it is sensible to solve the pure advection
equation
\begin{equation}
  \frac{\pd\theta}{\pd\time} + \uv\cdot\grad\theta = 0
  \label{eq:A}
\end{equation}
and try to extract some measure of mixing from~$\theta$; it's just not
possible to use variance as a measure.  The key is to use a different
norm that downplays the role of small scales.  For example, a simple
choice is~$\lVert\grad^{-1}\theta\rVert_2^2$.  The
operator~$\grad^{-1}$ will be defined more carefully later, but for
now it suffices to understand that it `smooths out' $\theta$, so that
small-scale variations are not detected.  (We make sure the operator
is well-defined by restricting to functions with vanishing
mean,~$\savg{\theta}=0$.)

\begin{figure}
\begin{indented}
\item[]
\begin{center}
\includegraphics[width=.8\textwidth]{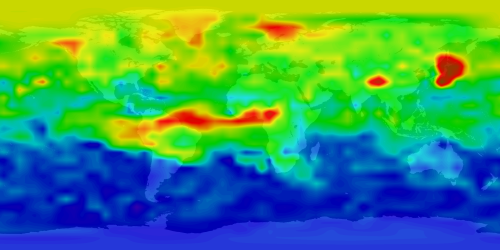}
\end{center}
\end{indented}
\caption{Satellite observations showing concentrations of carbon
  monoxide (CO) in the atmosphere. Red corresponds to high levels of
  CO (450 parts per billion) and blue to low levels (50 ppb). Note the
  immense clouds due to grassland and forest fires in Africa and South
  America (photo from NASA/NCAR/CSA).}
\label{fig:globalmar0323}
\end{figure}

So far our discussion has been of equation~\eref{eq:AD} when the total
quantity of scalar is conserved, \ie, $\savg{\theta}=\text{const}$.
This is the case most often discussed in the literature, and in recent
years much work has gone into understanding the factors affecting the
decay rate of the concentration and its variance.  There is also a
vast literature on the case where there are sources and sinks, either
within the fluid domain or at its boundaries.  In geophysical and
industrial situations this is often more relevant.
\Fref{fig:globalmar0323} shows a satellite image of the concentration
%
of carbon monoxide in the atmosphere.  Observe that there are several
large and many small sources, distributed in a complex manner
throughout the globe.  This begs the question: what do we mean by
well-mixed in a case as in \fref{fig:globalmar0323}?  Unmixed features
are persistent, since they are replenished by an inhomogeneous source.
But we can still use a norm-measure such as the variance: it will not
tend towards zero with time, but a smaller value of variance still
indicates that the passive scalar is getting mixed to some degree.  A
stirring flow that is effective at mixing will thus presumably tend to
reduce the variance of the concentration field.

\begin{figure}
\begin{indented}
\item[]
\begin{center}
\subfigure[]{
\includegraphics[width=.35\textwidth]{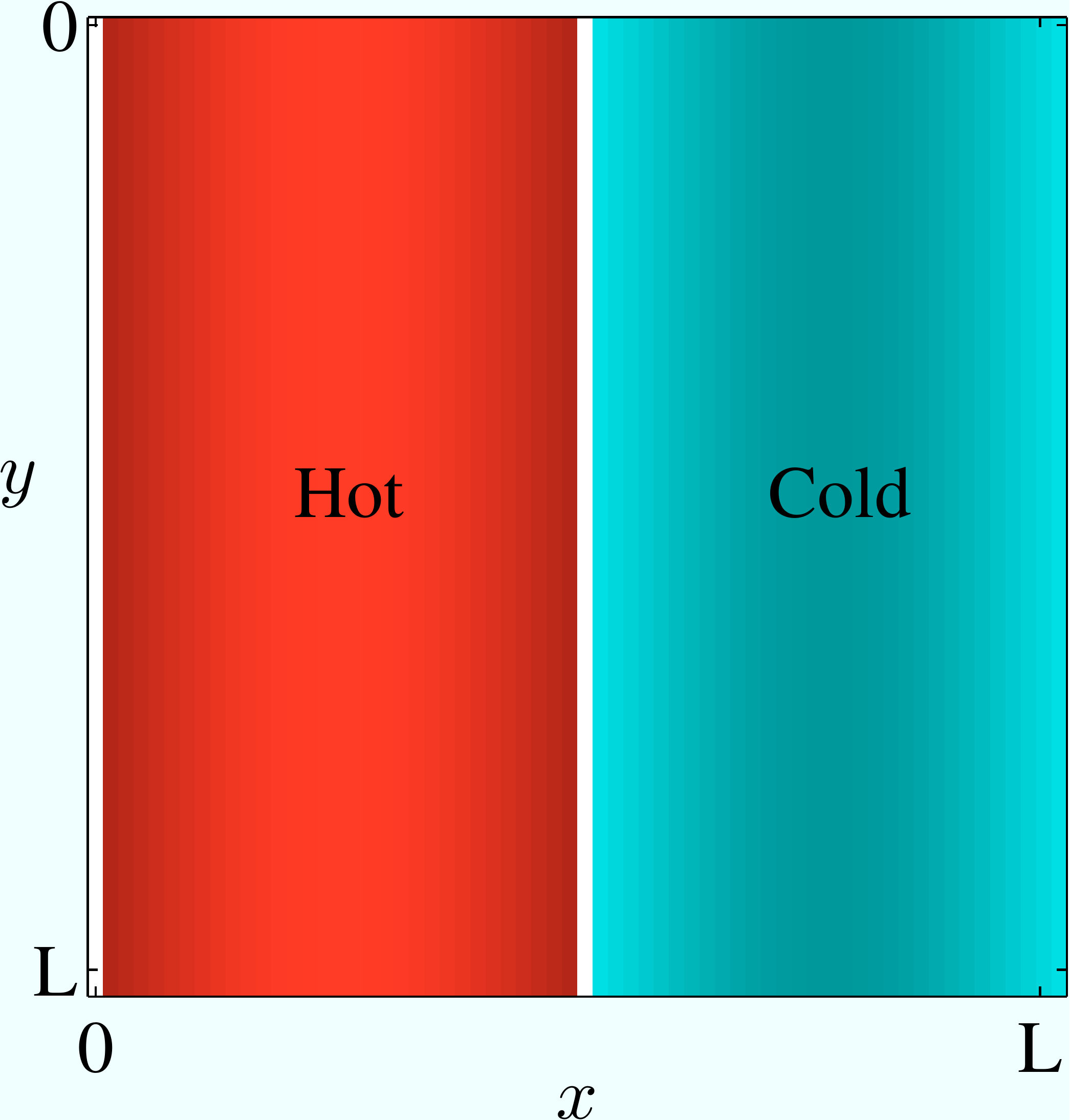}
\label{fig:hotcold_src}
}\hspace{.01\textwidth}
\subfigure[]{
\includegraphics[width=.35\textwidth]{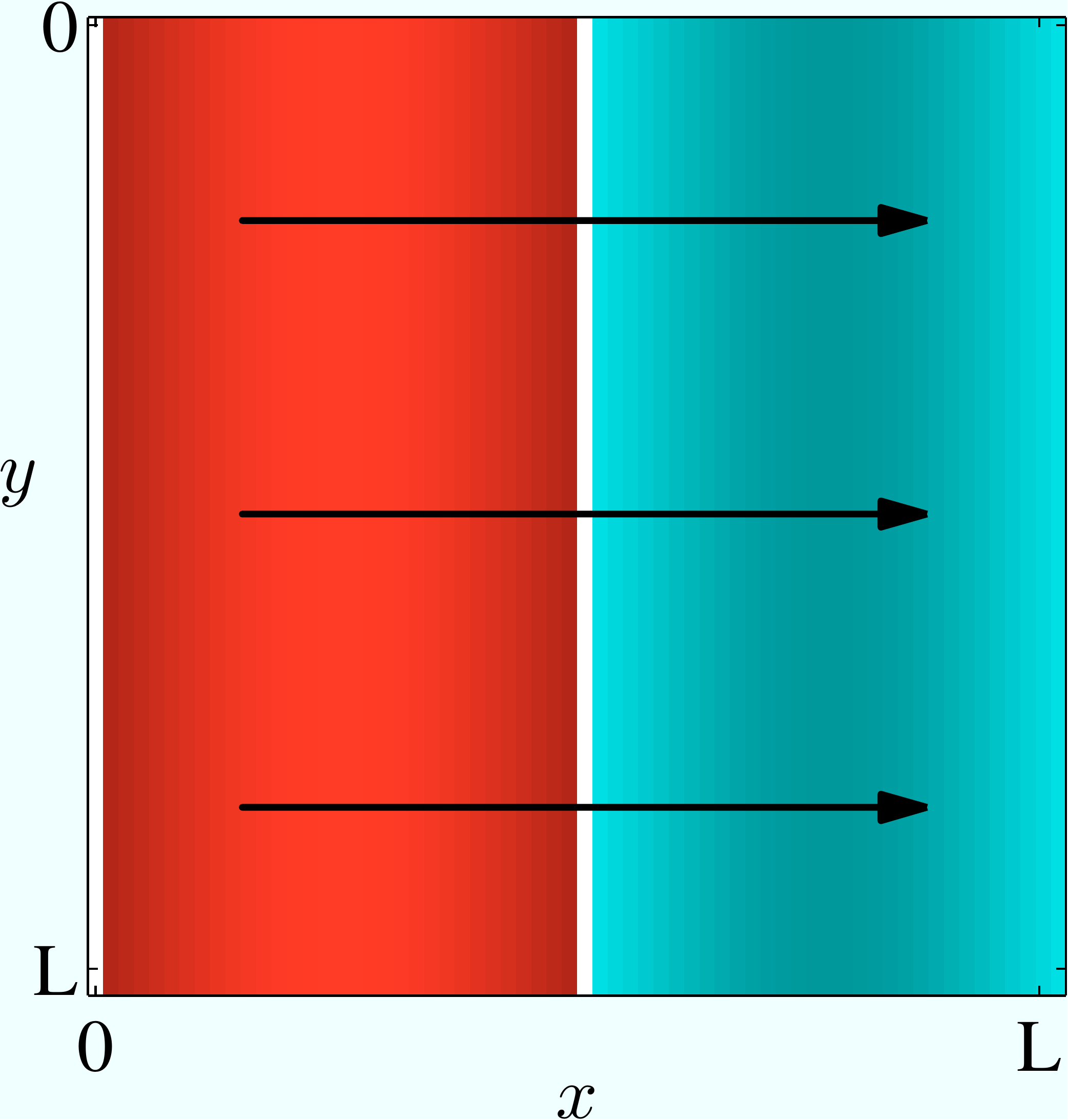}
\label{fig:hotcold_flow}
}
\end{center}
\end{indented}
\caption{(a) A source-sink distribution in a two-dimensional periodic
  square domain.  (b) Sketch of the velocity field that most
  effectively reduces the concentration variance of the source.}
\label{fig:hotcold}
\end{figure}

It has become apparent in recent years that the quality of mixing we
obtain will depend strongly on the source-sink distribution, in
addition to depending on the stirring flow itself.  Let us give a
simple but extreme example of this, discussed by Plasting \&
Young~\cite{Plasting2006} and Shaw~\etal\cite{Shaw2007}.  Consider a
two-dimensional biperiodic square domain where sources and sinks are
present, as in \fref{fig:hotcold_src}.
The equation to be solved is the analogue of~\eref{eq:AD} with a
body source term:
\begin{equation}
  \frac{\pd\theta}{\pd\time} + \uv\cdot\grad\theta =
  \kappa\lapl\theta + \src(\xv,\time),
  \label{eq:ADs0}
\end{equation}
with periodic boundary conditions.  Here we choose a simple
source-sink distribution, $\src(\xv)=\sin(2\pi\xc/\Lsc)$.  If we fix
the kinetic energy of the flow,
$\tfrac12\Ltnorm{\rho\uv}^2$,\footnote{We consider incompressible
  flows with constant density throughout, so that fixing the kinetic
  energy is equivalent to fixing the~$\Ltwo$-norm $\Ltnorm{\uv}^2$.
  For that reason, we shall often refer to~$\Ltnorm{\uv}^2$ itself as
  the kinetic energy.}  what is the incompressible stirring velocity
field that will most effectively reduce the variance?  The answer
sketched in \fref{fig:hotcold_flow} is somewhat surprising: the most
effective stirring consists of a uniform (constant) flow.  This flow
carries the hot fluid onto the cold sink, and cold onto hot, thereby
reducing the concentration variance as much as possible.

The reason this is surprising is that `common wisdom' in mixing
assumes that the best stirring is either turbulent or exhibits chaotic
trajectories~\cite{Aref1984,Ottino}.  Such complex behaviour increases
concentration gradients and thus allows diffusion to act more
effectively.  However, this particular source-sink configuration is
best mixed not by creating small scales, but rather by transporting
fluid appropriately.  One might not call this `mixing' in the strict
sense, but it is dependent on diffusion, as \eref{eq:ADs0} does not
converge to a steady-state without it.

Mixing a scalar field whose fluctuations are constantly replenished by
spatially inhomogeneous sources and sinks is a problem with a long
history.  Townsend~\cite{Townsend1951,Townsend1954} was concerned with
the effect of turbulence and molecular diffusion on a line source of
temperature --- a heated filament.  The spatial localisation of the
source, imposed by experimental constraints, enhanced the role of
molecular diffusivity.  Durbin~\cite{Durbin1980} and
Drummond~\cite{Drummond1982} introduced stochastic particle models to
turbulence modelling, and these allowed more detailed studies of the
effect of the source on diffusion.  Sawford \& Hunt~\cite{Sawford1986}
pointed out that small sources, such as heated filaments, lead to an
explicit dependence of the variance on molecular diffusivity.  Many
refinements to these models followed, see for
instance~\cite{Thomson1990, Antonsen1991, Borgas1994, Antonsen1996}
and the review by Sawford~\cite{Sawford2001}.  Chertkov
\etal\cite{Chertkov1995,Chertkov1995b,Chertkov1997,Chertkov1997b,%
  Chertkov1998} and Balkovsky \& Fouxon~\cite{Balkovsky1999} treated
the case of a random, statistically-steady source.

We now give a brief outline of the review.  In \sref{eq:AD} we recall
some basic properties of the advection-diffusion equation, to
complement the earlier material in this introduction.
\Sref{sec:norms} is devoted to a review of Sobolev norms.  The rest of
the review is divided into two parts.  Part I discusses norms as
measures of mixing for the freely-decaying passive scalar.
\Sref{sec:mixing} connects norms with the concept of mixing in the
sense of ergodic theory.  In \sref{sec:optdecay} we show how negative
Sobolev norms can be used to optimise flows to achieve rapid mixing.

Part II, which forms the bulk of the paper, is devoted to advection
and diffusion in the presence of sources and sinks.  In
\sref{sec:Meff} we introduce \emph{mixing efficiencies}, measures of
mixing based on norms.  We give some upper bounds on these
efficiencies in \sref{sec:mixing}.  In \sref{sec:SHIF} we investigate
the dependence of efficiencies on functional features of the
source-sink distribution.  We derive mixing efficiencies from a
homogenisation theory approach in \sref{sec:homo}.  In
\sref{sec:optsrc} we discuss optimisation of mixing efficiencies.
Finally, we offer some closing comments in \sref{sec:discussion}.

\section{The advection--diffusion equation}
\label{sec:AD}

The main equation discussed in this review is the advection--diffusion
equation for a passive scalar with concentration
field~$\theta(\xv,\time)$,
\begin{equation}
  \frac{\pd\theta}{\pd\time} + \uv\cdot\grad\theta =
  \kappa\lapl\theta + \src(\xv,\time),
  \qquad \div\uv=0.
  \label{eq:ADs}
\end{equation}
Following the usual route, we will assume that the domain of
interest~$\Vol$ is a periodic square box with dimension~$\sdim$, with
spatial period~$\Lsc$ in each direction.  Of course, everything
discussed in this paper can be repeated for a more general closed
domain with no-flux boundary conditions, but this adds little to the
discussion.  In particular, by using a periodic domain we can use
Fourier series expansions, which makes many calculations explicit.

The mean~$\savg{\theta}$ satisfies
\begin{equation}
  \frac{d}{d\time}\savg{\theta} = \savg{\src},
\end{equation}
with solution
\begin{equation}
  \savg{\theta}\!(\time) =\savg{\theta}\!(0) +
  \int_0^\time\savg{\src}\!(\time')\dint\time'.
\end{equation}
Thus, if we replace~$\theta$ by a new variable
\begin{equation}
  \theta'(\xv,\time) = \theta(\xv,\time)
  - \savg{\theta}\!(\time)
\end{equation}
then~$\theta'$ obeys the modified equation
\begin{equation}
  \frac{\pd\theta'}{\pd\time} + \uv\cdot\grad\theta' =
  \kappa\lapl\theta' + \src'(\xv,\time),
  \qquad \src'(\xv,\time) = \src(\xv,\time)
  - \frac{d}{d\time}\savg{\theta}.
\end{equation}
The new concentration field~$\theta'$ and source~$\src'$ have spatial
mean zero.  For the remainder of this paper, we drop the primes and
assume without loss of generality that both~$\theta$ and~$\src$ have
zero spatial mean.

\section{Norms}
\label{sec:norms}

\subsection{Definitions and basic properties}

In this section we introduce the measures of mixing we'll be using for
the rest of the paper.
The Sobolev norm we use for the space~$\Sobo^\qq(\Vol)$ is
\begin{equation}
  \norm{f}_{\Sobo^\qq} = \l(\frac{1}{\mVol}\int_\Vol
  \lvert (1 - \Lsc^2\lapl)^{\qq/2}f\rvert^2 \dVol\r)^{1/2},
  \label{eq:Sobo}
\end{equation}
or in terms of Fourier series,
\begin{equation}
  \norm{f}_{\Sobo^\qq} = \Bigl(
  \sum_\kv(1 + \km^2\Lsc^2)^{\qq}\,\lvert {\hat f}_\kv\rvert^2\Bigr)^{1/2},
  \label{eq:Sobok}
\end{equation}
where~$\km\ldef\lvert\kv\rvert$ and~${\hat f}_\kv$ are the Fourier
coefficients.  In this review we will prefer to use the seminorm on
the homogeneous space~$\hSobo^\qq(\Vol)$ (note the dot over~$\Sobo$),
\begin{equation}
  \norm{f}_{\hSobo^\qq}
  = \l(\frac{1}{\mVol}\int_\Vol \lvert\mlapl^{\qq/2} f\rvert^2\dVol\r)^{1/2}
  = \Ltnorm{\mlapl^{\qq/2} f},
  \label{eq:hSobo}
\end{equation}
or in terms of Fourier series
\begin{equation}
  \norm{f}_{\hSobo^\qq} = \Bigl(
  \sum_\kv\km^{2\qq}\, \lvert {\hat f}_\kv\rvert^2\Bigr)^{1/2}.
  \label{eq:hSobok}
\end{equation}
If we take a function~$f$ with Fourier coefficients behaving
asymptotically as~$\lvert {\hat f}_\kv\rvert \sim \km^p$, $\km\gg1$,
then the norms~\eref{eq:Sobo}--\eref{eq:hSobok} converge (exist)
for~$\qq+p<-\sdim/2$, where~$\sdim$ is the dimension of space.

The norm~\eref{eq:hSobo} has a more intimate connection with solutions
of the advection--diffusion equation than~\eref{eq:Sobo}, as will be
described later (\sref{sec:timeevol}).  Note that the manner in which
we have defined~\eref{eq:Sobo} and~\eref{eq:hSobo} allows for~$\qq$
positive, negative, or even fractional.  In \sref{sec:AD} we showed
that we could assume~$\savg{\theta}=0$, so we restrict attention to
functions~$f$ with mean zero.  In that case~\eref{eq:hSobo} becomes a
true norm since~$\norm{f}_{\hSobo^\qq}=0$ if and only if~$f$ is zero.

In fact it does not matter much which of the two norms~\eref{eq:Sobo}
or~\eref{eq:hSobo} we use, since they are equivalent for zero-mean
functions: by Poincar\'e's inequality, we have~$\Ltnorm{\mlapl^{\qq/2}
  f} \ge (2\pi/\Lsc)^\qq\Ltnorm{f}$ for~$\qq\ge 0$, so that
\begin{equation}
  \norm{f}_{\hSobo^\qq} \le \Lsc^{-\qq}\norm{f}_{\Sobo^\qq}
  \le (1 + (2\pi)^{-2})^{\qq/2}\norm{f}_{\hSobo^\qq},
  \qquad
  \qq \ge 0,
\end{equation}
for all zero-mean functions~$f$.  For~$\qq<0$ Poincar\'e's inequality
is reversed, so we have
\begin{equation}
  (1 + (2\pi)^{-2})^{\qq/2}\norm{f}_{\hSobo^\qq}
  \le \Lsc^{-\qq}\norm{f}_{\Sobo^\qq}
  \le \norm{f}_{\hSobo^\qq},
  \qquad
  \qq < 0.
\end{equation}
Equivalence means that if one of the two equivalent norms goes to
zero, then the other must as well, and they must do so at the same
rate~\cite{Mathew2005,Mathew2007}.

For the mathematically minded, we can give a rigorous definition of
what kind of functions live in the negative Sobolev
space~$\hSobo^\qq$, with~$\qq<0$, given that we understand the
space~$\hSobo^{-\qq}$.  The space~$\hSobo^\qq$ is defined as the space
dual to~$\hSobo^{-\qq}$ with respect to the standard pairing
\begin{equation}
  \pair{f}{g} = \frac{1}{\mVol}\int_\Vol f(\xv)\, g(\xv)\dVol,
  \label{eq:pair}
\end{equation}
with the dual norm
\begin{equation}
  \norm{f}_{\hSobo^{-\qq *}} = \sup_{g \in \hSobo^{-\qq}}
  \frac{\pair{f}{g}}{\norm{g}_{\hSobo^{-\qq}}}\,.
  \label{eq:dualSobo}
\end{equation}
The norm~\eref{eq:hSobo} is equal to the dual norm~\eref{eq:dualSobo}.
To show this, first observe that by using the Cauchy--Schwarz
inequality,
\begin{equation}
  \frac{\pair{f}{g}}{\norm{g}_{\hSobo^{-\qq}}}
  =
  \frac{\pair{\mlapl^{\qq/2}f}{\mlapl^{-\qq/2}g}}
  {\norm{g}_{\hSobo^{-\qq}}}
  \le
  \norm{f}_{\hSobo^\qq}
\end{equation}
independent of~$g$, so~$\norm{f}_{\hSobo^{-\qq *}} \le
\norm{f}_{\hSobo^\qq}$.  Now let~$g=\mlapl^{2\qq}f$:
\begin{equation}
  \frac{\pair{f}{g}}{\norm{g}_{\hSobo^{-\qq}}}
  =
  \frac{\pair{f}{\mlapl^{2\qq}f}}
  {\norm{\mlapl^{2\qq}f}_{\hSobo^{-\qq}}}
  =
  \frac{\Ltnorm{\mlapl^{\qq}f}^2}{\Ltnorm{\mlapl^{\qq}f}}
  =
  \norm{f}_{\hSobo^\qq}.
\end{equation}
Since the dual norm is defined as a~$\sup$ over~$g$, we
have~$\norm{f}_{\hSobo^{-\qq *}} \ge \norm{f}_{\hSobo^\qq}$.  We
conclude that~$\norm{f}_{\hSobo^{-\qq *}} = \norm{f}_{\hSobo^\qq}$.
The same argument can also be used to show~$\norm{f}_{\Sobo^{-\qq *}}
= \norm{f}_{\Sobo^\qq}$ for the inhomogeneous spaces.

\subsection{Evolution in time}
\label{sec:timeevol}

To get a feel for what these norms are telling us about mixing, it is
helpful to examine how they evolve in time.  That is, given
that~$\theta$ obeys~\eref{eq:ADs}, what
is~$d\norm{\theta}_{\hSobo^\qq}/d\time$?  We start from
\begin{equation}
  \tfrac12\frac{d}{d\time}\norm{\theta}_{\hSobo^\qq}^2
  = \savg{\mlapl^{\qq}\theta\,\pd_\time\theta}
\end{equation}
where~$\savg{\cdot}$ denotes an average over the periodic
domain~$\Vol$.  Inserting~\eref{eq:ADs} for~$\pd_\time\theta$,
\begin{align*}
  \tfrac12\frac{d}{d\time}\norm{\theta}_{\hSobo^\qq}^2
  &= \savg{\mlapl^{\qq}\theta\,(-\uv\cdot\grad\theta +
  \kappa\lapl\theta + \src)} \\
  &= -\savg{\mlapl^{\qq}\theta\,\uv\cdot\grad\theta}
   - \kappa\Ltnorm{\mlapl^{(\qq+1)/2}\theta}^2
   + \savg{\mlapl^{\qq}\theta\,\src}.
\end{align*}
The case~$\qq=0$ gives the evolution of the variance, for which the
velocity term on the right integrates away:
\begin{equation}
  \tfrac12\frac{d}{d\time}\norm{\theta}_{\hSobo^0}^2
  =
  - \kappa\Ltnorm{\grad\theta}^2 + \savg{\theta\,\src}.
   \label{eq:q0}
\end{equation}
There are two other cases that give particularly nice equations.  The
case~$\qq=1$ gives the evolution of scalar concentration gradients:
\begin{equation}
  \tfrac12\frac{d}{d\time}\norm{\theta}_{\hSobo^1}^2
  = -\savg{\grad\theta\cdot\grad\uv\cdot\grad\theta}
   - \kappa\Ltnorm{\lapl\theta}^2
   - \savg{\lapl\theta\,\src}.
   \label{eq:q1}
\end{equation}
The first term on the right is the familiar `stretching' term, which
says that gradients are increased or decreased proportionally to their
alignment with the principal axes of the rate-of-strain tensor.  A
direction of positive strain will decrease gradients, whilst a
direction of negative strain will increase gradients.

The final case of interest to us is~$\qq=-1$, for which
\begin{equation}
  \tfrac12\frac{d}{d\time}\norm{\theta}_{\hSobo^{-1}}^2
  = \savg{\grad^{-1}\theta\cdot\grad\uv\cdot\grad^{-1}\theta}
   - \kappa\Ltnorm{\theta}^2
   - \savg{\lapl^{-1}\theta\,\src}.
   \label{eq:q-1}
\end{equation}
Here we interpret~$\grad^{-1}$ via its action on Fourier modes
\begin{equation}
  (\grad^{-1}\theta)_{\kv} = -\frac{\imi\kv}{\km^2}\,{\hat\theta}_\kv,
  \label{eq:gradinv}
\end{equation}
so that~$\grad\cdot\grad^{-1}\theta=\theta$.  (Recall that we are
restricting to functions with vanishing mean.)

Compare the first term on the right-hand side of~\eref{eq:q-1} to the
same term for~\eref{eq:q1}: velocity gradients have the opposite
effect on~$\frac{d}{d\time}\norm{\theta}_{\hSobo^{-1}}$ as they do
on~$\frac{d}{d\time}\norm{\theta}_{\hSobo^{1}}$.  This is intuitively clear:
the creation of concentration gradients will tend to make~$\theta$
very filamented.  We will see in~\sref{sec:mixing} that this will
cause it to converge weakly to zero, and that this implies that any
negative Sobolev norm must go to zero.  Another argument that the two
norms should evolve with opposite trends arises from
\begin{equation}
  \Ltnorm{\theta}^2
  = -\savg{\grad\theta\cdot\grad^{-1}\theta}
  \le \norm{\theta}_{\hSobo^{1}}\norm{\theta}_{\hSobo^{-1}}.
  \label{eq:varbound}
\end{equation}
Since the variance~$\Ltnorm{\theta}^2$ is conserved
when~$\kappa=\src=0$, inequality~\eref{eq:varbound} implies that if
the norm~$\norm{\theta}_{\hSobo^{-1}}$ converges to zero,
then~$\norm{\theta}_{\hSobo^{1}}$ must diverge.

%
%
\section*{\Large Part I: The decaying problem}

\section{Mixing in the sense of ergodic theory}
\label{sec:mixing}

A divergence-free velocity field~$\uv(\xv,\time)$ generates a
time-dependent function~$\theta(\xv,\time)$ via the advection equation,
\begin{equation}
  \frac{\pd\theta}{\pd\time} + \uv\cdot\grad\theta = 0,
  \qquad \div\uv=0,
  \label{eq:pureA}
\end{equation}
with initial condition~$\theta(\xv,0)=\theta_0(\xv) \in \Ltwo(\Vol)$.
Note that~\eref{eq:pureA}
preserves~$\Ltnorm{\theta(\cdot,\time)}=\Ltnorm{\theta_0}$ for all
time, so~$\Ltnorm{\theta(\cdot,\time)}$ is uniformly bounded in time
by~$\Ltnorm{\theta_0}$.  We have left out diffusion
in~\eref{eq:pureA}: in the present section we will discuss how we can
define mixing without appealing to diffusion, and how the
norms~\eref{eq:Sobo} and~\eref{eq:hSobo} can be related to this type
of mixing.  The definition of mixing we will use is the one from
ergodic theory.

The property of mixing in the sense of ergodic theory is a little
opaque when described mathematically.  The rigorous definition is as
follows:
\begin{quote}\emph{%
  Let~$(X,\Asigma,\mu)$ be a normalised measure space
  and~$S^\time:X\rightarrow X$ be a measure-preserving flow.
  $S^\time$ is called \emph{mixing} if
  \begin{equation*}
    \lim_{\time\rightarrow\infty}
    \mu(A \cap S^{-\time}(B)) = \mu(A)\,\mu(B),
    \qquad\text{for all } A, B \in \Asigma.
  \end{equation*}
}
\end{quote}
\noindent
Here~$X$ is our domain~$\Vol$, $\Asigma$ is a so-called
$\sigma$-algebra over~$X$, and~$\mu$ is a measure.  The elements
of~$\Asigma$ are measurable sets, which we can think of as patches or
`blobs' in~$\Vol$.  The measure~$\mu$ assigns a positive real number
to a set in~$\Asigma$.  We have been using Lebesgue measure ---
the~$\dVol$ that appears in our integrals --- which means the measure
of a blob is just its volume.  The flow~$S^\time$ for us
takes~$\theta_0$ to~$\theta(\cdot,\time)$ subject to~\eref{eq:pureA}.
The flow~$S^\time$ preserves volume, and hence Lebesgue measure.

Intuitively, the definition of mixing works as follows.  As we only
deal with reversible systems here, we can replace~$S^{-\time}(B)$
by~$S^\time(B)$ in the definition above, since such a system should be
mixing both forward and backward in time (this makes things
conceptually easier).  Think of~$A$ as a fixed reference patch, and
$B$ as a blob that gets stirred.  Since the transformation is
volume-preserving, $S^\time(B)$ might stretch and filament, but it
does not change its volume.  However, if it fills the domain
`uniformly,' then the volume of its intersection with~$A$ just ends up
being proportional to~$A$ and~$B$'s volume.  Crucially, this is true
for every set~$A$ and~$B$, so everything ends up everywhere.  (Note
that this is stronger than ergodicity, which only requires sets to
visit every point, but not necessarily be `everywhere at the same
time.')

Mathew~\etal\cite{Mathew2005,Mathew2007} have introduced a norm,
called the \emph{mix-norm}, which captures the property of mixing in
the sense of ergodic theory.  The mix-norm is somewhat cumbersome to
define on the torus, but to give an idea of its flavor we will
describe it for the one-dimensional periodic interval~$[0,\Lsc]$.
First, define
\begin{equation}
  \mixnd(\theta,\mixnp,\mixns) \ldef
    \frac{1}{\mixns}\int_{\mixnp-\mixns/2}^{\mixnp+\mixns/2}
  \theta(\mixnxint)\dint\mixnxint
\end{equation}
for all~$\mixnp$, $\mixns \in [0,\Lsc]$.  The function
$\mixnd(\theta,\mixnp,\mixns)$ is the mean value of the concentration
$\theta$ in an interval of width~$\mixns$ centred on~$\mixnp$.  The
mix-norm~$\mixnPhi(\theta)$ is then obtained by averaging~$\mixnd^2$
over~$\mixnp$ and~$\mixns$:
\begin{equation}
  \mixnPhi^2(\theta) \ldef
  \frac{1}{\Lsc^2}
  \int_0^\Lsc\int_0^\Lsc
  \mixnd^2(\theta,\mixnp,\mixns)\dint\mixnp\dint\mixns.
  \label{eq:mixnorm}
\end{equation}
In words, the mix-norm averages the concentration over an interval of
width~$\mixns$, then averages the square of this over all intervals
and all widths.  In dimensions greater than one, the definition of the
mix-norm involves integrals over balls of varying sizes instead of
intervals.

For our purposes here it suffices that the mix-norm~\eref{eq:mixnorm}
and its higher-dimensional generalisation are equivalent to the
norm~\eref{eq:Sobo} with~\hbox{$\qq=-1/2$}~\cite{Mathew2005}.  We will
also see below that all Sobolev norms with~$\qq<0$ capture the
property of mixing in the same sense as the mix-norm, as shown by
Lin~\etal\cite{Lin2011b}.  For that reason, we extend Mathew \etal's
terminology and often refer to any negative Sobolev norm as a
mix-norm, not just for the case~$\qq=-1/2$.  The term \emph{multiscale
  norm} encompasses the Sobolev norms for any value of~$\qq$ ---
positive, negative, or zero.

The connection between negative Sobolev norms and mixing involves the
property of \emph{weak convergence}.  We refer the reader to the book
by Lasota \& Mackey~\cite{Lasota} for a more complete discussion of
the relation between weak convergence and mixing.  Weak convergence is
defined as follows:
\begin{quote}\emph{%
  A time-dependent function $\fw(\xv,\time)$, $\fw(\cdot,\time) \in
  \Ltwo(\Vol)$, is \emph{weakly convergent} to~$\fw_\infty \in
  \Ltwo(\Vol)$ if
\begin{equation*}
  \lim_{\time\rightarrow\infty} \langle \fw(\cdot,\time),\gt\rangle
  = \langle \fw_\infty,\gt\rangle,
  \qquad
  \text{for all } \gt \in \Ltwo(\Vol).
\end{equation*}
}
\end{quote}
\noindent
Note that if~$\fw(\cdot,\time)$ converges weakly to~$\fw_\infty$,
then~$\fw(\cdot,\time)-\fw_\infty$ converges weakly to zero.

Instead of a time-dependent function, we can also use a discrete
sequence~$\{\cw_m\}$, where~$m$ is like time.  A simple example of a
sequence that converges weakly to zero is given
by~\hbox{$\cw_m=\sin(2\pi m\,\xc/\Lsc)$}, since
\begin{equation*}
  \lim_{m\rightarrow\infty}
  \int_0^\Lsc\sin(2\pi m\,\xc/\Lsc)\,\gt(\xc)\dint\xc = 0
\end{equation*}
for all~$\gt\in \Ltwo(\Vol)$, by the Riemann--Lebesgue lemma.  The
connection to mixing is that under stirring a passive scalar usually
develops finer and finer scales, much like the function~$\sin(2\pi
m\,\xc/\Lsc)$ with increasing~$m$.  In practice diffusion smooths out
these large gradients and the concentration field tends to zero at
every point.  However, if we ignore diffusion and retain the small
scales we can still detect this mixing process by `projecting' onto
test functions such as~$\gt$.  We now discuss the connection between
mixing in the sense of ergodic theory and the norms~\eref{eq:hSobo}.

The following is a slightly more general version of the mix-norm
theorem by Mathew~\etal\cite{Mathew2005} (the case~$\qq=-1/2$ is
equivalent to their theorem):
\begin{quote}\emph{%
  A time-dependent function $\fw(\xv,\time)$, where $\fw(\cdot,\time)
  \in \Ltwo(\Vol)$ has mean zero and is bounded in the~$\Ltwo$ norm
  uniformly in time, is weakly convergent to zero if and only if
\begin{equation*}
  \lim_{\time\rightarrow\infty} \norm{\fw(\cdot,\time)}_{\Sobo^\qq}=0,
  \qquad
  \text{for any } \qq < 0.
\end{equation*}
}
\end{quote}
\noindent The proof given by Lin~\etal\cite{Lin2011b} is reproduced
in~\ref{apx:weakconv}.

Another theorem from Mathew~\etal\cite{Mathew2005} now implies that the
dynamics generated by~$\uv(\xv,\time)$ are mixing in the sense of
ergodic theory if and only if~$\lim_{\time\rightarrow\infty}
\norm{\theta(\cdot,\time)}_{\hSobo^\qq}=0$, for any $\qq<0$. This is a
direct consequence of the mix-norm theorem above.

Note that the equivalence of the norms~$\norm{\cdot}_{\hSobo^\qq}$,
$\qq<0$, with mixing in the sense of ergodic theory is only a useful
concept for the freely-decaying problem.  In the presence of sources
and sinks, diffusion plays the essential role of making an asymptotic
state possible (in its absence solutions can diverge), so we cannot
simply solve~\eref{eq:pureA} with a source term on the right and
expect to get anything sensible.  

We close this section with a rule of thumb to interpret the decay of
mix-norms.  Consider a function~$f$ with Fourier coefficients~${\hat
  f}_\kv$, where the coefficients vanish when~$\kv$ contains odd
wavenumbers.  Now define~$f'$ by~${\hat f}'_\kv = {\hat f}_{(\kv/2)}$,
that is,~$f'$ is the same as~$f$ but with all scales divided by two.
From definition~\eref{eq:hSobok}, we have
\begin{equation}
  \norm{f'}_{\hSobo^\qq} = \Bigl(
  \sum_\kv\km^{2\qq}\, \lvert {\hat f}_{\kv/2}\rvert^2\Bigr)^{1/2}
  = \Bigl(
  \sum_{\kv'}(2\km')^{2\qq}\, \lvert {\hat f}_{\kv'}\rvert^2\Bigr)^{1/2}
  = 2^{\qq}\norm{f}_{\hSobo^\qq}\,.
\end{equation}
Thus, a refinement of scales by a factor of two leads to a decrease
in~$\norm{\cdot}_{\hSobo^\qq}$ of a factor~$2^{\qq}$ ($\qq<0$).
For~$\qq=-1$, the norm decreases by half when scales are refined by
half.  For~$\qq=-1/2$, the
\emph{mix-variance}~$\norm{\cdot}_{\hSobo^\qq}^2$ decreases by half
when scales are refined by half.  In both cases, the norms decrease at
a rate which reflects the creation of small scales.

\section{Optimisation for decaying problem}
\label{sec:optdecay}

\subsection{Optimal control}
\label{sec:optcontr}

Mathew~\etal\cite{Mathew2007} have used optimal control
techniques~\cite{Slemrod1978, Jurdjevic1978, Ball1982, Kirk} to find
velocity fields that rapidly reduce the norm~$\Sobo^{-1/2}$ of a
concentration field.  (See also~\cite{Sharma1997, DAlessandro1999,
  Vikhansky2002, Balogh2005, Stremler2006, Cortelezzi2008, Liu2008,
  Gubanov2010}.)  The energy of the flow is held fixed (more
precisely, its total action over a time interval).  They assume that
the velocity field can be expressed as a linear combination of steady
incompressible velocity fields~$\uv_i(\xv)$ as
\begin{equation}
  \uv(\xv,\time) = \sum_{i=1}^\nc \ac_i(\time)\,\uv_i(\xv),\qquad
  \div\uv_i=0.
  \label{eq:uvcontrol}
\end{equation}
The coefficients~$\ac_i(\time)$ the controls that are adjusted to
achieve the optimisation.  It is assumed that the
flow~\eref{eq:uvcontrol} can be realised in practice for a given set
of functions~$\ac_i(\time)$.  The quantity to be optimised is the
Sobolev norm~$\normt{\theta}_{\Sobo^{-1/2}}$ of a concentration field
satisfying the advection equation~\eref{eq:A}, for some initial
condition~$\theta_0(\xv)$.

In this formalism, the time-integrated energy (\ie, the
action~$\Act$) and the advection equation~\eref{eq:A} itself enter an
augmented functional as constraints:
\begin{multline}
  \afunc[\acv,\theta,\elm,\Alm] = \normt{\theta(\cdot,\timef)}_{\Sobo^{-1/2}}^2
  - \elm\l(\Act
  - \int_0^{\timef}\acv(\time)\cdot\RRv\cdot\acv(\time)\dint\time\r)\\
  - \int_0^{\timef}
  \savg{\Alm(\xv,\time)\l(\frac{\pd\theta(\xv,\time)}{\pd\time}
    + \sum_{i=1}^\nc\ac_i(\time)\,
    \uv_i(\xv)\cdot\grad\theta(\xv,\time)\r)}\dint\time,
  \label{eq:afunc}
\end{multline}
where~$\elm$ and~$\Alm(\xv,\time)$ are Lagrange multipliers, and the
matrix~$\RRv$ describes the kinetic energy for the individual velocity
fields in~\eref{eq:uvcontrol},
\begin{equation}
  \RRc_{ij} \ldef \tfrac12\savg{\uc_i(\xv)\,\uc_j(\xv)}.
  \label{eq:Rij}
\end{equation}
Note that the functional~\eref{eq:afunc} involves the norm of the
concentration field only at the final time~$\timef$.  However, the
constraints involve the entire history of~$\theta(\xv,\time)$.

Taking the functional (Fr\'echet) derivatives of~\eref{eq:afunc} with
respect to~$\acv$, $\theta$, $\elm$, $\Alm$ and equating to zero leads
to a two-point boundary value problem: $\theta(\xv,\time)$ is
specified at the initial time~$\time=0$, but $\Alm(\xv,\time)$ is
specified at the final time~$\timef$ (see Eq.~(3.2)
in~\cite{Mathew2007}).  Both~$\theta$ and~$\Alm$ (the `costate field')
satisfy advection equations of the form~\eref{eq:A}, so they can be
solved by following particles on Lagrangian trajectories (backwards in
time for~$\Alm$), following the velocity field given by the current
best guess for~$\acv(\time)$.  This guess can be varied following an
iterative procedure, such as the conjugate gradient method.

\begin{figure}
\begin{indented}
\item[]
\begin{center}
\includegraphics[width=.8\textwidth]{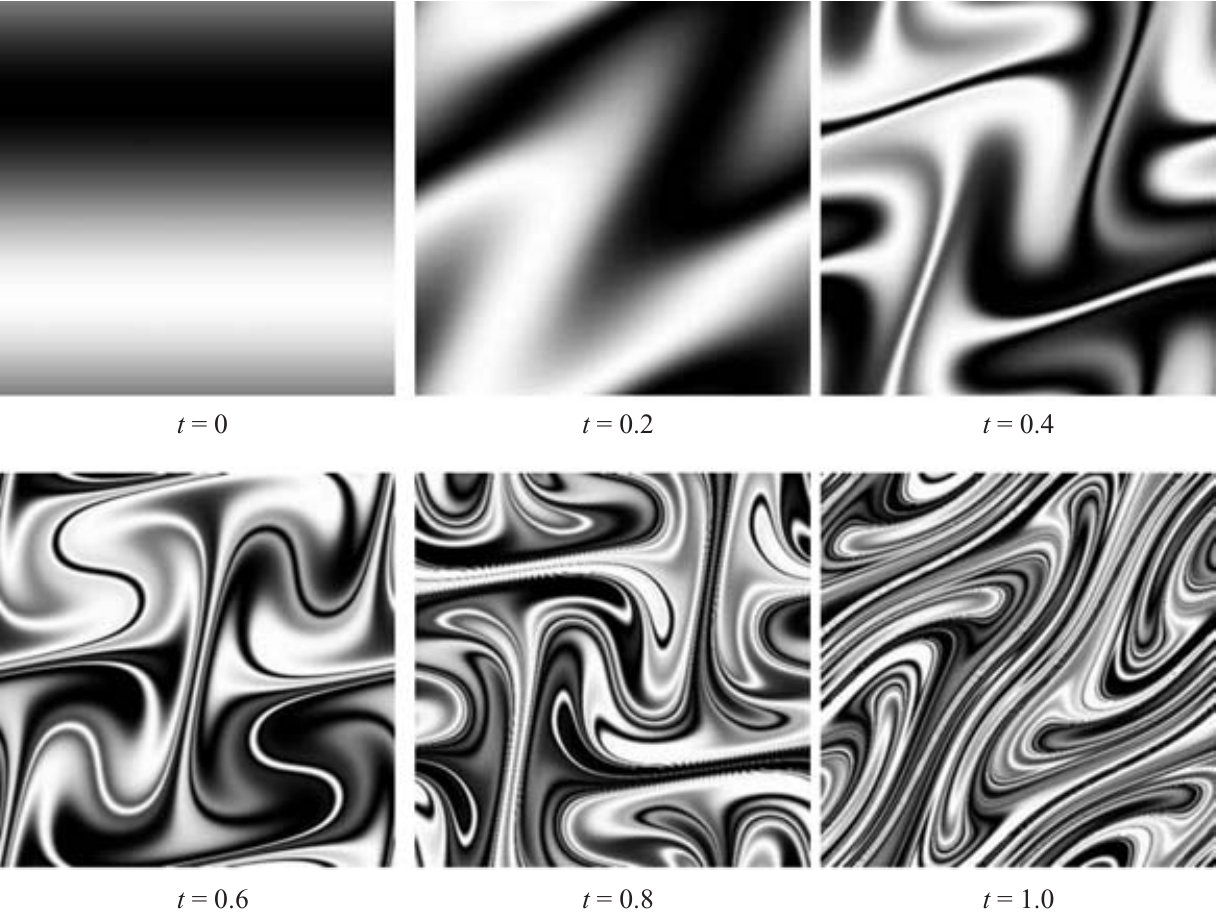}
\end{center}
\end{indented}
\caption{Snapshots of the concentration field~$\theta(\xv,\time)$
  advected by the optimal solution for the velocity fields defined
  by~\eref{eq:Mathew2007_vel} (from Mathew~\etal\cite{Mathew2007}).}
\label{fig:Mathew2007_fig4}
\end{figure}

As an illustration of the method, Mathew~\etal\cite{Mathew2007} apply
it to the velocity fields~$\uv_1$ and~$\uv_2$ with streamfunctions
\begin{equation}
  \psi_1(\xc,\yc) = \sin \xc\, \sin \yc,\qquad
  \psi_2(\xc,\yc) = \cos \xc\, \cos \yc.
  \label{eq:Mathew2007_vel}
\end{equation}
They set the time interval~$\timef=1$ and action~$\Act=1/4$, with
initial concentration~$\theta_0(\xv)=\sin\yc$.
\Fref{fig:Mathew2007_fig4} shows the concentration field evolved by
%
%
their optimal solution, which is a time-dependent linear combination
of~\eref{eq:Mathew2007_vel}.  Note that individually the two
flows~\eref{eq:Mathew2007_vel} lead to poor mixing, since they are
steady and two-dimensional.  \Fref{fig:Mathew2007_fig4} exhibits finer
and finer scales as time evolves, a hallmark of chaotic advection, as
well as a roughly exponential decay of the Sobolev
norm~$\normt{\theta}_{\Sobo^{-1/2}}$ (\Fref{fig:Lin2011_fig1a}),
dotted line).

\subsection{Local-in-time optimisation}

Lin~\etal\cite{Lin2011b} proposed an alternative to the global optimal
control approach of Mathew~\etal\cite{Mathew2007}.  Instead of
focusing on the~$\hSobo^{-1/2}$ norm of the concentration field, they
instead examined~$\hSobo^{-1}$.  As the theorem in \sref{sec:mixing}
tells us, any negative Sobolev norm will capture mixing in the sense
of ergodic theory, so there is no profound difference in using either
norm.  However, the rate at which a different norms decrease will in
general be different, so different optimal solutions can be obtained.
The advantage of~$\hSobo^{-1}$ arises from the evolution
equation~\eref{eq:q-1}, in the absence of sources and diffusion:
\begin{equation}
  \frac{d}{d\time}\norm{\theta}_{\hSobo^{-1}}^2
  = 2\savg{\grad^{-1}\theta\cdot\grad\uv\cdot\grad^{-1}\theta}.
   \label{eq:q-1nos}
\end{equation}
The right-hand side is a simple expression that can easily be
extremised \emph{instantaneously}, in the sense that given~$\theta$ at
any instant we tweak the velocity to cause the norm to decay as fast
as possible.  This local-in-time approach can never do better than
global optimal control, but it is often good enough, as we will see.
But most importantly, it is much less computationally expensive since
we do not have to `peer into the future' and evolve the system forward
in time to determine the optimal current velocity field.

In order to formulate an optimisation problem, we must impose some
constraints on the velocity field.  In \sref{sec:optcontr} we imposed
fixed total kinetic energy through~\eref{eq:Rij}.  Now we will consider
two types of constraints, fixed energy or fixed power.  For a Newtonian
fluid, the power is proportional to the~$\Ltwo$
norm~$\Ltnorm{\grad\uv}^2$, but we shall refer to this integral as
`the power' even if the fluid is not of this type.  The two
constraints, then, are to respectively fix
\begin{alignat}{2}
  \Ltnorm{\uv}^2 &= \Uc^2
  \qquad&\text{(fixed energy)}
  \label{fixecons} \\
\intertext{or}
  \Ltnorm{\grad\uv}^2 &=
  \sum_{i,j=1}^\sdim \savg{(\pd_i \uc_j)^2} = \frac{1}{\dunorm^2}
  \qquad&\text{(fixed power)}.
  \label{fixpcons}
\end{alignat}
These define the root-mean-square velocity $\Uc$ and rate of strain
$\dunorm^{-1}$ of the stirring.

We now proceed with the optimisation technique, that is, to maximise
the right-hand side of~\eref{eq:q-1nos}.  With a few integrations by
parts we recast~\eref{eq:q-1nos} in the form
\begin{equation}
  \frac{d}{d\time}\norm{\theta}_{\hSobo^{-1}}^2
  = -2\savg{\theta \, \uv\cdot\nabla (\lapl^{-1} \theta)}
  = -2\savg{\uv\cdot\Proj(\theta\grad\ftheta)}
  \label{mainfcnal}
\end{equation}
where~$\ftheta$ is the filtered scalar field,
\begin{equation}
  \ftheta(\xv,t) \ldef \left( \lapl^{-1} \theta \right)(\xv,t),
\end{equation}
and~$\Proj(\cdot)$ is the projector onto divergence-free fields defined by
\begin{equation}
\Proj(\bm{v}) \ldef \bm{v} - \grad \lapl^{-1} (\grad \cdot \bm{v}).
\end{equation}
Then with either the fixed energy~\eref{fixecons} or fixed
power~\eref{fixpcons} constraint the velocity field maximising the
decay rate of~$\hSobo^{-1}$ is
\begin{alignat}{2}
  \uve &= \Uc \,
  \frac{\Proj(\theta\grad\ftheta)}{\Ltnorm{\Proj(\theta\grad\ftheta)}}
  \qquad&\text{(fixed energy)}
  \label{elfixesol}\\
\intertext{or}
  \uvp &= -\frac{1}{\dunorm}\,
  \frac{\lapl^{-1}\Proj(\theta\grad\ftheta)}
  {\norm{\Proj(\theta\grad\ftheta)}_{\hSobo^{-1}}}
  \qquad&\text{(fixed power)}
  \label{elfixpsol}
\end{alignat}
as long as the denominator does not vanish.  Hence, $\uve$ or $\uvp$
is the best stirring velocity fields to use at any instant in time,
unless the denominator vanishes.  However, if either of the norms in
the denominators vanishes then $\Proj(\theta\grad\ftheta) = 0$
throughout the domain and {\it no} incompressible flow can
instantaneously decrease the $\hSobo^{-1}$ norm.  For example, this
will happen if the concentration field satisfies~$\lapl
\theta=F(\theta)$, which includes cases where $\theta$ is an
eigenfunction of the Laplacian.  If this situation arises in the
course of the time-evolution of~$\theta$, then some other optimisation
strategy must be adopted.

The most natural alternative when $\Proj(\theta\grad\ftheta) = 0$ is to
carry to optimisation to the order, that is, find the velocity field
that minimises
\begin{equation}
  \frac{d^2}{d^2\time}\norm{\theta}^2_{\hSobo^{-1}}
  = 2\savg{\left[\uv\cdot\grad\ftheta \, \grad\theta \cdot\uv
      - (\uv\cdot\grad\theta)\lapl^{-1}(\uv\cdot\grad\theta)\right]}.
\end{equation}
Then the optimal incompressible flow $\uv$ solves the eigenvalue problem
\begin{equation}
  \ev\,\uv = \Proj\Big((\uv\cdot\grad\theta)\grad\ftheta
  + (\uv\cdot\grad\ftheta)\grad\theta
  - 2[\lapl^{-1}(\uv\cdot\grad\theta)]\grad\theta\Big)
  \label{eigfixe}
\end{equation}
for the fixed energy constraint~\eref{fixecons} or
\begin{equation}
  \ev\,\uv = -\lapl^{-1} \Proj\Big((\uv\cdot\grad\theta)\grad\ftheta
  + (\uv\cdot\grad\ftheta)\grad\theta
  - 2[\lapl^{-1}(\uv\cdot\grad\theta)]\grad\theta\Big)
  \label{eigfixp}
\end{equation}
for the fixed power constraint~\eref{fixecons}.  In either case we
seek the eigenfunction corresponding to the minimum
eigenvalue~\hbox{$\ev_- < 0$} to use as the stirring field
momentarily, until~\hbox{$\Proj(\theta\grad\ftheta) \ne 0$}.  The
eigenvalue problems in~\eref{eigfixe} and~\eref{eigfixp} are generally
difficult; see~\cite{Lin2011b} for a discussion.  In practice, we may
need to solve one of the eigenvalue problems if we choose a `bad' initial
condition (such as an eigenfunction of the Laplacian, which is
commonly done), but one it has started the optimisation procedure does
not seem to get stuck very often.

This local-in-time optimal stirring strategy is a limiting case of the
short-horizon optimisation studied by
Cortelezzi~\etal\cite{Cortelezzi2008} when the horizon becomes
infinitesimal, but the locality and simplicity allows a much broader
class of flows to be used.  In order to implement it in practice the
full scalar field must be monitored so that the optimal flow field can
be computed at each instant.

\begin{figure}
\begin{indented}
\item[]
\begin{center}
  \includegraphics[width=.5\textwidth]{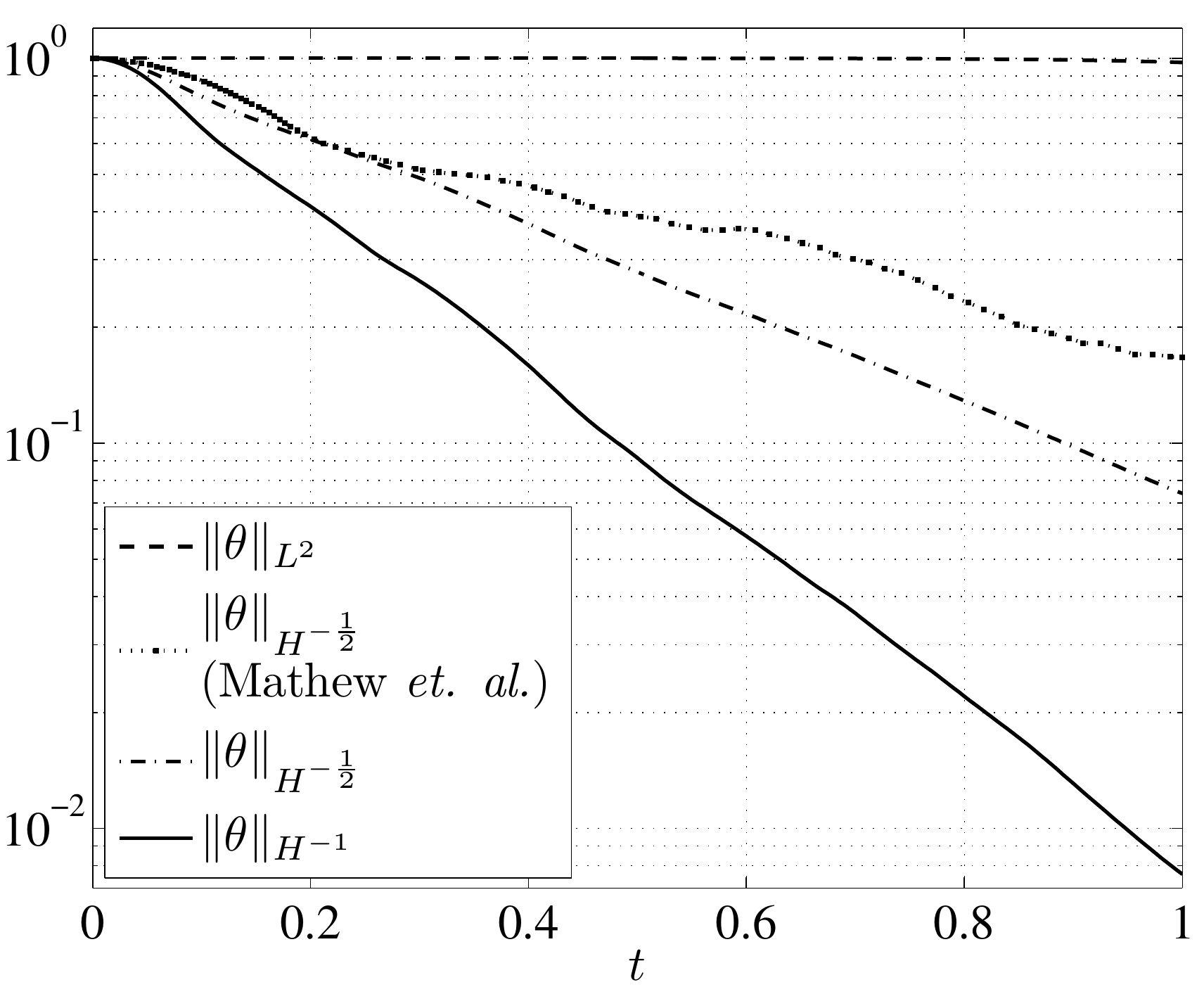}
  \label{fig:Lin2011_fig1a}
\end{center}
\end{indented}
\caption{Evolution of norms with the fixed power
  constraint~\eref{fixpcons} for $\theta_0(\xv)=\sin\xc$.  All norms
  are rescaled by their initial values, and the conserved $L^2$ norm
  is monitored as a numerical check.  The optimisation is over all
  possible velocity fields satisfying the power constraint, which is
  why the local-in-time optimisation outperforms the optimal control
  approach of Mathew~\etal\cite{Mathew2007}.  Snapshots of the
  velocity field are shown in \fref{Lin2011b_fig2}, and in
  \fref{fig:Mathew2007_fig4} for the Mathew \etal solution (from
  Lin~\etal\cite{Lin2011b}).}
\end{figure}

\begin{figure}
\begin{indented}
\item[]
\begin{center}
  \includegraphics[width=.8\textwidth]{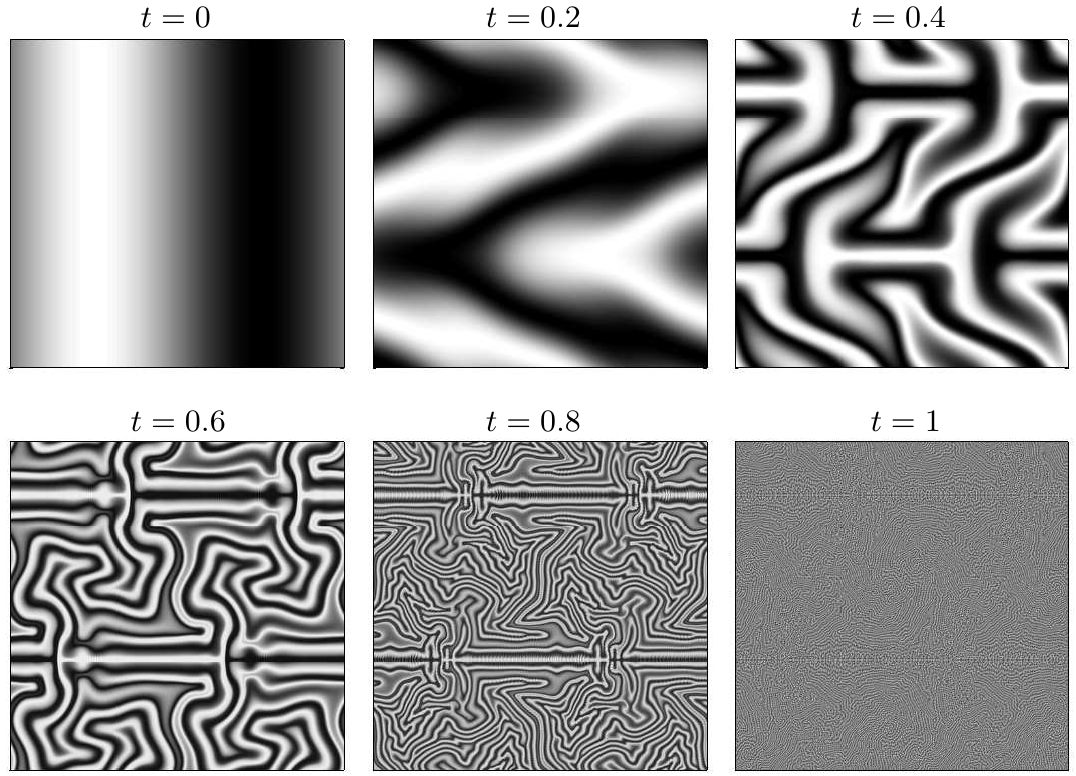}
\end{center}
\end{indented}
\caption{Snapshots of the scalar field in $[0,2\pi]^2$ for the
  fixed-power optimal mixing strategy \eref{elfixpsol}, with initial
  condition $\theta_0(\xv)=\sin \xc$ (from
  Lin~\etal\cite{Lin2011b}).}
\label{Lin2011b_fig2}
\end{figure}

We reproduce here the tests of this optimal stirring strategy
presented in Lin~\etal\cite{Lin2011b}.  They used initial scalar
distribution $\theta_0(\xv)=\sin \xc$ in a domain of size $\Lsc=2 \pi$
in $\sdim=2$ spatial dimensions, for the fixed power
constraint~\eref{fixpcons} with $\dunorm^{-1}=6.25\times(2\pi)^2$,
equivalent to the amplitude of the bi-component control used by
Mathew~\etal\cite{Mathew2007}.  The results for various norms are
shown in \Fref{fig:Lin2011_fig1a}.  The optimisation was performed
for~$\hSobo^{-1}$, but the $\hSobo^{-1/2}$ norm is also plotted to
allow a direct comparison with Mathew~\etal\cite{Mathew2007}.  The
local-time-optimisation seems to outperform the global optimal
control, but this is because the former has access to all possible
incompressible velocity fields.  The difference is evident when
comparing \fref{fig:Mathew2007_fig4} and \fref{Lin2011b_fig2}, which
shows the concentration field.  The Lin~\etal\cite{Lin2011b} solution
uses much smaller velocity scales (though always at fixed power, so
the flow must slow down).  Note also the the optimal flow in
\fref{Lin2011b_fig2} is suggestively self-similar in time.

%

%
%
\section*{\Large Part II: The source-sink problem}

\section{Mixing efficiencies}
\label{sec:Meff}

There are several ways to `calibrate' the norms to measure some
efficiency of mixing.  For the freely-decaying case (no sources or
sinks), one can normalise the norm by its initial condition to obtain
the ratio~$\norm{\theta}_{\hSobo^\qq}/\norm{\theta_0}_{\hSobo^\qq}$.
For the case~$\qq=0$ this is Danckerts' `intensity of
segregation'~\cite{Danckwerts1952}.  The goal of optimisation is then
to reduce~$\norm{\theta}_{\hSobo^\qq}/\norm{\theta_0}_{\hSobo^\qq}$ as
rapidly as possible.  This is a strategy that has been used by many
authors for the case of variance ($\qq=0$)~\cite{Constantin2008} and
for negative Sobolev norms~\cite{Mathew2007, Gubanov2010, Lin2011b},
as we discussed in \sref{sec:optdecay}.

In the presence of sources and sinks, the
norms~$\norm{\theta}_{\hSobo^\qq}$ typically reach an asymptotic
steady state (or at least a statistically-steady state).  In that case
normalising the norms by their initial values is not helpful, since
the asymptotic state is usually independent of initial condition.
Instead, a convenient measure of mixing efficiency is to normalise the
time-asymptotic norm~$\norm{\theta}_{\hSobo^\qq}$ by the value it
would have in the absence of stirring.  We define \emph{mixing
  efficiencies} (or \emph{mixing enhancement factors}) by
\begin{equation}
  \Eff_\qq \ldef {\normt{\thetaz}_{\hSobo^\qq}}\,\bigr/\,
  {\norm{\theta}_{\hSobo^\qq}}
  \label{eq:mixeff}
\end{equation}
where~$\theta$ is the steady solution to~\eref{eq:ADs}
and~$\thetaz$ is the steady solution to the diffusion equation
\begin{equation}
  \frac{\pd\thetaz}{\pd\time} =
  \kappa\lapl\thetaz  + \src\,.
  \label{eq:Ds}
\end{equation}
(If the velocity field or source are explicitly time-dependent, then an
appropriate long-time average must be added to the norms
in~\eref{eq:mixeff}; we will see this in \sref{sec:bounds}.)

The efficiencies measure the amount by which a norm is
\emph{decreased} by stirring.  If stirring decreases a
norm~$\norm{\theta}_{\hSobo^\qq}$ over its purely-diffusive value,
then~$\Eff_\qq$ is larger.  For~$\qq\le0$, an increase in efficiency
is associated with better mixing, since the flow has suppressed
fluctuations.

We might expect that stirring should always decrease the norms from
their purely-diffusive value.  For~$\Eff_1$, this is easily shown to
be the case~\cite{Shaw2007}.  From the definition~\eref{eq:gradinv} of
the inverse gradient of a mean-zero function, we have
\begin{align}
  \kappa\norm{\theta}_{\hSobo^1}^2
  &= \savg{\theta\src}
  = \savg{\theta\,\grad\cdot\grad^{-1}\src}
  = -\savg{\grad\theta\cdot\grad^{-1}\src} \nonumber\\
  &\le \norm{\theta}_{\hSobo^1}\Ltnorm{\grad^{-1}\src}
  \label{eq:e1bound}
\end{align}
where we used the Cauchy--Schwarz inequality.  The steady-state
solution of~\eref{eq:Ds} is~$\thetaz=-\kappa^{-1}\lapl^{-1}\src$,
so~$\grad\thetaz=-\kappa^{-1}\grad^{-1}\src$.  We conclude
from~\eref{eq:e1bound} that~$\norm{\theta}_{\hSobo^1} \le
\normt{\thetaz}_{\hSobo^1}$, or
\begin{equation}
  \Eff_1 \ge 1.
  \label{eq:Eff1lower}
\end{equation}
Thus, the efficiency defined with gradients of~$\theta$ is always
decreased by stirring.  This is somewhat counter-intuitive, since we
expect stirring to create gradients, but this result holds only for
the steady state (or a long-time average).  In fact the conventional
wisdom in mixing holds that stirring creates sharp gradients, and that
those sharp gradients are responsible for good mixing (see
introduction).  The bound~\eref{eq:Eff1lower} shows that this
viewpoint must be qualified when sources and sinks are present:
stirring may indeed create small scales, but the gradients are never
as sharp overall as those that would build up if we didn't stir at
all.

Perhaps even more surprising is that the efficiencies~$\Eff_0$
and~$\Eff_{-1}$ are \emph{not} always increased by stirring.  The
possibility of this was mentioned by Shaw~\etal\cite{Shaw2007}.
Indeed, the combination of flow and source-sink distribution given by%
\footnote{The question of whether such `unmixing' flows exist was
  posed by Charles R. Doering at the \emph{Workshop on Transport and
    Mixing in Complex and Turbulent Flows}, Institute for Mathematics
  and its Applications, Minneapolis, in April 2010.  The
  form~\eref{eq:unmix} is derived from a solution suggested by Jeffrey
  B. Weiss by the end of the workshop.}
\begin{subequations}
\begin{align}
  \uv &= (\sin 2\xc\cos2\yc\,,\,-\cos2\xc\sin2\yc),
  \label{eq:unmix_vel} \\
  \src &= \cos2\xc\sin\yc,
  \label{eq:unmix_src}
\end{align}
\label{eq:unmix}%
\end{subequations}
with~$\Pe=10$ has~$\Eff_0\simeq .991$,
$\Eff_{-1} \simeq .638$, both less than unity.  Here the P\'eclet
number~$\Pe$ is defined in terms of the~$\Ltwo$ norm of~$\uv$ as
\begin{equation}
  \Pe \ldef \Uc\Lsc/\kappa,\qquad
  \Uc = \Ltnorm{\uv},
  \label{eq:Pe}
\end{equation}
and~$\Lsc=2\pi$, $\Uc=1/\sqrt{2}$ for the velocity
field~\eref{eq:unmix_vel}.  (These unmixing flows may be related to
flows that create `hotspots'~\cite{Iyer2010,MattFinn2011}.)

\begin{figure}
\begin{indented}
\item[]
\begin{center}
\subfigure[]{
\includegraphics[width=.25\textwidth]{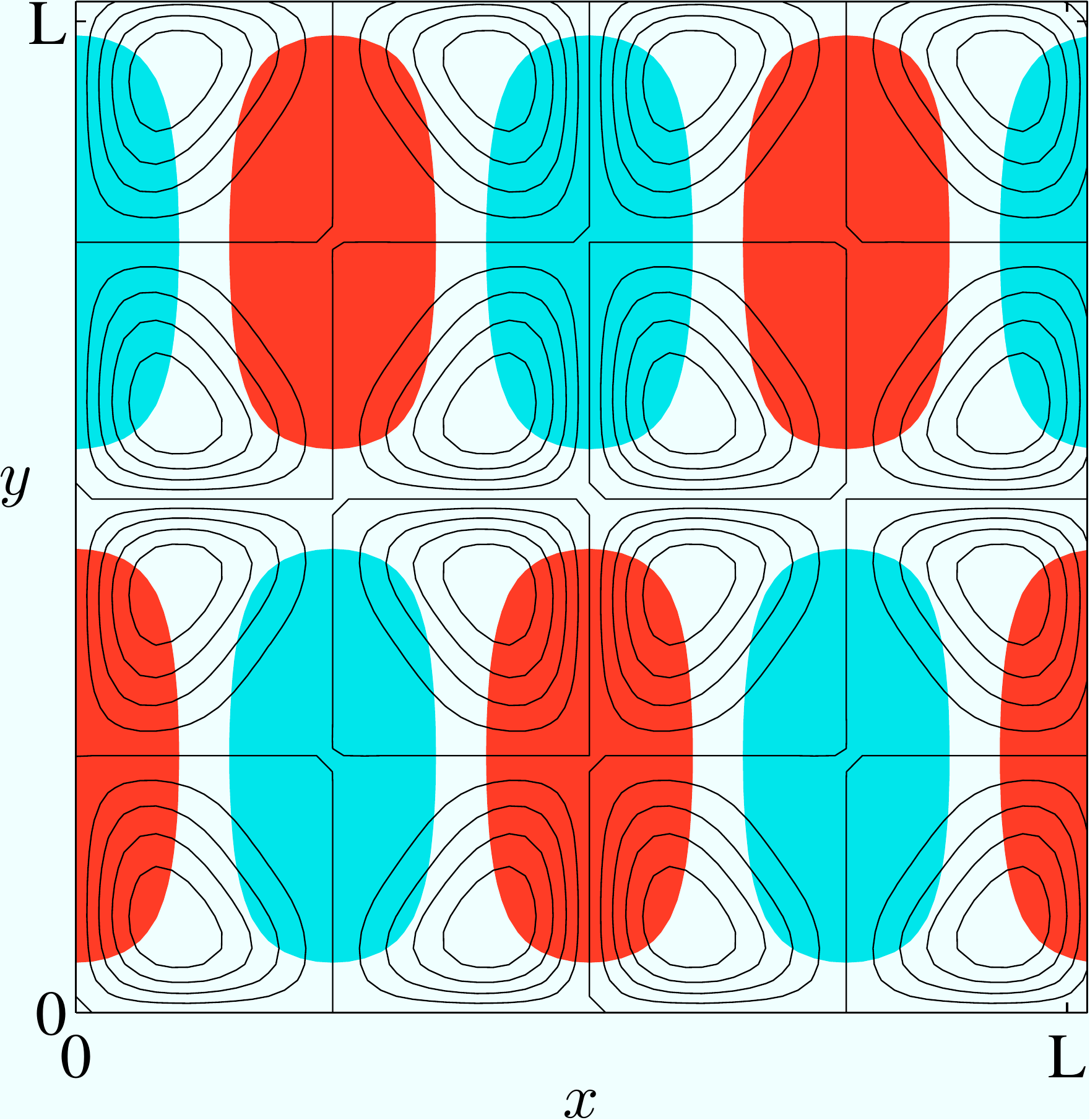}
\label{fig:unmix_psi_src}
}\hspace{.01\textwidth}
\subfigure[]{
\includegraphics[width=.25\textwidth]{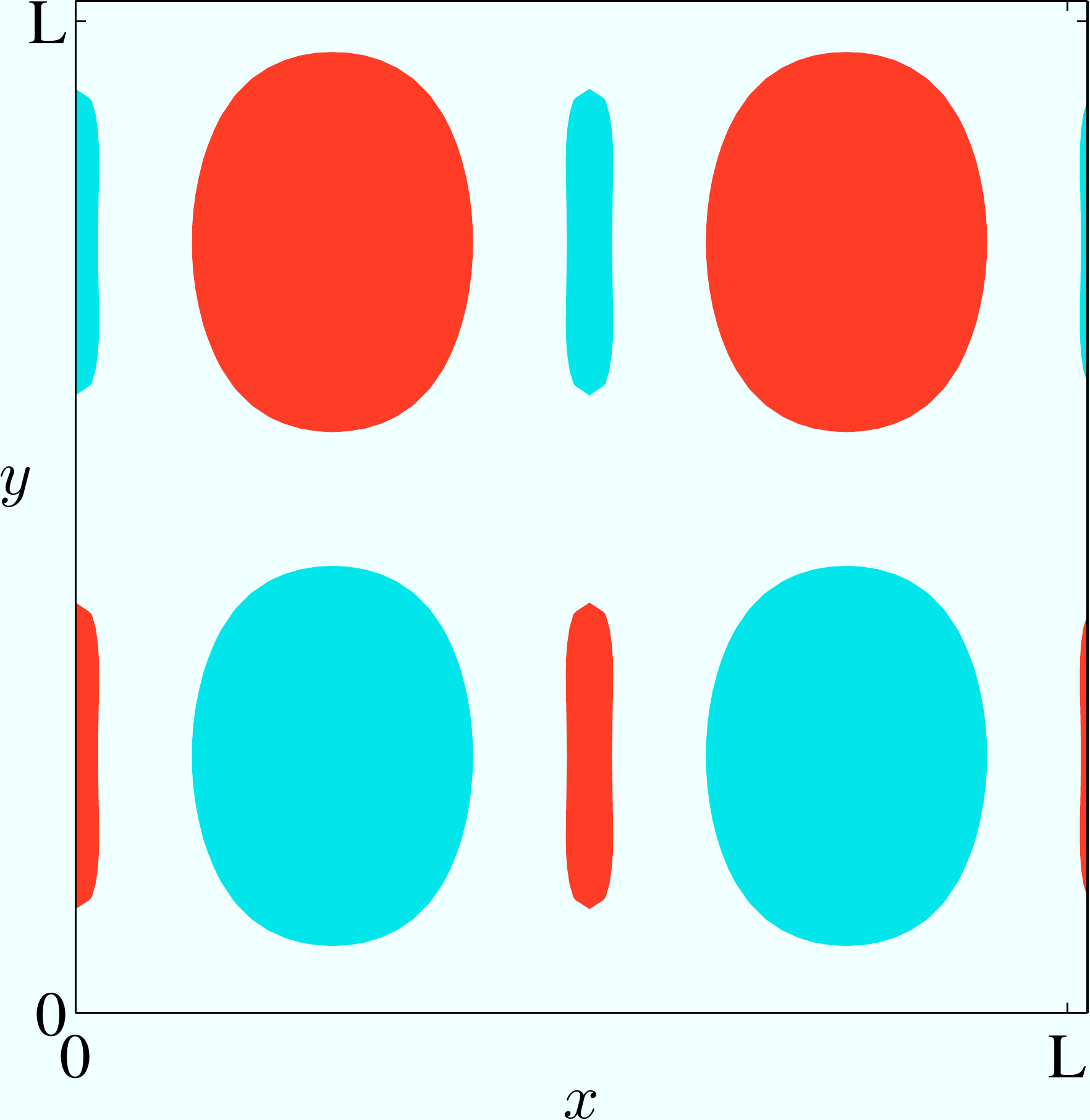}
\label{fig:unmix_theta}
}\hspace{.01\textwidth}
\subfigure[]{
\includegraphics[width=.25\textwidth]{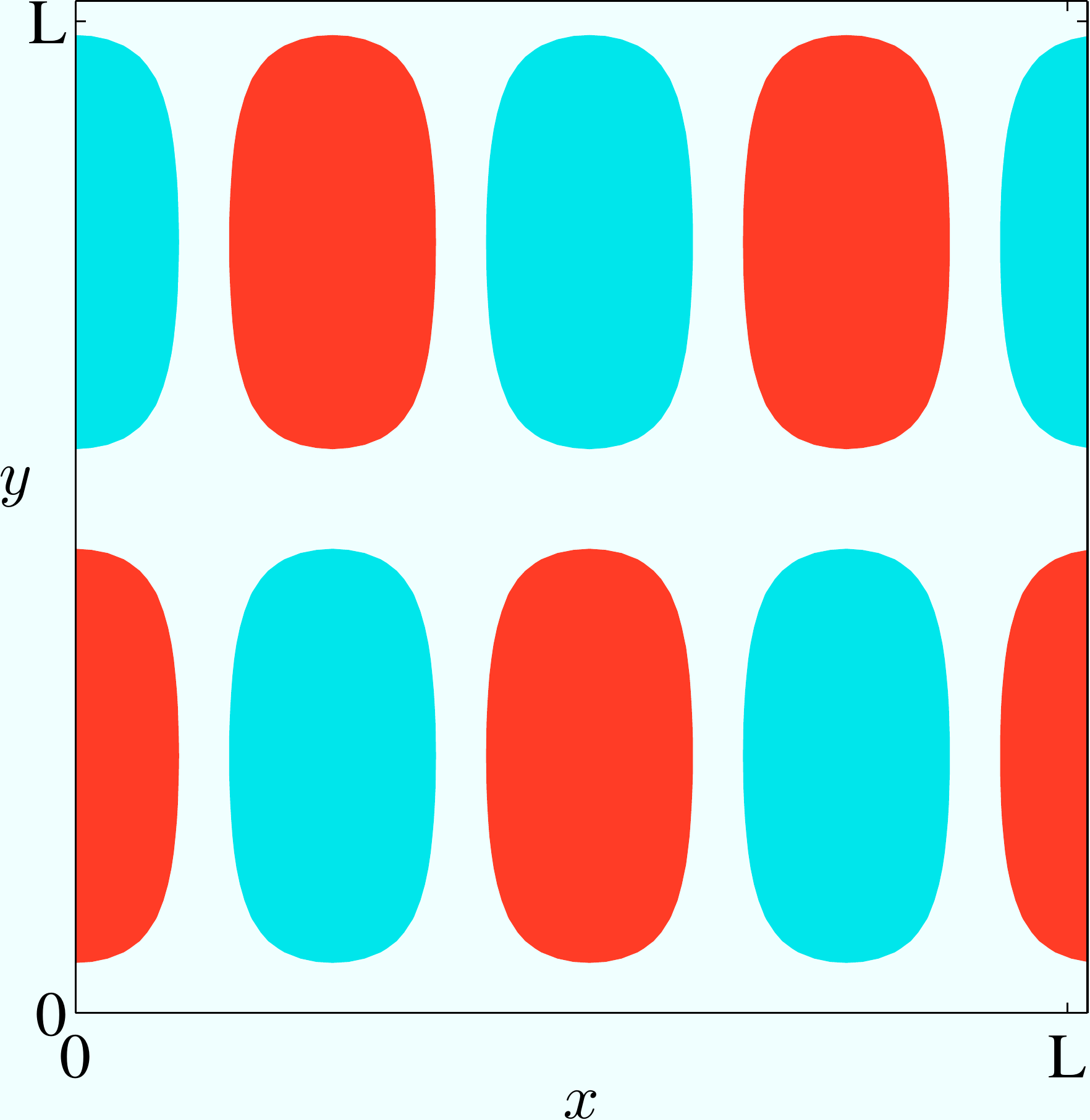}
\label{fig:unmix_theta0}
}
\end{center}
\end{indented}
\caption{(a) Streamlines of the optimised \emph{unmixing} flow for the
  source-sink distribution~\eref{eq:unmix_src} (shown in background)
  at~$\Pe=10$.  (b)--(c) Corresponding temperature field with and
  without stirring, respectively.}
\label{fig:unmix}
\end{figure}

We can optimise the stirring velocity field to give the \emph{worst}
possible mixing efficiency for the source~\eref{eq:unmix_src}, using
the same techniques as in \sref{sec:velopt}.  \Fref{fig:unmix} shows
the resulting streamfunction for~$\Pe=10$, as well as the
solutions~$\theta$ and~$\thetaz$ to the advection--diffusion and
diffusion equation, respectively.  The optimised unmixing flow
has~$\Eff_0\simeq .945$, $\Eff_{-1} \simeq .642$, a modest improvement
over~\eref{eq:unmix_vel} (in fact~$\Eff_{-1}$ went up, since the flow
was optimised for smallest~$\Eff_0$).  At larger P\'eclet number the
optimised solution has lower suboptimal efficiency.
The flow appears to achieve this low efficiency by stretching parts of
the source along the vertical direction, creating thin peaks, but
avoiding concentrating it in other places.  Such `unmixing flows' are
fairly rare and delicate to construct.  It remains true that for a
given source most velocity fields will have efficiency greater than
one, though the specific details of this question have not been
thoroughly investigated.

In fact it was thought that sources of the form~\eref{eq:unmix_src},
where the source is an eigenfunction of the Laplacian operator, could
not lead to~$\Eff_0$ or~$\Eff_{-1}$ less than unity.  However, the
argument in~\cite{Shaw2007} contains a flaw.  To obtain a lower
bound on~$\Eff_0$, the authors solved the constrained optimisation
problem
\begin{equation}
  \langle \theta^2 \rangle \le \max_{\vartheta} \,
  \{ \langle \vartheta^2 \rangle \, | \,
  \kappa \langle |\nabla \vartheta|^2 \rangle = 
  \langle s \vartheta \rangle
  \}
  \label{eq:H0max}
\end{equation}
using the Euler--Lagrange equation
\begin{equation}
  2 \vartheta_{*} + 2\mu \kappa \lapl \vartheta_{*} + \mu s = 0
  \label{eq:H0maxEL}
\end{equation}
where~$\mu$ is the Lagrange multiplier enforcing the constraint
in~\eref{eq:H0max}.  In terms of the Fourier coefficients the
solution of~\eref{eq:H0maxEL} is straightforward,
\begin{equation}
  {\hat{\vartheta}_*}{}_\kv = \tfrac12 \mu\,\srck/
  ({\mu \kappa \km^2 - 1}),
  \label{eq:H0maxELsol}
\end{equation}
but $\mu$ is the solution of the generally-difficult problem
\begin{equation}
  \sum_{{\kv}} \frac{2 - \mu \kappa \km^2}{(\mu \kappa \km^2 - 1)^2}\,
    |\srck|^2 = 0.
  \label{eq:H0maxELmu2}
\end{equation}
However, a convexity argument shows
that~\eref{eq:H0maxELsol}--\eref{eq:H0maxELmu2} is only a maximum if
\begin{equation}
  \mu\kappa\K^2  \ge 1
  \label{eq:maxcriterioneq}
\end{equation}
where~$\K=2\pi/\Lsc$ is the magnitude of the smallest wavenumber.

When the source is an eigenfunction of the Laplacian, with
eigenvalue~$-\kms^2$, we can solve for~$\vartheta_*$ and~$\mu$
explicitly:
\begin{equation}
  \mu = 2/(\kappa \kms^2),
  \qquad \vartheta_* = \src/(\kappa \kms^2),
\end{equation}
for which the maximum criterion~\eref{eq:maxcriterioneq} reads
\begin{equation}
  \kms^2 \le 2\K^2.
  \label{eq:monomax}
\end{equation}
Thus, only eigenfunction sources with~$\kms^2=\K^2$ and~$\kms^2=2\K^2$
are guaranteed to have~$\Eff_0\ge1$.  The source~\eref{eq:unmix_src}
has~$\kms^2=1^2+2^2=5$ (with~$\K=1$), so it can lead to~$\Eff_0<1$, as
we found numerically.  The same criterion~\eref{eq:monomax} also hold
for~$\Eff_{-1}$ to be bounded below by~$1$.

Shaw~\etal\cite{Shaw2007} also derive the rigorous lower bounds
\begin{equation}
  \Eff_0^2 \ge
  \frac{\sum_\kv(\km/\K)^{-4}\lvert\srck\rvert^2}
  {\sum_\kv(\km/\K)^{-2}\lvert\srck\rvert^2},
  \qquad
  \Eff_{-1}^2 \ge
  \frac{\sum_\kv(\km/\K)^{-6}\lvert\srck\rvert^2}
  {\sum_\kv(\km/\K)^{-2}\lvert\srck\rvert^2}.
  \label{eq:Eff0m1bounds}
\end{equation}
However, these are always less than or equal to unity, so in principle
they do not rule out the possibility than any source could be rendered
inefficient by \emph{some} flow.  Charles~R.\ Doering comments
(private communication):
\begin{quote}
  While the lower bounds~\eref{eq:Eff0m1bounds} may be less than one
  they're greater than zero \emph{uniformly} in~$\Pe$.  That is,
  ``unstirring'' or ``herding'' is not something that can be enhanced
  in an unlimited manner by stirring (really a somewhat curious
  situation in my opinion!).
\end{quote}
The bounds~\eref{eq:Eff0m1bounds}
are also very permissive: for the source~\eref{eq:unmix_src} they read
$\Eff_0 \ge 1/5$ and~$\Eff_{-1} \ge 1/25$, whereas the optimised
unmixing flow in \fref{fig:unmix_psi_src} has~$\Eff_0\simeq .945$,
$\Eff_{-1} \simeq .642$ (though the unmixing flow could be a local
minimum, or could decrease the efficiency further at some higher
P\'eclet number).

\section{Upper bounds on mixing efficiencies}
\label{sec:bounds}

It is a simple matter to obtain estimates on the various mixing
efficiencies.  Thiffeault \& Doering~\cite{Thiffeault2004} used an
idea of Doering \& Foias~\cite{Doering2002} to obtain a simple bound
on the mixing efficiency~$\Eff_0$: multiply~\eref{eq:ADs} by an
arbitrary smooth, spatially periodic `comparison
function'~$\varphi(\xv)$, integrate, then integrate by parts to find
\begin{equation}
  \stavg{\theta\,(\uv\cdot\grad + \kappa\lapl)\varphi}
  = -\stavg{\varphi\src}.
  \label{eq:projeq}
\end{equation}
Here we introduced the double-bracket notation
\begin{equation}
  \stavg{F} \ldef \tavg{\savg{F(\xv,\time)}}
\end{equation}
for a space and time average, the latter defined by
\begin{equation}
  \tavg{F}(\xv) \ldef
  \lim_{\time\rightarrow\infty}\frac{1}{\time}\int_0^\time
  F(\xv,\time')\dint\time'.
\end{equation}
(We always assume that such time-averages exist.)  The time derivative
term from the advection-diffusion equation~\eref{eq:ADs} has vanished
from~\eref{eq:projeq}, since~$\tavg{\varphi\,\pd_\time\theta} =
\tavg{\pd_\time(\varphi\,\theta)} =
\varphi(\xv)\lim_{\time\rightarrow\infty}\theta(\xv,\time)/\time=0$,
since~$\theta$ is bounded.  Then apply the Cauchy--Schwarz inequality
to~\eref{eq:projeq}, to obtain
\begin{equation}
  \stavg{\theta^2} \ge \max_{\varphi}\,\,
  \stavg{\varphi\src}^2/
  \stavg{\l(\uv\cdot\grad\varphi + \kappa\lapl\varphi\r)\!{}^2}
     \label{eq:Varbound}
\end{equation}
where the maximisation is over smooth functions~$\varphi(\xv)$.  At
the cost of some sharpness we can take the square root and then use
the Minkowski inequality in the denominator,
\begin{equation}
  \stavg{\theta^2}^{1/2} \ge \max_{\varphi}\,\,
  \lvert\stavg{\varphi\src}\rvert/\Bigl(
  \stavgt{(\uv\cdot\grad\varphi)^2}^{1/2}
  + \kappa\Ltnorm{\lapl\varphi}\Bigr).
\end{equation}
We then apply H\"older's inequality and find
\begin{equation}
  \stavg{\theta^2}^{1/2} \ge \max_{\varphi}\,\,
  \lvert\stavg{\varphi\src}\rvert/\bigl(
  \Uc\norm{\grad\varphi}_{\Linf}
  + \kappa\Ltnorm{\lapl\varphi}\bigr)
  \label{eq:Varbounduniform}
\end{equation}
where
\begin{equation}
  \Uc = \bigl(\tavg{\Ltnorm{\uv}\!{}^2}\bigr)^{1/2}
  \label{eq:Ucdef}
\end{equation}
is proportional to the time-averaged total kinetic energy.  The two
bounds~\eref{eq:Varbound} and~\eref{eq:Varbounduniform} both have
their uses: the former is tailored to a specific velocity field, but
the latter is a global bound valid for \emph{any} stirring velocity
field with bounded energy.

To illustrate the usefulness of these estimates, we shall
use~\eref{eq:Varbound} to bound the mixing efficiency~$\Eff_0$, now
defined to included a space-time average: where the efficiencies are
now defined with a time average,
\begin{equation}
  \Eff_\qq^2
  \ldef {\tavgnormt{\thetaz}}{}_{\hSobo^\qq}^2\,\bigr/\,
  {\tavgnormtwo{\theta}\!\!{}_{\hSobo^\qq}}\,.
  \label{eq:mixefft}
\end{equation}
Of course, this reduces to the earlier definition~\eref{eq:mixeff} for
time-independent functions.  We shall prove the following surprising
fact mentioned in the introduction:
\begin{quote}\emph{%
  An optimal way to stir a steady
  one-dimensional source~$\src(\xc)$ in a periodic box, given a fixed
  time-averaged energy, is to use a spatially-uniform constant flow in
  the~$\xc$ direction.
}\end{quote}
See~\fref{fig:hotcold} for the type of source-sink distribution and
flow that we have in mind.  Here, by optimal we mean a flow that
maximises~\eref{eq:mixefft} for~$\qq=0$, or equivalently minimises the
time-averaged variance norm~$\stavg{\theta^2}$.

Now for the proof.  First take~$\uv(\xv,\time)$ to be an arbitrary
divergence-free vector field.  Expand the denominator on the right
in~\eref{eq:Varbound}:
\begin{equation}
  \stavg{\l(\uv\cdot\grad\varphi + \kappa\lapl\varphi\r)\!{}^2}
  = \stavg{\l(\uv\cdot\grad\varphi\r)^2} +
  \kappa^2\stavg{\l(\lapl\varphi\r)^2}
  + 2\kappa\stavg{(\uv\cdot\grad\varphi)\lapl\varphi}.
  \label{eq:Varbounddenom}
\end{equation}
Given that the source~$\src(\xv)=\src(\xc)$ is a function of~$\xc$
only, choose~$\varphi(\xv)=\varphi(\xc)$.  Then the last term
in~\eref{eq:Varbounddenom} vanishes:
\begin{equation}
  \stavg{(\uv\cdot\grad\varphi)\lapl\varphi}
  = \stavg{\uc\,\varphi'(\xc)\varphi''(\xc)}
  = \tfrac12\stavgt{\uc\,({\varphi'}^2)'}
  = \tfrac12\stavgt{\uv\cdot\grad({\varphi'}^2)} = 0.
\end{equation}
We also have~$\stavg{\l(\uv\cdot\grad\varphi\r)^2} \le
\Uc^2\savg{\lvert\grad\varphi\rvert^2}$, where~$\Uc$ is defined
by~\eref{eq:Ucdef}.  Hence, from~\eref{eq:Varbound} we have the bound
\begin{equation}
  \stavg{\theta^2} \ge \max_{\varphi}\,
  \savg{\varphi\src}^2 / \l(\Uc^2\savg{\lvert\grad\varphi\rvert^2}
  + \kappa^2\savg{(\lapl\varphi)^2}\r).
  \label{eq:Varbound1}
\end{equation}
We can solve the variational problem~\eref{eq:Varbound1} using its
Euler--Lagrange equation, in an identical manner
to~\cite{DoeringThiffeault2006,Shaw2007}, to find
\begin{equation}
  \stavg{\theta^2} \ge
  \savg{\src\l\{\kappa^2\lapl^2 - \Uc^2\lapl\r\}^{-1}\src}.
  \label{eq:Varbound2}
\end{equation}
However, the right-hand side of~\eref{eq:Varbound2} is exactly the
variance of the periodic zero-mean solution to
\begin{equation}
  \Uc \theta'(\xc) = \kappa\,\theta''(\xc) + \src(\xc),
  \label{eq:ADconst}
\end{equation}
that is, the steady advection-diffusion equation for a constant flow.
Hence, the constant flow is optimal, in the sense that any other flow
with velocity norm~$\Uc^2$ cannot decrease the variance further.
(This optimal solution might not be unique.)  This is a surprising
fact: it means that any other process, even turbulence, cannot `stir'
the source-sink better.  A constant flow also minimises
the~$\hSobo^{1}$ norm at fixed kinetic energy, since the bound for
this norm is
\begin{equation}
  \tavgnormtwo{\theta}\!\!{}_{\hSobo^1} \ge
  \savg{\src\l\{\kappa^2\lapl^2 - \Uc^2\lapl\r\}^{-1}\src}
  \label{eq:hSobo1bound}
\end{equation}
which is also saturated for the periodic zero-mean solution
of~\eref{eq:ADconst}.  The third norm, $\hSobo^{-1}$, associated
with~$\Eff_{-1}$, is \emph{not} optimised by a constant flow,
reflecting the norm's preference for small scales.  A simple bound
such as~\eref{eq:Varbound2} and~\eref{eq:hSobo1bound} which does not
depend on the details of~$\uv(\xv,\time)$ cannot be derived in this
case, except for particular classes of flows (see \sref{sec:SHIF}).
Nevertheless, a bound can be derived from~\cite{DoeringThiffeault2006,
  Shaw2007}
\begin{equation*}
  \stavg{\lvert\grad^{-1}\theta\rvert^2} \ge \max_{\varphi}
  {\stavg{\varphi\src}^2} \Bigl/
       {\stavg{\lvert\grad\uv\cdot\grad\varphi +
	 \uv\cdot\grad\grad\varphi +\kappa\lapl\grad\varphi\rvert^2}}\,,
\end{equation*}
which after using the Minkowski inequality gives for the denominator
gives what is likely a terrible bound:
\begin{equation}
  \stavg{\lvert\grad^{-1}\theta\rvert^2} \ge \max_{\varphi}
  {\stavg{\varphi\src}^2} \Bigl/
       {\l(\dunorm^{-1}\Ltnorm{\grad\varphi}
         + \Uc\Ltnorm{\lapl\varphi}
         + \kappa\Ltnorm{\grad\lapl\varphi}\r)^2}\,,
  \label{eq:badbound}
\end{equation}
where~$\dunorm^{-2} = \tavg{\Ltnorm{\grad\uv}\!{}^2}$ is a time-averaged
version of~\eref{fixpcons}.  This could in principle be maximised
over~$\varphi$, but this is much harder than for the other norms.  The
important fact about~\eref{eq:badbound} is that it depends on the
gradient norm~$\Ltnorm{\grad\uv}$ of the flow, and clearly the bound
can be made arbitrarily small by increasing this norm.  Thus flows
that minimise this norm are likely mixing.

If one further constrains the problem other solutions are possible
(see for example the discussion of Plasting \&
Young~\cite{Plasting2006} below).  Nevertheless, it is surprising that
the optimal answer in this case could be so different from a `mixing'
flow, that is, one that amplifies gradients of concentration (in the
sense of ergodic theory -- see \sref{sec:mixing}).  It is an open
question whether there exists source-sink configurations for which the
flow that maximises~$\Eff_0$ is also mixing.

\begin{figure}
\begin{indented}
\item[]
\begin{center}
\includegraphics[width=.6\textwidth]{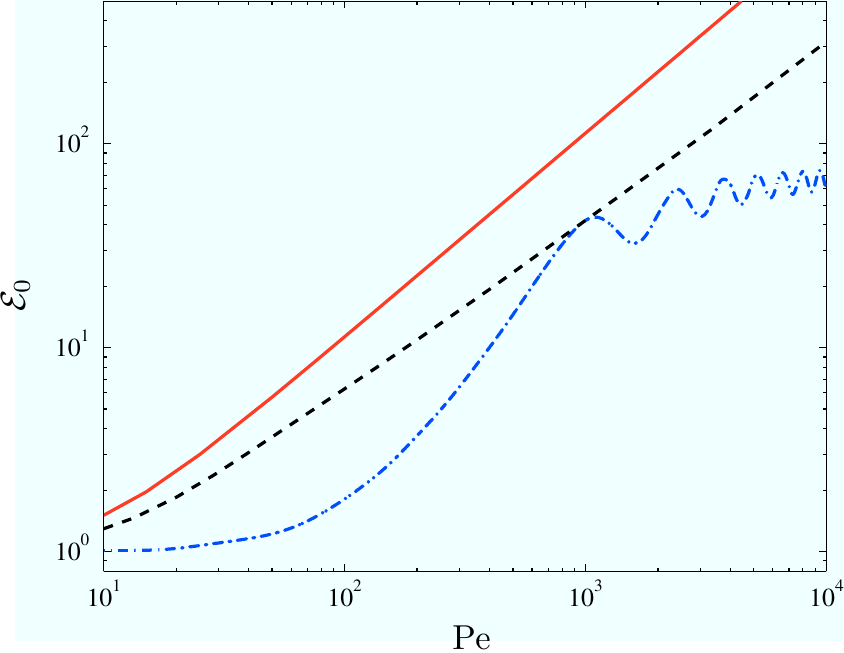}
\end{center}
\end{indented}
\caption{Mixing efficiency~$\Eff_0$ for the random sine flow.  The
  solid line is the upper bound~\eref{eq:sineboundopt}.  The dashed
  line is the result of direct numerical simulations with~$\Uc$
  and~$\tau$ fixed.  The dashed-dot curve plots simulation data with
  $\kappa$ and $\tau$ held constant while varying~$\Uc$ (after
  Thiffeault \etal\cite{Thiffeault2004}).}
\label{fig:rw_Deff}
\end{figure}

To check how sharp the bound~\eref{eq:Varbound} is for a model system,
Thiffeault \etal\cite{Thiffeault2004} considered the two-dimensional
`random sine flow' of
Pierrehumbert\cite{Pierrehumbert1994,Antonsen1996}.  This flow
consists of alternating horizontal and vertical sine shear flows, with
phase angles~$\zeta_1(\time)$ and~\hbox{$\zeta_2(\time) \in [0,2\pi]$}
randomly chosen at each time period,~$\tau$.  In the first half of the
period, the velocity field is
\begin{subequations}
\begin{equation}
  \uv^{(1)}(\xv,\time)
  = \sqrt{2}\, \Uc \l(0\ ,\ \sin(2\pi\NN \xc/\Lsc + \zeta_1(\time))\r);
    \label{eq:sineflow1}
\end{equation}
and in the second half-period it is
\begin{equation}
  \uv^{(2)}(\xv,\time)
  = \sqrt{2}\, \Uc \l(\sin(2\pi\NN \yc/\Lsc + \zeta_2(\time))\ ,\ 0\r),
    \label{eq:sineflow2}
\end{equation}
\label{eq:sineflow}%
\end{subequations}
where~$\NN$ is an integer indicating the scale of the flow.  The
source function used is \hbox{$\src(\xv) = \sqrt2\sin(2\pi \xc/\Lsc)$},
and we set the integer~$\NN=1$ for now.  For the simple
choice~$\varphi=\src$, we have the efficiency bound
\begin{equation}
  \Eff_0 \le \sqrt{\frac{\Pe^2}{8\pi^2} + 1}\ .
  \label{eq:sineboundopt}
\end{equation}
This bound is plotted in \fref{fig:rw_Deff} against numerical
simulations for the random sine flow.
The P\'eclet number is varied in two ways: by varying the
diffusivity~$\kappa$ and the amplitude~$\Uc$.  The bound captures the
trend of the numerical simulation, especially as~$\kappa$ is varied.
The oscillations as $\Uc$ gets larger are due to the spatial
periodicity of the domain.

\begin{figure}
\begin{indented}
\item[]
\begin{center}
\includegraphics[width=.6\textwidth]{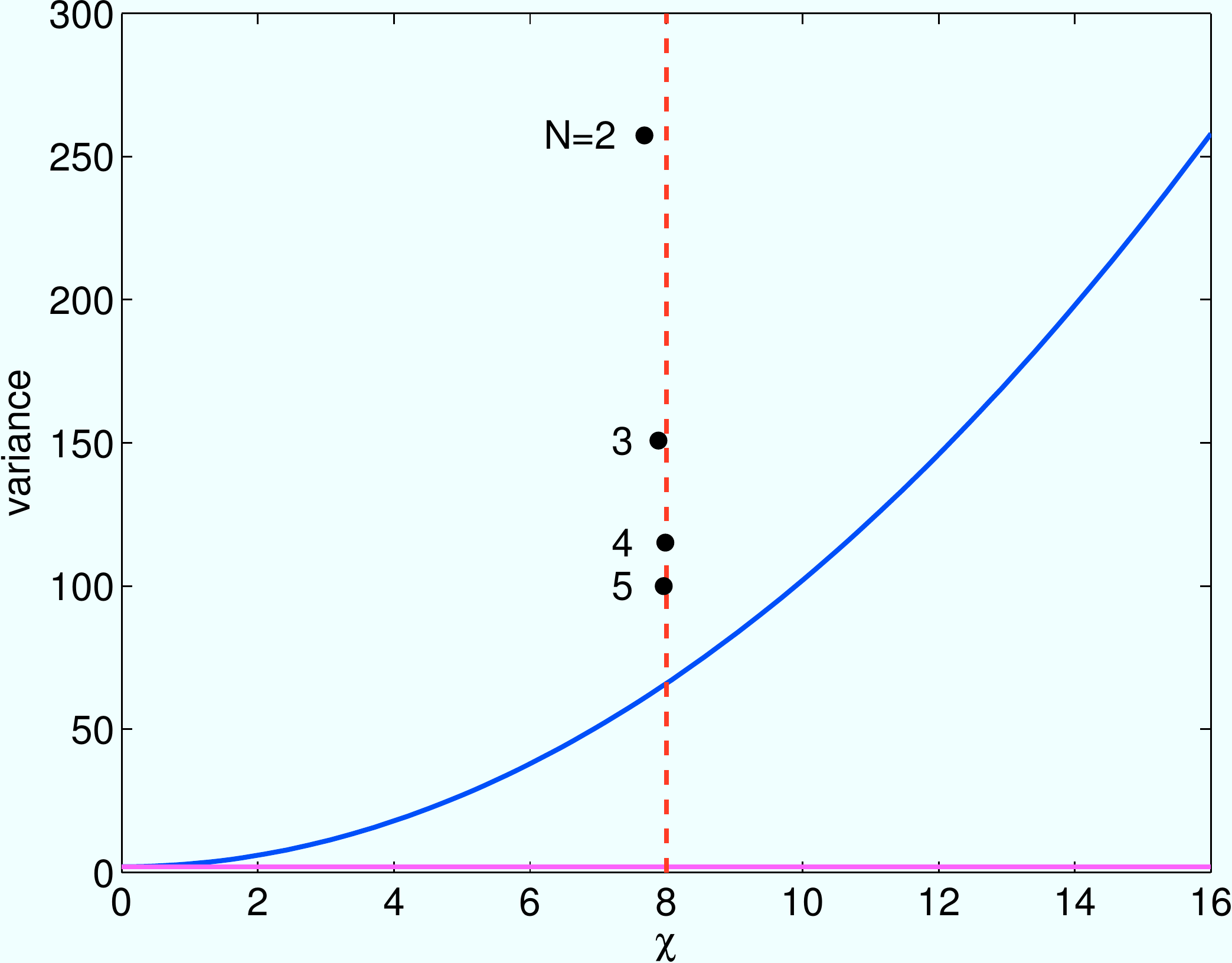}
\end{center}
\end{indented}
\caption{For the sine flow~\eref{eq:sineflow}: the lower
  bound~\eref{eq:PYbound} on variance~$\stavg{\theta^2}$ (solid
  parabolic line) as a function of the dissipation~$\entr$.  The
  horizontal solid line is the bound~\eref{eq:sineboundopt}, which
  assumes no knowledge of~$\entr$.  The vertical dashed line is
  the~$\NN\rightarrow\infty$ limit for~$\entr$, and the dots are
  numerical simulation results (after Plasting \&
  Young~\cite{Plasting2006}).}
\label{fig:PYbound}
\end{figure}

Plasting \& Young\cite{Plasting2006} enhanced the bound by including
the scalar dissipation rate as a constraint.  They define the entropy
production (or half the variance dissipation rate) as
\begin{equation}
  \entr \ldef \kappa\stavg{\lvert\grad\theta\rvert^2}
\end{equation}
which is of course proportional to the time-average of the
$\hSobo^1$-norm of~$\theta$.  The entropy production satisfies the
power integral
\begin{equation}
  \entr = \stavg{\theta\,\src}.
  \label{eq:entrpower}
\end{equation}
Plasting \& Young minimise the variance subject to
both~\eref{eq:projeq} and~\eref{eq:entrpower}, taking~$\entr$ as given.
They find the bound
\begin{equation}
  \stavg{\theta^2} \ge
  \frac{\stavg{\Aa^2}\entr^2 + 2\stavg{\src\,\Aa}\stavg{\src\,\varphi}\entr
  + \stavg{\src\,\varphi}^2\stavg{\src^2}}
  {\stavg{\Aa^2}\stavg{\src^2} - \stavg{\src\,\Aa}^2}.
  \label{eq:PYbound}
\end{equation}
Their bound takes into account the creation of scalar gradients
through the constraint~\eref{eq:entrpower}.  For the sine
flow~\eref{eq:sineflow}, their bound is plotted in \fref{fig:PYbound}
in the~$\stavg{\theta^2}$--$\entr$ plane (parabolic solid curve).  The
horizontal curve at the bottom is the bound~\eref{eq:sineboundopt}.
Notice that, if we know~$\entr$, the lower bound of Plasting \& Young is
a vast improvement over\eref{eq:sineboundopt}.  The problem is that we
usually don't know~$\entr$.  

However, for the sine flow we can find~$\entr$ in the
limit~$\NN\rightarrow\infty$.  First, for the sine
flow~\eref{eq:sineflow} we can compute the effective
diffusivity~$\Deff$ explicitly~\cite{Fannjiang1994,Majda1999}:
\begin{equation}
  \Deff = \tfrac18\Uc^2\tau
  \label{eq:Deffsine}
\end{equation}
where we neglect the small molecular diffusivity.  For~$\NN$ large
in~\eref{eq:sineflow}, we can solve for a `mean-field' temperature
field~\cite{Plasting2006}, and obtain
\begin{equation}
  \entr \simeq \frac{\Lsc^2}{4\pi^2\Deff} = \frac{2\Lsc^2}{\pi^2\Uc^2\tau},
  \qquad \NN\gg1,
  \label{eq:chihomo}
\end{equation}
for the same source~$\src(\xv) = \sqrt2\sin(2\pi \xc/\Lsc)$ used
previously.  This is the `homogenisation limit,' where the scale of
the source is much larger than the scale of the flow (see
\sref{sec:homo}).  The dashed line in \fref{fig:PYbound} shows the
large $\NN$ form~\eref{eq:chihomo}, and the dots are numerical
simulations by Plasting \& Young for various~$\NN$ ($\NN=1$ is off
scale).  The numerical results approach the vertical line for
remarkably small~$\NN$.  They conjecture that for the sine flow as the
scale separation ($\NN$ is made larger) is increased the lower bound
is approached.  It is an open question whether the
bound~\eref{eq:PYbound} can be realised for more general classes of
flow than the single-wavenumber sine flow.

Recently, Alexakis \& Tzella~\cite{Alexakis2011_preprint} addressed
the issue of getting bounds that reflect mixing instead of transport
in a different way.  They focused on a dissipation length scale~$\ld$
and its inverse~$\kd$ defined by
\begin{equation}
  \kd^2 = \ld^{-2} \ldef
  {\tavgnormtwo{\theta}\!\!_{\hSobo^1}} /
  {\tavgnormtwo{\theta}}
  = {\Eff_0^2}/{\Eff_1^2}
  = {\entr} / {(\kappa\stavg{\theta^2})}\,.
\end{equation}
(This length scale was denoted~$\lambda$ in Thiffeault
\etal\cite{Thiffeault2004}.)  This length scale characterises the
scale or variation of the passive scalar.  They find that~$\ld$
scale is \emph{not} always equivalent to the Batchelor length scale
\begin{equation}
  \lB^2 \ldef {\kappa\lu}/{\Uc}
\end{equation}
where~$\lu$ is the typical length scale of the velocity field.  They
introduce the ratio
\begin{equation}
  \luls \ldef \lu/\ls
  \label{eq:rhodef}
\end{equation}
where~$\ls$ is a typical length scale of the source.  For example, for
the sine flow~\eref{eq:sineflow} we have~$\lu=\Lsc/\NN$, and for the
source~$\src(\xv) = \sqrt2\sin(2\pi \xc/\Lsc)$ we have~$\ls=\Lsc$,
so~$\luls=\NN^{-1}$.  The homogenisation limit is
when~$\luls\rightarrow 0$, used by Plasting \& Young to
obtain~\eref{eq:chihomo}.  This represents the ideal of scale
separation between a small-scale stirring velocity field and a
large-scale source.  (See \sref{sec:homo}.)

Alexakis \& Tzella define the correlation~$\xits$ between the source
and the concentration field by
\begin{equation}
  \xits^2 \ldef
  \frac{\stavg{\theta\,\src}^2}{\stavg{\src^2}\stavg{\theta^2}},
  \qquad
  0 \le \xits \le 1,
  \label{eq:xits}
\end{equation}
from which
\begin{equation}
  \stavg{\theta^2} = \xits^2\,\ld^4\,\kappa^{-2}\,
  \stavg{\src^2}.
\end{equation}
Given~$\kappa$ and~$\stavg{\src^2}$, there are then two ways to reduce
the variance: decrease~$\xits$ or decrease~$\ld$ (equivalently,
increase~$\kd$).  Decreasing~$\xits$ is best achieved by
\emph{transport}, that is, by having a flow that rapidly carries
source onto sink and vice-versa.  Decreasing~$\ld$ relies on
\emph{mixing}, that is, by creating small scales of the concentration
field.  Both achieve the same thing in the end, but in very different
ways.  Thus, one can target whichever variance-minimising method one
prefers by focusing on~$\xits$ or~$\ld$.  So far, our emphasis for the
advection-diffusion problem with sources and sinks has been on
decreasing~$\xits$.  One advantage of aiming instead to decrease~$\ld$
is that the flows obtained can be good at reducing the variance
\emph{regardless of the precise structure of the source-sink
  configuration}.

\begin{figure}
\begin{indented}
\item[]
\begin{center}
\includegraphics[width=.6\textwidth]{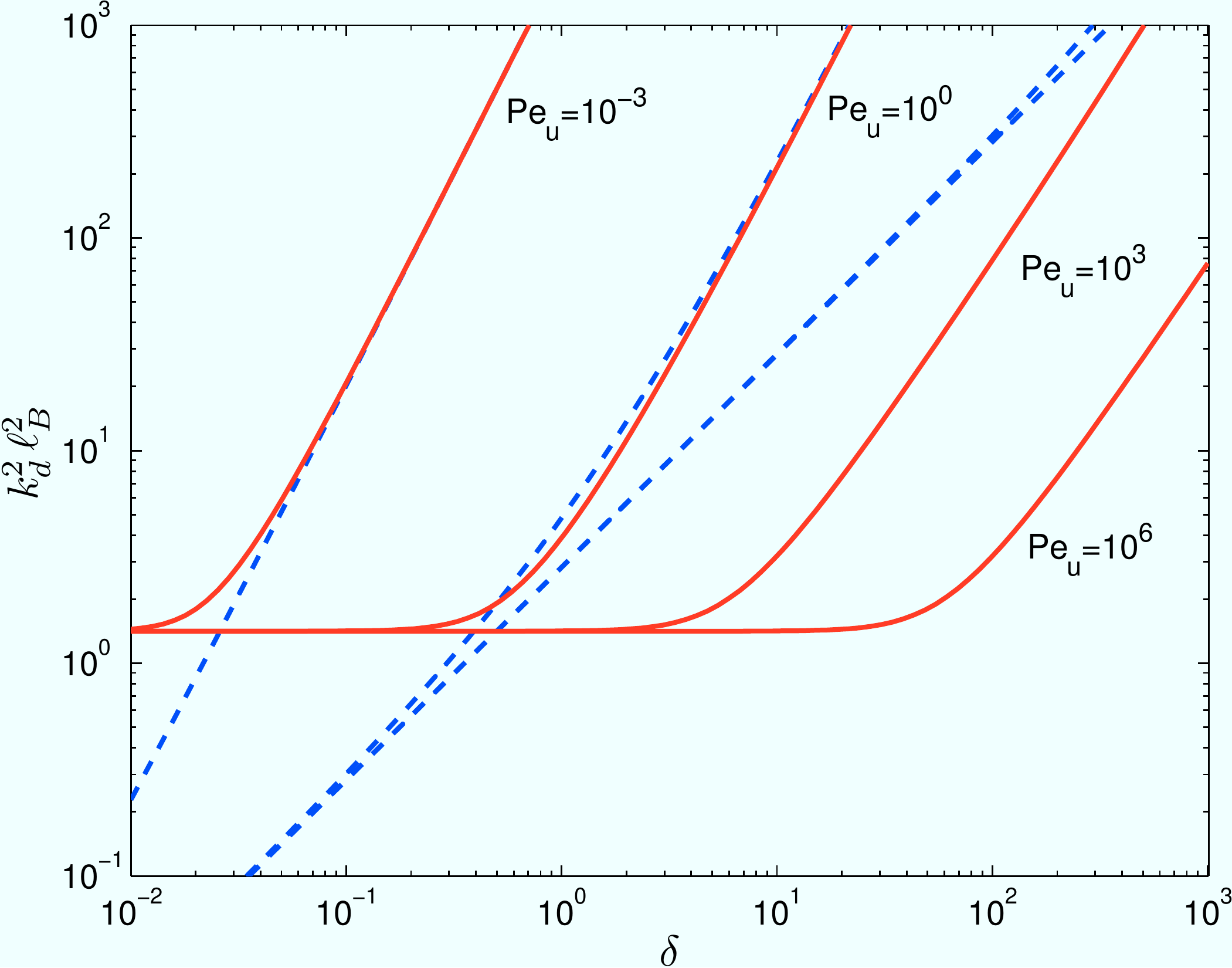}
\end{center}
\end{indented}
\caption{The upper bound~\eref{eq:kdboundTDG} (dashed line) and the
  improved bound~\eref{eq:kdboundAT} (solid line), for various values
  of the P\'eclet number~$\Peu$.  The constants were chosen for the
  sine flow with a sinusoidal source: $\ci=2\sqrt{2}$, $\cii=2$,
  $\ciii=\sqrt{2}$.  Note that for~$\Peu\gtrsim 1$ the
  bounds~\eref{eq:kdboundTDG} `bunch up' and become independent
  of~$\Peu$ (after Alexakis \& Tzella~\cite{Alexakis2011_preprint}).}
\label{fig:ATbound}
\end{figure}

A simple bound on~$\kd$ was given in Thiffeault
\etal\cite{Thiffeault2004}, and after being adapted to the two
scales~$\lu$ and~$\ls$ it reads
\begin{equation}
  \kd^2\lB^2 \le \luls\l(\ci + \cii\luls\,\Peu^{-1}\r),
  \qquad
  \Peu \ldef \Uc\lu/\kappa,
  \label{eq:kdboundTDG}
\end{equation}
where we used a new version of the P\'eclet number based on the source
scale, and the dimensionless constants~$\ci$ and~$\cii$ depend on the
shape of the source and velocity field.  By working directly from the
time-evolution equation for~$\grad\theta$, Alexakis \&
Tzella~\cite{Alexakis2011_preprint} improve this to
\begin{equation}
  \kd^2\lB^2 \le \tfrac12\ciii
  + \tfrac12\sqrt{\ciii^2 + 4\luls^3\Pe^{-1}\cii(\ci + \cii\luls\Peu^{-1})}\,.
  \label{eq:kdboundAT}
\end{equation}
where~$\ciii$ is a dimensionless constant that depends on the shape of
the velocity field.  At large~$\luls$ and large~$\Pe$, this is a vast
improvement over~\eref{eq:kdboundTDG}, as can be seen in
\fref{fig:ATbound}.  The constants were chosen for the sine flow with
a sinusoidal source: $\ci=2\sqrt{2}$, $\cii=2$, $\ciii=\sqrt{2}$
\cite{Alexakis2011_preprint}.  At smaller~$\luls$ the crude
bound~\eref{eq:kdboundTDG} does better, which suggests a further
improved bound could be derived which captures both.  At
larger~$\luls$ the bounds~\eref{eq:kdboundTDG} `bunch up' and do not
improve further, whereas the bound~\eref{eq:kdboundAT} continues to
decrease, achieving an asymptotic value~$\kd^2\lB^2\le\ciii$
as~$\Peu\rightarrow\infty$.

From the upper bounds~\eref{eq:kdboundTDG} and~\eref{eq:kdboundAT},
Alexakis \& Tzella~\cite{Alexakis2011_preprint} investigate the
behaviour of~$\kd$ as $\luls$ varies.  They identify five regimes,
which are summarised in \tref{tab:kdscaling}.  In regime I diffusion
is very fast, so scalar gradients are set by the source scale and the
flow is irrelevant.  Regime II is a transitory regime where~$\Uc$
and~$\kappa$ now appear explicitly, but the length scale~$\lu$ is
still absent.  Regime III is the classical Batchelor
regime~\cite{Batchelor1959}, where the gradients of~$\theta$ scale
as~$\lB^{-1}$.  Regime IV is also Batchelor-like, in that $\kd^2$ is
proportional to $\Uc/\kappa$, but where the length scale $\lu$ has
been replaced by $\ls$, since the source now has larger scales than
the velocity field.  It is indeed remarkable that all these regimes
can be captured by \eref{eq:kdboundTDG}--\eref{eq:kdboundAT}.

There is a fifth regime not captured by these bounds: this is the
homogenisation regime when the source has much larger scale than the
velocity field ($\luls\ll1$).  We discussed this regime earlier in
connection with the Plasting \& Young bound; see also Majda \&
Kramer~\cite{Majda1999} for an extensive review, or Kramer \&
Keating~\cite{Kramer2009} and Keating \etal\cite{Keating2010} for a
treatment explicitly involving sources and sinks.  In homogenisation
theory the resulting effective diffusivity~$\Deff$ usually scales
as~$\Deff \sim \Pe^\alpha\kappa$, where $\alpha=2$ for shear flows
(Taylor--Aris dispersion~\cite{Taylor1953,Aris1956}), $\alpha=1$ for
perfect chaotic mixing ($\Deff$ is then independent of~$\kappa$, as
for the sine flow in \eref{eq:Deffsine}), and $\alpha=1/2$ for
cellular flows.  Note that, as pointed out by Alexakis \& Tzella, the
range of validity in $\luls$ of regime V in \tref{tab:kdscaling} may
strictly speaking be beyond homogenisation theory: Lin
\etal\cite{Lin2010} have shown that~$\luls\ll\Pe^{-1}$ is required for
the theory to apply.

\begin{table}
  \caption{The different regimes deduced from the bounds
    \eref{eq:kdboundTDG}--\eref{eq:kdboundAT} (I--IV) and by
    homogenisation theory (V).}
\begin{indented}
\item[]\begin{tabular}{llcl}
\br
regime & $\kd^2$ estimate & range of validity & note \\
\mr
I & $\le\cii/\ls^2$ & $\luls \gg \Pe$ & diffusion-dominated \\[2pt]
II &$\le(\ci\cii\Uc/\kappa\ls^3)^{1/2} $ & $\Pe^{1/3} \ll \luls \ll \Pe$ & \\[2pt]
III & $\le\ciii/\lB^2$ & $\Order{1} \lesssim \luls \ll \Pe^{1/3}$ & 
  Batchelor regime \\[2pt]
IV & $\le\ci\Uc/\kappa\ls$ & $\luls \lesssim \Order{1}$ & \\[2pt]
V & $\sim \luls^2\Pe^{\alpha-1}/\lB^2 $ & $\luls \ll
\min(1,\Pe^{1-\alpha})$ &
  homogenisation regime\\
\br
\end{tabular}
\end{indented}
\label{tab:kdscaling}
\end{table}

\section{Dependence of norms on source-sink structure}
\label{sec:SHIF}

In this section we discuss the results of Doering \&
Thiffeault~\cite{DoeringThiffeault2006} and Shaw, Thiffeault, \&
Doering~\cite{Shaw2007}, who derive bounds on the dependence of the
mixing efficiencies with P\'eclet number.  The `classical' scaling for
a smooth source-sink distribution is linear in~$\Pe$.  But if the
source has complicated small-scale structures (`roughness'), then the
efficiencies can scale anomalously with~$\Pe$, with exponents less
than unity, or even logarithmic corrections.  The specific behaviour
depends on the degree of roughness, as characterised by the rate of
decay of the power spectrum for large wavenumbers, as well as the
dimensionality of space.

As usual, we consider~$\uv(\xv,\time)$ it to be a specified
divergence-free vector field.  In addition, we assume the following
equal-time single-point statistical properties shared by statistically
homogeneous isotropic flows (SHIFs):
\begin{equation}
\begin{split}
\overline{\uc_{i}(\xv,\cdot)} &= 0,
\quad
\overline{\uc_{i}(\xv,\cdot) \uc_{j}(\xv,\cdot)}
  = \frac{\Uc^2}{\sdim} \delta_{ij}
\\
\overline{\uc_{i}(\xv,\cdot)
  \frac{\partial \uc_{j}(\xv,\cdot)}{\partial \xc_{k}}} &= 0,
\quad
\overline{\frac{\partial \uc_{i}(\xv,\cdot)}{\partial \xc_{k}}
\frac{\partial \uc_{j}(\xv,\cdot)}{\partial \xc_{k}}} =
\frac{\Oms^2}{\sdim}\, \delta_{ij}
\end{split}
\label{eq:shif}
\end{equation}
where overbar represents the long-time average (assumed to exist) at
each point in space.  (See~\cite{Shaw2007} for a derivation of these
properties.) The r.m.s.\ velocity~$\Uc$ measures the strength of the
stirring and~$\Oms$ indicates the flow field's strain or shear
content.  The ratio~$\Tms = \Uc/\Oms$ corresponds to the Taylor
microscale for homogeneous isotropic turbulence.  The P\'eclet number
for the flow is $\Pe = \Uc\Lsc/\kappa$.  Note that there are flows,
such as the `random sine flow'~\eref{eq:sineflow}, that satisfy the
SHIF conditions~\eref{eq:shif} but are not genuinely
isotropic~\cite{DoeringThiffeault2006, Shaw2007, Birch2007}.
Nevertheless, we will refer to flows satisfying as SHIFs for
expediency.

The reason for introducing SHIFs as a class of flows, in addition to
their simplicity and physical relevance, is that the maximisation
over~$\varphi$ in~\eref{eq:Varbound} is particularly simple, for then
the denominator in that equation becomes
\begin{equation}
      \stavg{\l(\uv\cdot\grad\varphi + \kappa\lapl\varphi\r)\!{}^2 }
      = \savg{\frac{\Uc^2}{d}\lvert\grad\varphi\rvert^2 +
      \kappa^2\l(\lapl\varphi\r)^2 }.
     \label{eq:den1}
\end{equation}
Assuming a time-independent source, the simple variational
problem~\eref{eq:Varbound} then gives~\cite{DoeringThiffeault2006,
  Shaw2007}
\begin{equation}
  \Eff_0^2 \le \frac{\savg{\src\,\lapl^{-2}\src}}
       {\savg{\src\l\{\lapl^2 - (\Uc^2/\kappa^2\sdim)\,
\lapl\r\}^{-1}\src}}\,.
  \label{eq:ME0HIT}
\end{equation}
This bound depends on the spatial structure of the source function,
but not its amplitude; the stirring velocity field only enters through
the length scale $\kappa/\Uc = \Pe^{-1}\Lsc$.

We can bound the small scale and large scale efficiencies $\Eff_{\pm
  1}$ from~\eref{eq:projeq} in the same manner after integrations by
parts and application of the Cauchy--Schwarz inequality.  For~$\Eff_1$,
\begin{equation*}
  \stavg{\varphi\src}^2 =
  \stavg{(\uv\varphi + \kappa\grad\varphi)\cdot \grad\theta }^2
  \le \stavg{\lvert\uv\varphi + \kappa\grad\varphi\rvert^2}
  \stavg{\lvert\grad\theta\rvert^2 }
\end{equation*}
so
\begin{equation}
  \stavg{\lvert\grad\theta\rvert^2} \ge \max_{\varphi}
  \stavg{\varphi\src}^2/
  \stavg{\l(\uv\varphi + \kappa\grad\varphi\r)\!{}^2}\,.
  \label{eq:gradphiest}
\end{equation}
A potentially sharper bound involving the full two-point correlation
function for the velocity field can be obtained by formally minimising
over $\theta$~\cite{Shaw2007}, but for our purposes the
estimate~\eref{eq:gradphiest} suffices.  For SHIFs the denominator
in~\eref{eq:gradphiest} is $\l<\varphi [-\kappa^2\lapl + \Uc^2]
\varphi \r>$ and optimisation over $\varphi$ leads
to~\cite{DoeringThiffeault2006, Shaw2007}
\begin{equation}
  \Eff_1^2 \le \frac{\savg{\src\,(-\lapl^{-1})\src}}
     {\savg{\src\l\{-\lapl + \Uc^2/\kappa^2\, \r\}^{-1}\src}}
  \label{eq:ME1HIT}
\end{equation}
for a time-independent source.

We can obtain a bound on~$\Eff_{-1}$ from~\eref{eq:projeq} by
using~$\theta=\grad\cdot\grad^{-1}\theta$, integrating by parts, and
using Cauchy--Schwarz:
\begin{multline*}
  \stavg{\varphi\src}^2 =
  \stavg{\grad(\uv\cdot\grad\varphi +
  \kappa\,\lapl\varphi)\cdot(\grad^{-1}\theta)\,}^2 \\
  \le \stavg{\lvert\grad\uv\cdot\grad\varphi +
  \uv\cdot\grad\grad\varphi +\kappa\lapl\grad\varphi\rvert^2}
  \stavg{\lvert\grad^{-1}\theta\rvert^2 }
\end{multline*}
so that
\begin{equation*}
  \stavg{\lvert\grad^{-1}\theta\rvert^2} \ge \max_{\varphi}
  \frac{\stavg{\varphi\src}^2}
       {\stavg{\lvert\grad\uv\cdot\grad\varphi +
	 \uv\cdot\grad\grad\varphi +\kappa\lapl\grad\varphi\rvert^2}}\,.
\end{equation*}
For SHIFs the denominator is $\l<\varphi [-\kappa^2\lapl^3 +
(\Uc^2/\sdim)\,\lapl^2 - (\Oms^2/\sdim)\, \lapl] \varphi \r>$ so
that
\begin{equation}
  \Eff_{-1}^2 \le \frac{\savg{\src\,(-\lapl^{-3})\src}}
       {\savg{\src\l\{-\lapl^{3} + (\Uc^2/\kappa^2\sdim)\,
           \lapl^2 - (\Oms^2/\kappa^2\sdim)\,\lapl \r\}^{-1}\src}}
       \label{eq:MEm1HIT}
\end{equation}
for a time-independent source.

Assuming again that the fluid domain is periodic and that the source
is time-independent, it will be helpful to rewrite the mixing
efficiency bounds~\eref{eq:ME0HIT}, \eref{eq:ME1HIT}, and
\eref{eq:MEm1HIT} in Fourier space:
\begin{subequations}
\begin{gather}
 \Eff_{1}^2  \le \frac {\sum_\kv
   \lvert\srck\rvert^2/\km^2}
  {\sum_\kv
  \lvert\srck\rvert^2/(\km^2 + \Pe^2)}\,,
  \label{eq:MEHITb}
\\
  \Eff_0^2 \le \frac {\sum_\kv
  \lvert\srck\rvert^2/\km^4}
  {\sum_\kv \lvert\srck\rvert^2/ (\km^4 + \Pe^2\km^2/\sdim)}\,,
  \label{eq:MEHITm}
\\
  \Eff_{-1}^2 \le \frac {\sum_\kv \lvert\srck\rvert^2/\km^{6}}
  {\sum_\kv \lvert\srck\rvert^2/
  (\km^6 + \Pe^2\km^4/\sdim + \Pe^2\km^2/\Tms^2\sdim)}\,,
  \label{eq:MEHITp}
\end{gather}%
\label{eq:MEHIT}%
\end{subequations}
where we have rescaled $[0,\Lsc]^{\sdim}$ to $[0,1]^{\sdim}$ so that
wavevector components are integer multiples of $2\pi$.  Now we
investigate the large P\'eclet number behaviour of these bounds for a
variety of classes of sources.

\subsection{Eigenfunction sources.} The simplest class consists of
sources that depend only on a single wavenumber~$\kms$, \ie, that are
eigenfunctions of the Laplacian~$\lapl$ with eigenvalue~$-\kms^2$.
The bounds~\eref{eq:MEHIT} then simplify to
\begin{subequations}
\begin{gather}
  \Eff_{1} \le \sqrt{1 + \Pe^2/\kms^2}\,, \\
  \Eff_{0} \le \sqrt{1 + \Pe^2/\kms^2\sdim}\,, \\
  \Eff_{-1} \le \sqrt{1 + \Pe^2/\kms^2\sdim + 
  \Pe^2/\Tms^2\kms^4\sdim}\,.
\end{gather}
\end{subequations}
Observe that each efficiency is asymptotically proportional to $\Pe$
for large~$\Pe$, corresponding to the expected suppression of variance
if the molecular diffusivity $\kappa$ is replaced by an eddy
diffusivity proportional to $\Uc\Lsc$.  Moreover these upper bounds
are sharp: they may be realised by uniform flow fields whose direction
varies appropriately in time to satisfy the weak statistical
homogeneity and isotropy conditions used in the
analysis~\cite{Shaw2007,Plasting2006}.  Each estimate also exhibits a
decreasing dependence on the length scale of the source: for
large~$\Pe$ the bounds for the small- and intermediate-scale
efficiencies $\Eff_{1}$ and $\Eff_{0}$ are proportional
to~$\Pe/\kms$.

\subsection{Square-integrable sources and sinks.}
The next simplest case is when the Fourier coefficients of the
source-sink distribution are such that the sums in the denominators
of~\eref{eq:MEHIT} converge in the limit as~$\Pe\rightarrow\infty$.
For example, the Fourier coefficients of smooth sources decay
exponentially for large~$\km$, so convergence is guaranteed.  We can
then use the asymptotic $\Pe \rightarrow \infty$ behaviour of the
mixing efficiency bounds to find
\begin{subequations}
\begin{gather}
  \Eff_{1} \le \Pe\, \sqrt{ 
  \frac{\sum_\kv \lvert\srck\rvert^2/\km^2}
  {\sum_\kv {\lvert\srck\rvert^2}}}\,,
  \label{eq:smoothMEboundb} \\
  \Eff_{0} \le \Pe\, \sqrt{ 
  \frac{\sum_\kv \lvert\srck\rvert^2/\km^4}
  {\sdim  \sum_\kv \lvert\srck\rvert^2/\km^2}}\,,
  \label{eq:smoothMEboundm} \\
  \Eff_{-1} \le \Pe\, \sqrt{ 
  \frac{\sum_\kv \lvert\srck\rvert^2/\km^6}
  {\sdim \sum_\kv \lvert\srck\rvert^2/
  (\km^4+\km^2/\Tms^2)}}\,.
    \label{eq:smoothMEboundp}
\end{gather}
\label{eq:smoothMEbound}%
\end{subequations}
\noindent
These are the same $\Pe$ scalings as above for eigenfunction sources,
but the prefactors now depend on different combinations of length
scales in the source.  For instance, the efficiency~$\Eff_1$ depends
more strongly on the high wavenumbers in the source than the other two
efficiencies, as would be expected.

\subsection{Rough sources.}
\label{sec:roughsources}

%
\begin{figure}
\begin{indented}
\item[]
\begin{center}
\includegraphics[width=.8\textwidth]{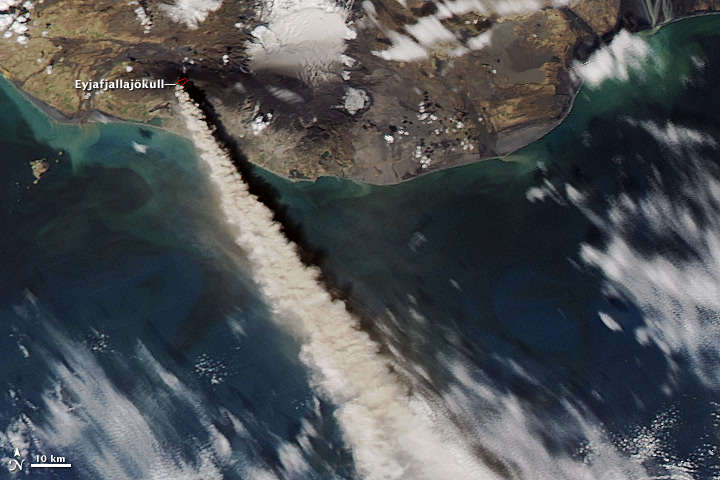}
\end{center}
\end{indented}
\caption{Plume of ash from Eyjafjallaj\"okull volcano, Iceland, 10 May
  2010 (NASA MODIS image).}
\label{fig:iceland_amo_2010130}
\end{figure}

Rough sources are very common in nature: oil gushing from an
underwater well is an infamous recent example of a point source, as is
the volcanic ash plume in \fref{fig:iceland_amo_2010130}.  We can
define `rough' sources as those for which some or all the sums in the
large-$\Pe$ bounds~\eref{eq:smoothMEbound} are divergent.  For
instance, if the source is not in~$\Ltwo$ then the denominator
in~\eref{eq:smoothMEboundb} will diverge.  For those cases, the
P\'eclet number {\it scaling} may change, resulting in anomalous
behaviour for some or all of the efficiencies~\eref{eq:MEHIT}.

The extreme case is the roughest physically-meaningful sources:
measure-valued sources such as $\delta$-functions with nondecaying
Fourier coefficients~$\lvert\srck\rvert = {\cal O}(1)$
as~$\km\rightarrow\infty$.  Then the sums in~\eref{eq:smoothMEboundb}
for~$\Eff_1$ and the denominator of~\eref{eq:smoothMEboundm}
for~$\Eff_0$ diverge in dimension $\sdim = 2$ or~$3$, rendering those
scalings invalid.  In this case the $\Pe$ dependence of $\Eff_{1}$
drops out completely and {\it all} finite-kinetic-energy stirring
velocity fields are completely ineffective at suppressing small scale
fluctuations.  To determine the high-$\Pe$ behaviour of $\Eff_{0}$ we
approximate sums by integrals.  The denominator of~\eref{eq:MEHITm} is
\begin{equation}
  \sum_\kv \frac{1}{\km^4 + (\Pe^2/\sdim)\,\km^2} \sim
  \int_{2\pi}^\infty
  \frac{\xx^{\sdim-1}\dint\xx}{\xx^4 + \Pe^2\xx^2/(4\pi^2\sdim)}.
  \label{eq:sum2int}
\end{equation}
For~$\sdim=2$ the integral in~\eref{eq:sum2int} is
\begin{equation}
  \int_{2\pi}^\infty
  \frac{\xx \dint\xx}{\xx^4 + \Pe^2\xx^2/8\pi^2}
  \sim
  \frac{\log \Pe}{\Pe^2}\,,
  \label{eq:anom2a}
\end{equation}
resulting in the asymptotic bound
\begin{equation}
  \Eff_0 \lesssim
  {\Pe}/{\sqrt{\log\Pe}}\,,
  \quad \sdim = 2.
  \label{eq:anom2}
\end{equation}
Hence in dimension two there is a logarithmic correction to $\Eff_0$
as compared to the square-integrable source case.

For~$\sdim=3$ the integral in~\eref{eq:sum2int} becomes
\begin{equation}
  \int_{2\pi}^\infty
  \frac{\xx^2\dint\xx}{\xx^4 + \Pe^2\,\xx^2/12\pi^2}
  \sim \frac{1}{\Pe}
\end{equation}
resulting in an anomalous scaling bound
\begin{equation}
  \Eff_0 \lesssim \sqrt{\Pe}\,, \quad \sdim = 3.
  \label{eq:anom3}
\end{equation}
This is a dramatic modification of the classical scaling.  A similar
analysis shows that the upper bound on the large scale mixing
efficiency $\Eff_{-1} \sim \Pe$ in~\eref{eq:smoothMEboundp} persists
even for these roughest sources.

We may also analyse anomalous scalings for more general rough sources
where the Fourier spectrum $\lvert\srck\rvert$ decays
as~$\km^{-\rexp}$ with~\hbox{$0 \le \rexp \le \sdim/2$}.  The
roughest measure--valued sources have~$\rexp=0$ while for $\rexp >
\sdim/2$ the source is square-integrable and thus effectively smooth
as far as these multiscale mixing efficiencies are concerned.  In
order to examine the high-P\'eclet-number asymptotics of the bounds on
the various $\Eff_\qq$ we estimate integrals similar
to~\eref{eq:sum2int} but with an extra factor of~$\xx^{-2\rexp}$ in
the numerator arising from~$\lvert\srck\rvert^2$.  The results are
summarised in \tref{tab:scalingsall}.
\begin{table}
  \caption{Scalings of the bound on the mixing efficiency~$\Eff_\qq$ as
    functions of the source roughness exponent~$\rexp$ in two and
    three dimensions.}
\begin{indented}
\item[]\begin{tabular}{lccc}
\br
$\sdim = 2$ & $\qq=1$ & $\qq=0$ & $\qq=-1$ \\
\mr
$\rexp = 0$       & 1 & $\Pe/(\log\Pe)^{1/2}$ & \Pe \\
$0 < \rexp < 1$ & $\Pe^\rexp$ & \Pe & \Pe \\
$\rexp = 1$       & \ \ \ $\Pe/(\log\Pe)^{1/2}$ & \Pe & \Pe \\
$\rexp > 1$       & \Pe & \Pe & \Pe \\[2pt]
\mr
$\sdim = 3$\\
\mr
$\rexp = 0$         & 1 & $\Pe^{1/2}$ & \Pe \\
$0 < \rexp < 1/2$   & 1 & $\Pe^{\rexp+1/2}$ & \Pe \\
$\rexp = 1/2$       & 1 & \quad $\Pe/(\log{\Pe})^{1/2}$ & \Pe \\
$1/2 < \rexp < 3/2$ & $\Pe^{\rexp - 1/2}$ & \Pe & \Pe \\
$\rexp = 3/2$       & \ $\Pe/(\log{\Pe})^{1/2}$ & \Pe & \Pe \\
$\rexp > 3/2$       & \Pe & \Pe & \Pe \\
\br
\end{tabular}
\end{indented}
\label{tab:scalingsall}
\end{table}
In $\sdim=2$ the scaling for $\Eff_{1}$ is anomalous for any degree of
roughness while $\Eff_{0}$ is anomalous only for the roughest sources
with~$\rexp=0$.  In $\sdim=3$, $\Eff_{1}$ is again anomalous for any
degree of roughness while $\Eff_{0}$ scales anomalously
for~$0\le\rexp<1/2$.  For both $\sdim=2$ and $3$ the bound on the
large scale mixing efficiency $\Eff_{-1}$ is always classical (\ie,
linear in~$\Pe$).  Of course these scalings neglect any large-$\km$
cutoff for the rough sources, as discussed in the next section.

\subsection{Rough sources with a cutoff.}  In nature, it can be
argued that rough sources are never truly encountered: physical
systems tend to be smooth beyond a certain small scale (as long as we
stay away from atomic scales, but that is a different story\dots), or
at least they are modelled that way.  With this in mind, how are the
scalings derived in the previous section realised by sources which are
only rough when seen `from afar' but are actually smooth upon closer
examination?  Answering this will help, for instance, in understanding
how such scalings can be observed in data for which the roughness
exponent is meaningful for a limited range of wavenumbers.  We will
focus on the roughest type of sources for which Fourier coefficients
do not decay, but the analysis is easily extended to any type of rough
source discussed in \sref{sec:roughsources}.

\begin{figure}
\begin{indented}
\item[]
\begin{center}
\includegraphics[width=.6\textwidth]{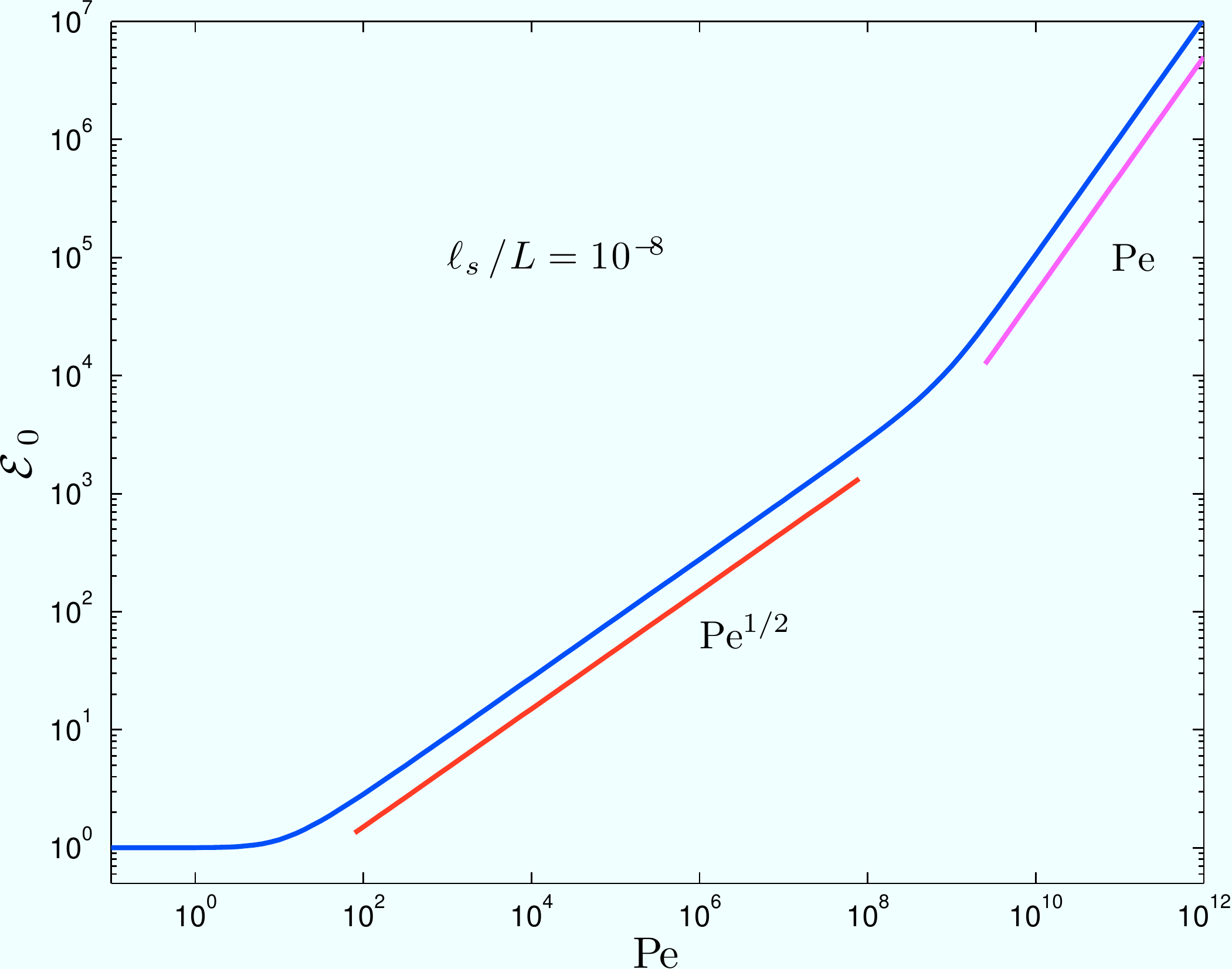}
\end{center}
\end{indented}
\caption{Upper bound for the mixing efficiency $\Eff_{0}$ as a
  function of P\'eclet number for a small source with $\ls = 10^{-8}
  \Lsc$ stirred by a three-dimensional statistically homogeneous and
  isotropic flow [computed from Eq.~\eref{eq:MEHITm}].  The
  intermediate $\Pe^{1/2}$ scaling for~$1 \ll \Pe \ll (\Lsc/\ls)$ is
  evident (after Doering \& Thiffeault~\cite{DoeringThiffeault2006}).}
\label{fig:mixeff3D}
\end{figure}

Point-like sources of small but finite size~$\ls$ have Fourier
coefficients~$\srck$ that are approximately constant in magnitude
up to a cutoff wavenumber of order~$2\pi/\ls$, beyond which the
spectrum decays as for a smooth source.  We may deduce the behaviour of
the bound on $\Eff_{0}$ for such sources by inserting an upper limit
at~\hbox{$\Lsc/\ls \gg 1$} into the integral in~\eref{eq:sum2int}.
For large but intermediate P\'eclet numbers satisfying~$1 \ll \Pe \ll
\Lsc/\ls$, the cutoff is irrelevant so the logarithmic
correction~\eref{eq:anom2} in $\sdim=2$ and the anomalous
scaling~\eref{eq:anom3} in $\sdim=3$ appear.  However for $\Pe \gg
\Lsc/\ls$, \ie, when the modified P\'eclet number based on the
smallest scale in the source $\Uc\ls/\kappa \gg 1$, the smooth source
results apply and we recover the mixing efficiency bounds linear
in~$\Pe$, as in~\eref{eq:smoothMEbound}.  Figure~\eref{fig:mixeff3D}
shows this scaling transition for the $\sdim=3$ case.  Even in the
ultimate regime where the source appears smooth, the \emph{prefactor}
in front of the high-$\Pe$ scaling bounds are significantly diminished
by the small scales in the source: $\Eff_0 \lesssim
\l[\log(\Lsc/\ls)\r]^{-1/2}\Pe$ in $\sdim = 2$, and $\Eff_0 \lesssim
[\ls/\Lsc]^{1/2}\Pe$ in $\sdim = 3$.

\subsection{Summary and numerical evidence.}

\begin{figure}
\begin{indented}
\item[]
\begin{center}
\includegraphics[width=.6\textwidth]{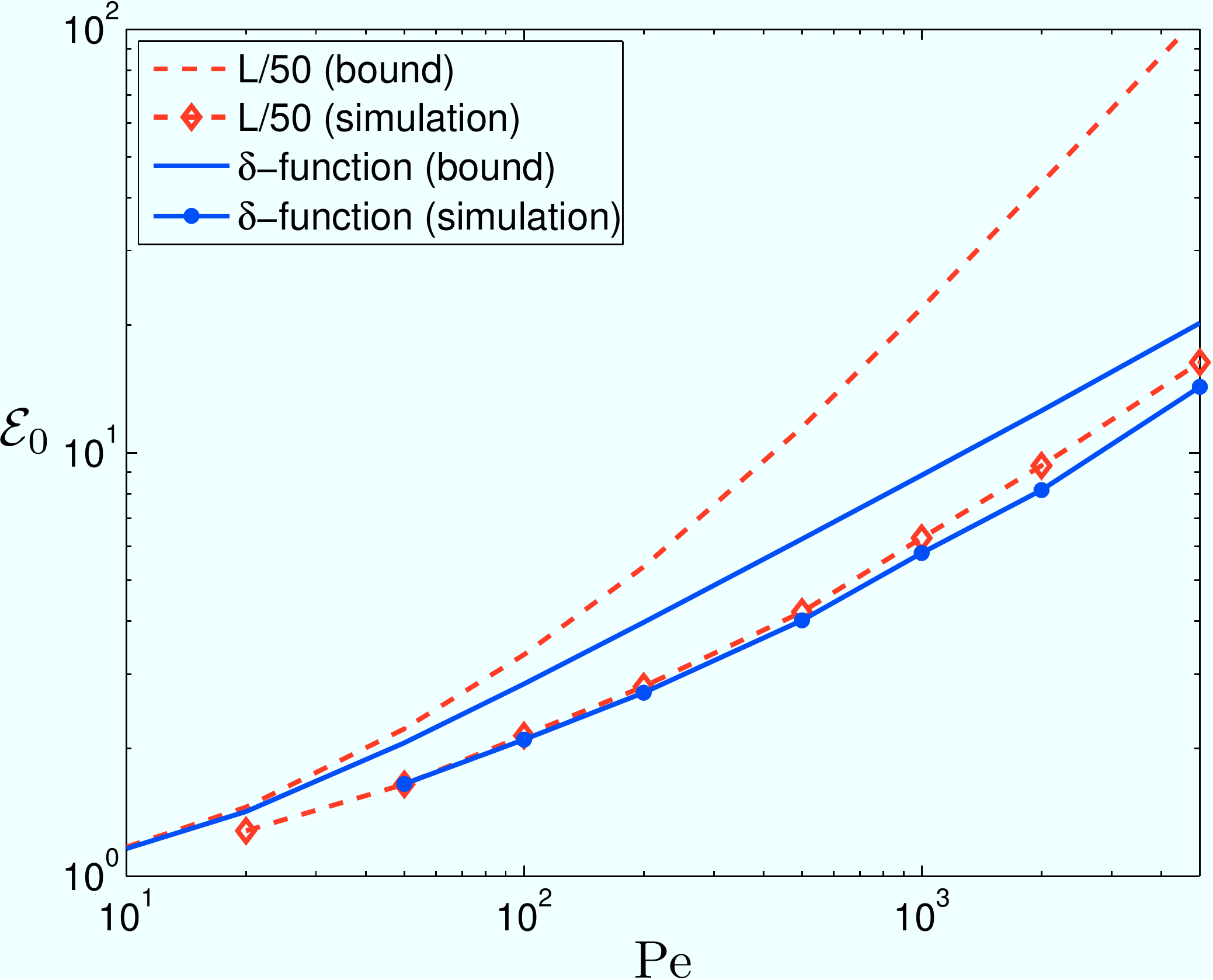}
\end{center}
\end{indented}
\caption{The theoretical upper bounds and simulation results for the
  mixing enhancement of a cubic source of size~$\ls=L/50$, and a
  $\delta$-function source.  The stirring velocity field is a 3D
  version of the random sine flow~\cite{Pierrehumbert1994} (after
  Okabe~\etal\cite{Okabe2008}).}
\label{fig:fig_OETD}
\end{figure}

An important aspect of the mixing efficiency for rough source-sink
distributions discussed above is that an anomalous large-$\Pe$ scaling
implies that molecular diffusivity is \emph{always} important, for any
SHIF.  To emphasise: there is no `residual' effective diffusivity due
to stirring in the limit of negligible molecular diffusion.  Since we
derived upper bounds, the actual scaling could be worse (see below for
numerical results).  This again highlights a theme of this review: the
source-sink distribution takes centre stage, and should not be treated
as a secondary aspect when compared to the stirring flow.

For actual SHIF flows, there is evidence that the upper bound scalings
are somewhat generous for finite-size sources.  \Fref{fig:fig_OETD}
shows results from Okabe~\etal\cite{Okabe2008}, who used a 3D
generalisation of Pierrehumbert's random sine
flow~\cite{Pierrehumbert1994} as a SHIF.  The dashed lines are for a
source of size~$\ls=L/50$, where~$L$ is the domain size.  In that
case, the upper bound scales is a rather poor indicator of the actual
flow efficiency.  However, as the source size is made smaller the
bound improves considerably: the solid lines in \fref{fig:fig_OETD}
are for a $\delta$-function source.
Chertock~\etal\cite{Chertock2010} have confirmed these results using
an accurate operator-splitting method.

Whether the results derived in this section for statistically
homogeneous and isotropic flows (SHIFs) generalise to a wider class of
flows is an important open question.  Another open question is whether
there exist SHIFs that saturate the scalings of \tref{tab:scalingsall}.

\section{Homogenisation theory with sources and sinks}
\label{sec:homo}

Homogenisation theory is a type of multiscale analysis that exploits a
large spatial scale separation between the stirring velocity field and
the source~\cite{Papanicolaou1979, Fannjiang1994, CioranescuDonato,
  Majda1999}.  We used this already in describing Plasting \& Young's
results for the sine flow~\eref{eq:sineflow} for $\NN\gg1$ in
\sref{sec:bounds}, and discussed it in the same section in connection
with regime V of Alexakis \& Tzella.

Here we review the results of Kramer \& Keating\cite{Kramer2009},
Keating \etal\cite{Keating2010}, and Lin \etal\cite{Lin2010}.  This
series of papers has its origin in a definition of an \emph{equivalent
  diffusivity} defined by Thiffeault \etal\cite{Thiffeault2004} and
generalised in Shaw \etal\cite{Shaw2007}:
\begin{equation}
  \Deq_\qq \ldef \kappa\,\Eff_\qq
  \label{eq:Deq}
\end{equation}
where~$\Eff_\qq$ is just the mixing efficiency~\eref{eq:mixefft}.
This is the diffusivity that would be required in the absence of
stirring to achieve the same level of suppression of the
norm~$\hSobo^\qq$ with stirring.  Homogenisation theory has its own
\emph{effective diffusivity}, $\Deff$, which we encountered before,
based on a mean-field approach and exploiting separation of scales.
(We note that the effective diffusivity is additive on top of the
molecular diffusivity, whereas~$\Deq_\qq$ includes the molecular
diffusivity.)  For large P\'eclet number, the effective diffusivity
satisfies the rigorous scaling bound~\cite{Avellaneda1989,
  Avellaneda1991, Fannjiang1994} $\Deff \le \Chomo\,\kappa\Pe^2$,
whereas~$\Deq_\qq$ satisfies~$\Deq_\qq \le \Chomo_2\,\kappa\Pe$
for~$\qq=1$, $0$, $-1$ (see \tref{tab:scalingsall}).  Since both
quantities claimed to measure essentially the same thing, both
large-$\Pe$ scalings could not be right.

The answer, of course, as explained in great detail for a model system
by Lin \etal\cite{Lin2010}, is that the homogenisation bound applies
only when the concept of an effective homogenisation diffusivity
exists, that is, in the homogenisation limit~$\luls \ll \Pe^{-1}$,
where the scale separation~$\luls=\lu/\ls$ was defined
in~\eref{eq:rhodef}.  Lin \etal point out that the homogenisation
limit~$\luls\rightarrow0$ does not commute with the large~$\Pe$ limit.
As a result, the large-$\Pe$ dependence of a mixing efficiency such
as~$\Eff_0$ has two distinguished regimes which cross over
when~$\luls$ is of order~$\Pe$.  They exhibit a specific example where
the efficiency~$\Eff_0 \sim \luls^{7/6}\,\Pe^{5/6}$ for fixed~$\luls$
and $\Pe\rightarrow\infty$, consistent with an upper bound linear
in~$\Pe$.  For fixed~$\Pe$ and~$\luls\rightarrow0$, they recover the
homogenisation scaling~$\Eff_0 \sim 1 + \Pe^2$.  They then introduce a
modification of Batchelor's dispersion theory~\cite{Batchelor1949},
called \emph{Dispersion-Diffusion Theory}, which successfully
reconciles effective diffusion in terms of particle dispersion and of
suppression of variance of a source-sink distribution.

A general treatment of homogenisation theory with sources and sinks
was given by Kramer \& Keating\cite{Kramer2009}.  The starting point
is a rescaled version of the advection-diffusion
equation~\eref{eq:ADs}: we let~$\xv'=\luls\xv$, $\time'=\luls^2\time$
and immediately drop the primes to get
\begin{equation}
  \luls^2\,\frac{\pd\theta}{\pd\time}
  + \luls\,\uv\!\l(\frac{\xv}{\luls},\frac{\time}{\luls^2}\r)
  \cdot\grad\theta
  - \luls^2\,\kappa\lapl\theta
  = \src\!\l(\frac{\xv}{\luls},\frac{\time}{\luls^2}\,;
  \xv,\time\r),
  \qquad \luls \ll 1,
  \label{eq:ADshomo0}
\end{equation}
where we have assumed that the source can vary on the small spatial
scale~$\xv/\luls$ and fast time scale~$\time/\luls^2$ as well as on
the larger spatial scale~$\xv$ and slower time scale~$\time$, whereas
the velocity field is confined to small spatial scales.  The
variable~$\xv/\luls$ is assumed periodic, with~$\uv$ having space-time
mean zero.  The concentration $\theta$ is expanded in the usual manner
\begin{equation}
  \theta(\xv,\time) = \theta^{(0)}(\xhv,\timeh;\xv,\time)
  + \luls\,\theta^{(1)}(\xhv,\timeh;\xv,\time)
  + \luls^2\,\theta^{(2)}(\xhv,\timeh;\xv,\time)
  + \ldots
  \label{eq:thetahomoexp}
\end{equation}
where~$\xhv\ldef\xv/\luls$, $\timeh\ldef\time/\luls^2$ (not to be
confused with the period~$\tau$ of the sine flow~\eref{eq:sineflow}).
With the fast and slow variables separated as
in~\eref{eq:thetahomoexp} we must write~$\pd_\time \rightarrow
\luls^{-2}\pd_\timeh + \pd_\time$, $\grad \rightarrow
\luls^{-1}\grad_{\xhv} + \grad_{\xv}$.  The advection-diffusion
operator on the left-hand side of~\eref{eq:ADshomo0} then splits into
three orders in~$\luls$:
\begin{subequations}
\begin{alignat}{2}
  \LL^{(0)} &= \pd_\timeh + \uv\cdot\grad_{\xhv} - \kappa\lapl_{\xhv},
  \qquad &\text{at order $\luls^0$}; \\
  \LL^{(1)} &= \uv\cdot\grad_{\xv} - 2\kappa\grad_{\xv}\cdot\grad_{\xhv},
  \qquad &\text{at order $\luls^1$}; \\
  \LL^{(2)} &= \pd_\time - \kappa\lapl_{\xv},
  \qquad &\text{at order $\luls^2$}.
\end{alignat}
\end{subequations}

We have not yet posited a magnitude for the source.  The simplest case
that will yield a nontrivial self-consistent solution is to take a
weak source,~$\src = \luls^2\ssrc$, where~$\ssrc$ is order one.  Then
at leading order we have~$\LL^{(0)}\theta^{(0)}=0$, which means
that~$\theta^{(0)}(\xhv,\timeh;\xv,\time) = \Theta^{(0)}(\xv,\time)$,
\ie, it depends only on the slow variables.  (This is a consequence
of the solvability condition and uniqueness results --- see Lemma 3.1
in~\cite{Kramer2009}.)  At the next order we
have~$\LL^{(0)}\theta^{(1)} = -\LL^{(1)}\theta^{(0)} =
-\uv\cdot\grad_{\xv}\Theta^{(0)}$.  The solvability condition for this
equation says that~$\uv\cdot\grad_{\xv}\Theta^{(0)}$ must average to
zero over the small scales, which it does since~$\uv$ averages to zero
and~$\Theta^{(0)}$ does not depend on the small scales.  Hence, we
have~$\theta^{(1)}=\Theta^{(1)}(\xv,\time) +
\bm{\chi}\cdot\grad_{\xv}\Theta^{(0)}$, where~$\bm{\chi}$ satisfies
the so-called \emph{cell problem}
\begin{equation}
  \LL^{(0)}\bm{\chi} = -\uv(\xhv,\timeh),
  \label{eq:cell}
\end{equation}
where~$\bm{\chi}$ has space-time mean zero.

The next and final order has~$\LL^{(0)}\theta^{(2)} =
-\LL^{(1)}\theta^{(1)} -\LL^{(2)}\theta^{(0)} + \ssrc$, but we only
require the solvability condition that the space-time average of the
right-hand side over the small spatial and temporal scales vanishes.
This solvability condition leads directly to the homogenised diffusion
equation
\begin{equation}
  \pd_\time\Theta^{(0)}
  = \grad_{\xv}\cdot\l(\Defft(\xv,\time)\,
  \grad_{\xv}\Theta^{(0)}\r) + \Src(\xv,\time)
  \label{eq:ADshomo}
\end{equation}
where~$\eye$ is the unit tensor, $\Src$ is $\ssrc$ averaged over small
space-time scales, and $\Defft$ is the tensor
\begin{equation}
  [\Defft]_{ij}(\xv,\time) \ldef
  \kappa\bigl(\eye
  + \stavg{\grad_{\xhv}\chi_i\grad_{\xhv}\chi_j}_{\xhv,\timeh}\bigr),
  \qquad
  \Src(\xv,\time) = \stavg{\ssrc}_{\xhv,\timeh}
  \label{eq:DefftSrc}
\end{equation}
where the subscripts~$\xhv,\timeh$ remind us that the average is over
small and fast scales~$\xhv$ and~$\timeh$.  If the system is
isotropic, the scalar effective diffusivity~$\Deff$ we introduced
earlier appears on the diagonal of~$\Defft$.

So far everything has proceeded as one would expect: the homogenised
equation~\eref{eq:ADshomo} is exactly the standard one with an
averaged source added.  How can things go wrong and become more
interesting?  The most obvious way is if the small-scale
average~$\Src$ in~\eref{eq:DefftSrc} vanishes identically.  In that
case we do not get a self-consistent equation involving the source,
and we must rescale the source differently.  Following Kramer \&
Keating, we set~$\src = \luls\ssrc$, which is a stronger source.  At
order~$\luls^0$ nothing changes from before, but at order~$\luls^1$ we
get~$\LL^{(0)}\theta^{(1)} = -\uv\cdot\grad_{\xv}\Theta^{(0)} +
\ssrc$, the source now making an appearance.  The solvability
condition is still satisfied, since the source averages to zero at the
small scales, by assumption.  We may then express the solution as
\begin{equation}
  \theta^{(1)}(\xhv,\timeh;\xv,\time)
  = \Theta^{(1)}(\xv,\time) + \bm{\chi}\cdot\grad_{\xv}\Theta^{(0)}
  + \theta_\src(\xhv,\timeh;\xv,\time)
\end{equation}
where~$\bm{\chi}$ is again the unique periodic mean-zero solution to
the cell problem~\eref{eq:cell}, and~$\theta_\src(\xhv,\timeh;\xv,\time)$
is the unique periodic mean-zero solution to
\begin{equation}
  \LL^{(0)}\theta_\src(\xhv,\timeh;\xv,\time) = \ssrc(\xhv,\timeh;\xv,\time)
\end{equation}
for every~$(\xv,\time)$.  Again, this `source cell problem' has a
solution because the source satisfies~$\stavg{\ssrc}_{\xhv,\timeh}=0$,
by assumption.  The term~$\theta_\src$ gives us one extra term in the
solvability condition for the next order, which now appears as a
source instead of the vanishing averaged source in~\eref{eq:ADshomo}:
\begin{equation}
  \pd_\time\Theta^{(0)} =
  \grad_{\xv}\cdot\l(\Defft(\xv,\time)\,
  \grad_{\xv}\Theta^{(0)}\r)
  - \grad_{\xv}\cdot\stavg{\uv\,\theta_\src}_{\xhv,\timeh}
  \label{eq:ADshomo2}
\end{equation}
where~$\Defft$ is defined as in~\eref{eq:DefftSrc}.  Thus in this case
because the small-scale mean of the source vanishes the source of
concentration arises from the small-scale correlations between~$\uv$
and~$\theta_\src$ that give rise to large-scale variations.  To
paraphrase Kramer \& Keating, $\theta_\src$ is exactly the local
response of the passive scalar field to the source on the small scale,
with the large-scale variation frozen at its local value.  Then
$\uv\,\theta_\src$ is the advective flux of passive scalar density
generated in response to the local behaviour of the source.  It would
be of great interest to generate examples of this type, and to study
how their efficiency scales.

The is a third case mentioned by Kramer \& Keating\cite{Kramer2009},
when~$\grad_{\xv}\cdot\stavg{\uv\,\theta_\src}_{\xhv,\timeh}$ vanishes as well
as~$\stavg{\theta_\src}_{\xhv,\timeh}$.  Then we must promote the
strength of the source again, so that it comes in at order zero.  Two
source cell problems must then be solved for the, but the resulting
homogenised equation looks very similar to~\eref{eq:ADshomo2}.

The mixing efficiencies associated with~\eref{eq:DefftSrc} can easily
be derived~\cite{Keating2010}, since in the absence of flow we just
set~$\Defft=\kappa$:
\begin{subequations}
\begin{align}
  \Eff_0^2 &= \stavg{\Src\lapl_{\xv}^{-2}\Src}
  \bigl/
  \stavg{\lvert\l(\grad_{\xv}\cdot
    (\Defft/\kappa)\cdot\grad_{\xv}\r)^{-1}\Src\rvert^2},\\[6pt]
  \Eff_{-1}^2 &= 
  \stavg{\Src\lapl_{\xv}^{-3}\Src}
  \bigl/
  \stavg{
    \lvert
    \grad^{-1}\l(\grad_{\xv}\cdot(\Defft/\kappa)\cdot\grad_{\xv}\r)^{-1}\Src
    \rvert^2},
\end{align}
\label{eq:Effshomo}%
\end{subequations}
where now the space-time averages are over large scales.  These
simplify considerably
if~$\Defft(\xv,\time)=\Deff\,\eye=\text{const.}$:
\begin{equation}
  \Eff_0 = \Deff/\kappa,\qquad
  \Eff_{-1} = \Deff/\kappa.
  \label{eq:Effshomodiag}
\end{equation}
Thus, in this homogenisation limit these two efficiencies are
identical, and the definition of equivalent diffusivity~\eref{eq:Deq}
for~$\qq=-1,0$ is the same as the effective diffusivity~$\Deff$ (if we
include the additive molecular value).

The astute reader will have noticed that we did not list~$\Eff_1$.
Indeed, Keating \etal\cite{Keating2010} showed that $\Eff_1$ expressed
as for~\eref{eq:Effshomo} directly in terms of~$\theta_0=\Theta_0$ is
not correct in the context of homogenisation theory.  The cause is
that the gradient now takes the scale-separated
form~$\grad=\luls^{-1}\grad_{\xhv}+\grad_{\xv}$, so the~$\luls^{-1}$
can promote the smaller-order term~$\theta_1$ to leading other.  We
thus have
\begin{equation}
  \grad\theta = \grad_{\xv}\theta^{(0)} + \grad_{\xhv}\theta^{(1)}
  + \Order{\luls^1}
\end{equation}
from which
\begin{equation}
  \stavg{\lvert\grad\theta\rvert^2}_{\xhv,\timeh}
  = \stavg{\lvert\grad_{\xv}\theta^{(0)}\rvert^2}_{\xhv,\timeh}
  + \stavg{\lvert\grad_{\xhv}\theta^{(1)}\rvert^2}_{\xhv,\timeh}
  + 2\stavg{\grad_{\xv}\theta^{(0)}\cdot\grad_{\xhv}\theta^{(1)}}_{\xhv,\timeh}
  + \Order{\luls^1}
\end{equation}
The last term vanishes since~$\grad_{\xv}\theta^{(0)}$ doesn't depend
on the fast variables, and then using~\hbox{$\theta^{(0)}=\Theta^{(0)}$}
and~$\theta^{(1)}=\Theta^{(1)}(\xv,\time) +
\bm{\chi}\cdot\grad_{\xv}\Theta^{(0)}$ we get
\begin{align*}
  \stavg{\lvert\grad\theta\rvert^2}_{\xhv,\timeh}
  &= \lvert\grad_{\xv}\Theta^{(0)}\rvert^2
  +
  \sum_{i,j}\stavg{\grad_{\xhv}\chi_i\cdot\grad_{\xhv}\chi_j}_{\xhv,\timeh}
  \frac{\pd\Theta^{(0)}}{\pd\xc_i}\frac{\pd\Theta^{(0)}}{\pd\xc_j}\\
  &= \grad_{\xv}\Theta^{(0)}\cdot(
  \Defft/\kappa)\cdot\grad_{\xv}\Theta^{(0)}
\end{align*}
in the limit as~$\luls\rightarrow0$.  Inserting the steady
solution~$\Theta^{(0)} =
\l(\grad_{\xv}\cdot(\Defft/\kappa)\cdot\grad_{\xv}\r)^{-1}\Src$, we
obtain the gradient norm efficiency
\begin{equation}
  \Eff_1^2 = 
  \stavg{\Src\lapl_{\xv}^{-1}\Src}
  \bigl/
  \stavg{
    \grad\l(\grad_{\xv}\cdot(\Defft/\kappa)\cdot\grad_{\xv}\r)^{-1}\Src
    \cdot
    (\Defft/\kappa)
    \cdot
    \grad\l(\grad_{\xv}\cdot(\Defft/\kappa)\cdot\grad_{\xv}\r)^{-1}\Src}.
  \label{eq:Effshomom1}
\end{equation}
The difference from directly trying to generalise~\eref{eq:Effshomo}
for the gradient norm mixing efficiency is that here there is an
extra~$(\Defft/\kappa)$ sandwiched in the denominator.  It is easier
to see how this differs from~\eref{eq:Effshomo} by specialising
to~$\Defft(\xv,\time)=\Deff\,\eye=\text{const.}$,
\begin{equation}
  \Eff_1 = (\Deff/\kappa)^{1/2},
  \label{eq:Effshomom1diag}
\end{equation}
and comparing to~\eref{eq:Effshomodiag}.

Keating \etal\cite{Keating2010} suggest defining a
small-scale-averaged version of the equivalent diffusivities,
\begin{equation}
  \Deq_\qq(\xv,\time) \ldef
  \kappa\,\stavgt{\lvert\grad^\qq\thetaz\rvert^2}_{\xhv,\timeh} \bigl/
  \stavgt{\lvert\grad^\qq\theta\rvert^2}_{\xhv,\timeh},
  \label{eq:Deqss}
\end{equation}
where as before~$\thetaz$ is the purely-diffusive solution.  The
advantage of definition~\eref{eq:Deqss} over an homogenised~$\Deff$ is
that it can be made \emph{even if there is no formal scale separation}
between the large scales and the small scales, that
is,~$\Deq_\qq(\xv,\time)$ characterises scalar dissipation due to
processes at `subgrid scales' whether or not there is a `gap' between
the small and large scales.  In fact for~$\qq=\pm1$ we can go a step
further and naturally generalise~\eref{eq:Deqss} to tensorial
quantities,
\begin{equation}
  [\Deqt_{\pm1}(\xv,\time)]_{ij} \ldef
  \kappa\,\stavgt{\lvert\grad^{\pm1}\thetaz\rvert^2}_{\xhv,\timeh} \bigl/
  \stavgt{(\grad^{\pm1}\theta)_i(\grad^{\pm1}\theta)_j}_{\xhv,\timeh},
  \label{eq:Deqtss}
\end{equation}
with~$\qq=-1$ being the preferred choice to relate to the effective
diffusivity, because of the different scaling for~$\Eff_1$
in~\eref{eq:Effshomom1diag}.

\section{Optimisation for the source-sink problem}
\label{sec:optsrc}

Given sources and sinks, there is an obvious optimisation problem: for
fixed energy, which incompressible velocity field has the highest
mixing efficiency~$\Eff_\qq$?  The stirring velocity field should in
principle satisfy a fluid equation such as Stokes or Navier--Stokes,
but we can also optimise over all incompressible velocity fields to
get an upper bound on efficiency.  This optimisation problem will be
discussed in~\sref{sec:velopt}.

The presence of sources and sinks implies a different optimisation
problem.  If the position of the sources and sinks is part of the
design process (as it often is for industrial applications), then we
may try to optimise the source-sink locations as well.  This is a less
familiar problem, but one that is easier to tackle because of the
structure of~\eref{eq:ADs}.  We discuss thus this problem first
in~\sref{sec:srcopt}.

\subsection{Source optimisation}
\label{sec:srcopt}

By far the easier optimisation problem is one that is less intuitive:
given a stirring velocity field, what is the source-sink distribution
which is \emph{best mixed} by the flow.  This problem was examined by
Thiffeault \& Pavliotis~\cite{Thiffeault2008}.  For example, suppose
we have a room whose temperature we wish to control, and that there
happens to be a predominant airflow in that room which is relatively
unaffected by the temperature distribution.  Then we can ask where to
put heaters (sources) and windows (sinks) so that the temperature is
as uniform as possible.

We illustrate the optimisation procedure on the time-independent
advection--diffusion equation,
\begin{equation}
  \uv(\xv) \cdot \grad \theta - \kappa \lapl \theta = \src(\xv),
  \qquad \div\uv = 0,
  \label{eq:ADsTI}
\end{equation}
in $\Vol=[0,\Lsc]^\sdim$ with periodic boundary conditions. The
velocity field $\uv(\xv)$ is specified. Both $\uv(\xv)$ and
$\src(\xv)$ are assumed to be sufficiently smooth.  As before, we
assume that the source and initial condition have spatial mean zero,
which implies that the scalar concentration also has mean zero.

Our goal is to maximise the efficiency~$\Eff_\pexp$ defined by
\eref{eq:mixeff},
\begin{equation*}
  \Eff_\pexp^2 = {\Ltnormt{\mlapl^{\pexp/2}\thetaz}^2} \bigl/
  {\Ltnormt{\mlapl^{\pexp/2}\theta}^2}\,,
\end{equation*}
where $\thetaz$ solves equation \eref{eq:ADsTI} in the absence of
advection, $-\kappa \lapl \thetaz = \src$.  In maximising the
efficiency~$\Eff_\pexp$, we fix the $\Ltwo$ norm of the velocity field
(or equivalently, the mean kinetic energy of the flow), and vary the
diffusivity through the P\'eclet number~$\Pe =
{\Ltnorm{\uv}\!\Lsc}/{\kappa}$.

Define the linear operators
\begin{equation*}
  \LL \ldef \uv(\xv) \cdot \grad - \kappa \lapl \qquad \text{and} \qquad \LLz
  \ldef - \kappa \lapl,
  \label{eq:LLdef}
\end{equation*}
from which we can write the solution to~\eref{eq:ADsTI} and to the
purely-diffusive problem as~$\theta = \LL^{-1}\src$ and~$\thetaz =
\LLz^{-1}\src$, respectively.  We can then rewrite the
efficiency~\eref{eq:mixeff} as
\begin{equation}
  \Eff_\pexp^2 =
  \frac{\Ltnormt{\mlapl^{\pexp/2} \LLz^{-1}\src}^2}
  {\Ltnormt{\mlapl^{\pexp/2} \LL^{-1}\src}^2}
  = \frac{\bigl\langle s\, \cAz_\pexp^{-1}\src \bigr\rangle}
  {\bigl\langle \src\, \cA_\pexp^{-1}\src \bigr\rangle}\,,
  \label{eq:mixeff2}
\end{equation}
where the self-adjoint operators~$\cA_\pexp$ and~$\cAz_\pexp$ are
\begin{equation}
  \cA_\pexp \ldef \LL \mlapl^{-\pexp} \LL^*\,,\qquad \cAz_\pexp \ldef \LLz
  \mlapl^{-\pexp} {\LLz}^* = \kappa^2 \mlapl^{2-\pexp}\,,
\label{eq:cA}
\end{equation}
and as before $\langle \cdot \rangle$ denotes an average over~$\Vol$.
To maximise~$\Eff_\pexp^2$, we compute its variation with respect to
$\src$ and set it equal to zero,
\begin{equation}
  \delta\Eff_\pexp^2 = \frac{2}{\bigl\langle \src\, \cA_\pexp^{-1}\src
    \bigr\rangle}\l\langle
  \l(\cAz^{-1}_\pexp \src - \Eff_\pexp^2\,\cA^{-1}_\pexp \src\r)\delta \src
  \r\rangle = 0,
  \label{eq:firstvar}
\end{equation}
which implies
\begin{equation}
  \cAz^{-1}_\pexp \src = \Eff_\pexp^2\,\cA^{-1}_\pexp \src\,.
  \label{eq:EuLagz}
\end{equation}
This is an eigenvalue problem for the operator $\cA_\pexp
{\cAz_\pexp}^{-1}$. Optimal sources are given by ground states of the
inverse of this operator, and the normalised variance is given by the
corresponding (first) eigenvalue.  The minimisation problem has a
unique minimum, though it may be realised by more than one source-sink
distribution, in particular when the flow has
symmetry~\cite{Thiffeault2008}.

The operators $\cA_\pexp^{-1}$ and $\cAz_\pexp^{-1}$ are self-adjoint
from~$\Ltwo(\Vol)$ to~$\Ltwo(\Vol)$; furthermore,
they are both positive operators in~$\Ltwo(\Vol)$
(restricted to functions with mean zero).  Consequently, the
generalised eigenvalue problem \eref{eq:EuLagz} has real positive
eigenvalues, and the eigenfunctions~$\src$ and~$\src'$ corresponding
to distinct eigenvalues are orthogonal with respect to the weighted
inner product~$(\src\,,\,\src') \ldef \langle
\src\,\cAz_\pexp^{-1}\src'\rangle$.  For
numerical implementation, it is preferable to solve the equivalent
self-adjoint eigenvalue problem
\begin{equation}
  (\cAz_\pexp^{-1/2}\cA_\pexp\cAz_\pexp^{-1/2})\,\rsrc = \Eff_\pexp^2\,\rsrc\,,
  \qquad
  \src \rdef \cAz_\pexp^{1/2}\,\rsrc\,,
  \label{eq:ge}
\end{equation}
for the eigenvector~$\rsrc$, which then yields the optimal source
distribution~$\src$.  The advantage of the form~\eref{eq:ge} is that
the self-adjoint structure of the operator is explicit.

Our goal now is to calculate the optimal source and the corresponding
mixing efficiency for some simple velocity fields.  Notice that the
operator $\cAz_\pexp^{-1}$ is a diagonal operator in Fourier space
with entries~$\kappa^2 \km^{2\pexp-4}$, where~$\km=\lvert\kv\rvert$ is
the magnitude of the wavevector.  For~$\pexp<2$, this operator acts as
a low-pass filter, suppressing high frequencies.

\subsubsection{Dependence on P\'eclet number}
\label{sec:diffdep}

\begin{figure}
\begin{indented}
\item[]
\begin{center}
\subfigure{\includegraphics[width=.25\textwidth]{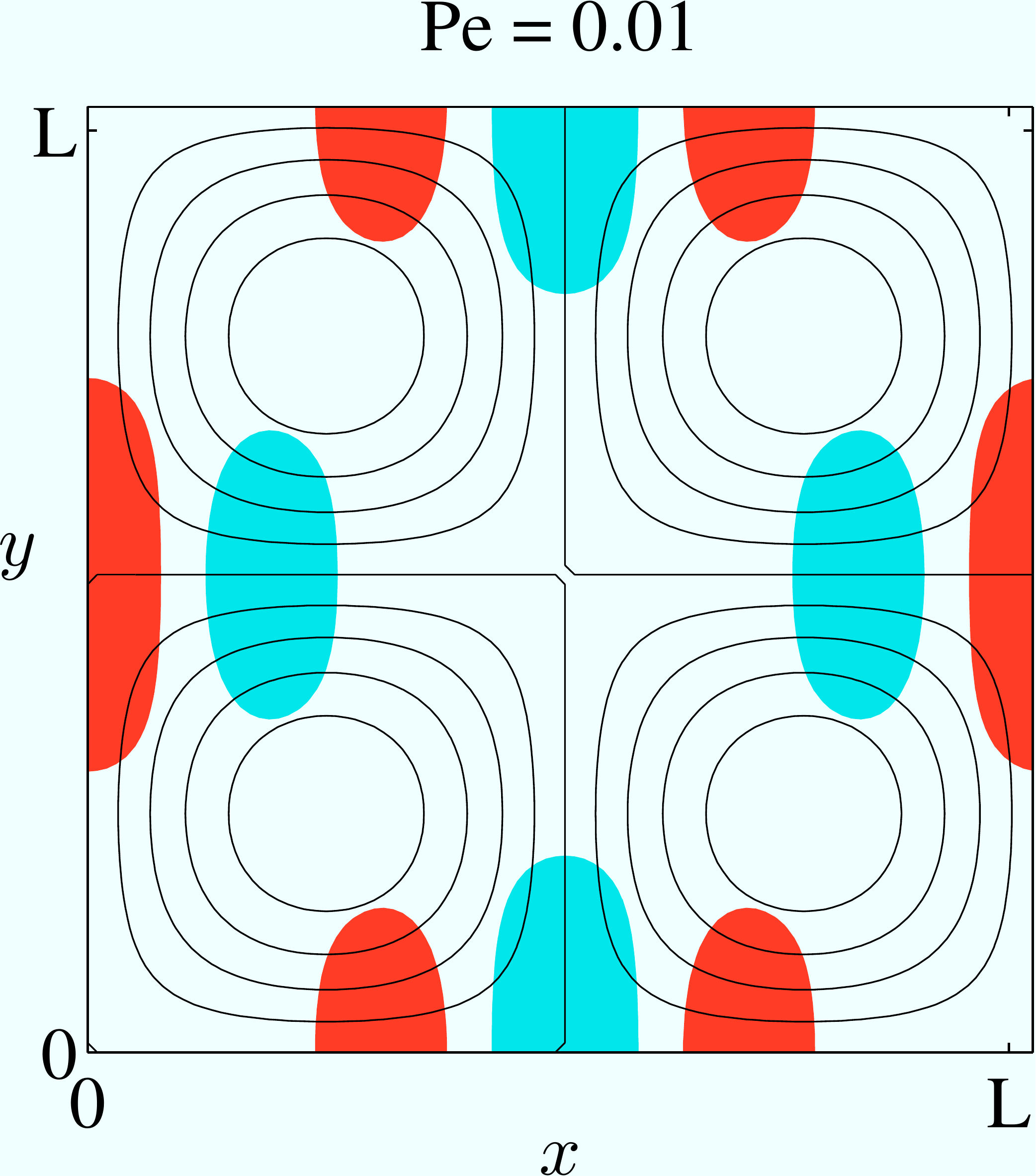}}
\hspace{.5em}
\subfigure{\includegraphics[width=.25\textwidth]{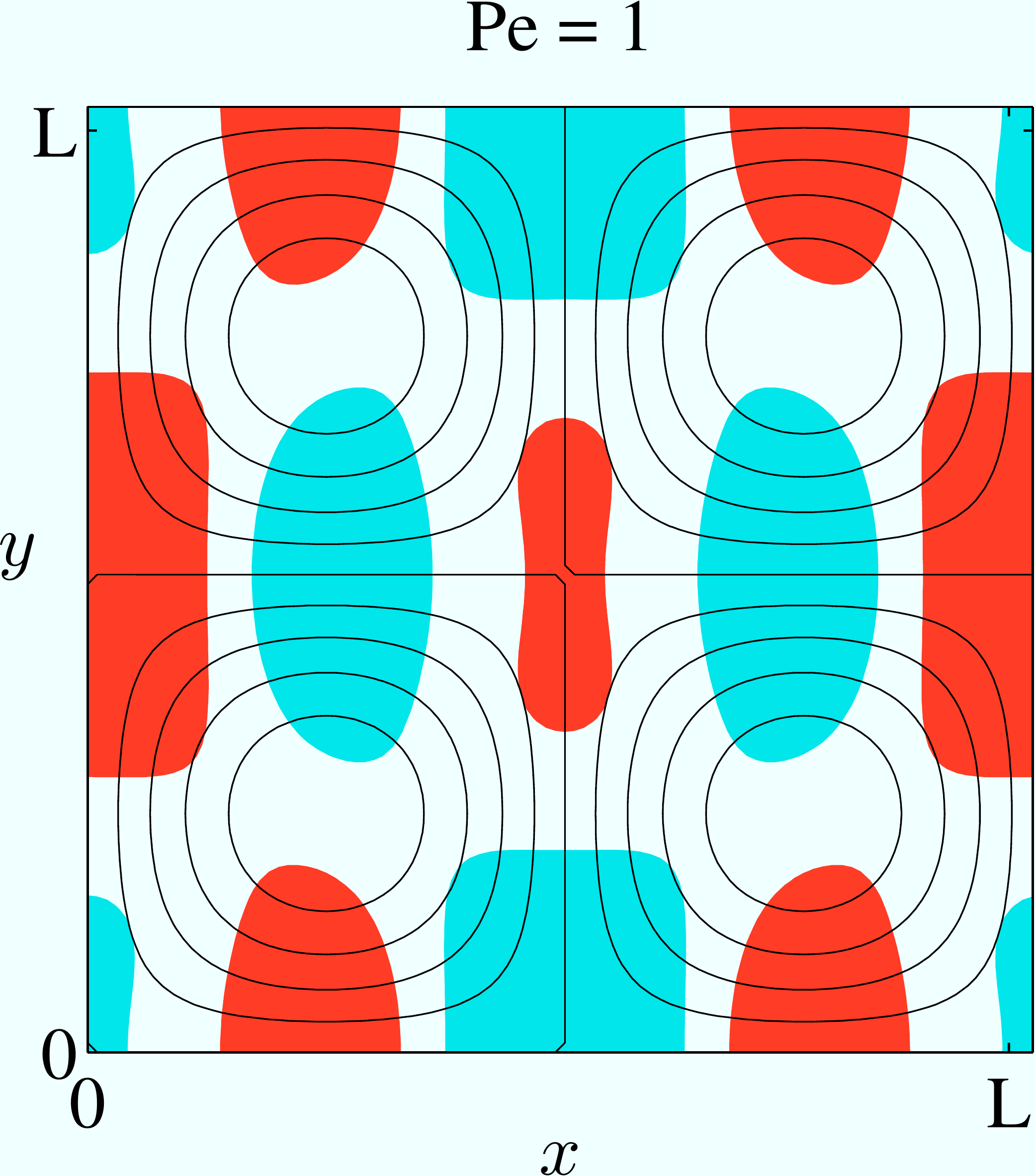}}
\hspace{.5em}
\subfigure{\includegraphics[width=.25\textwidth]{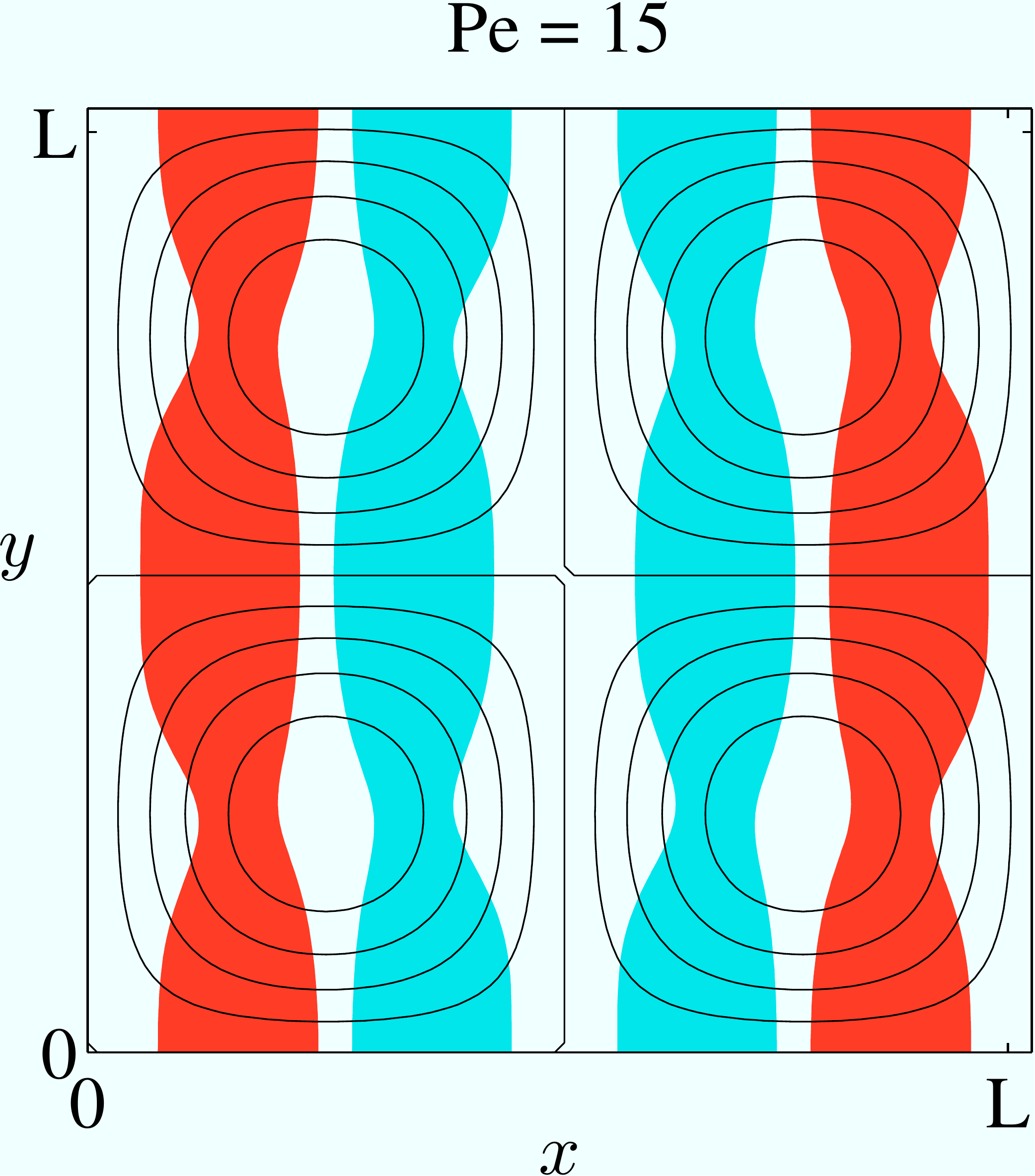}}
\end{center}
\end{indented}
\caption{Source distribution that optimises~$\Eff_0$ for the cellular
  flow~\eref{eq:sf}, for three values of the P\'eclet numbers.  Note
  how there is no source of heat over the stagnation points.  The
  background shading shows hot (red, or dark grey) and cold (blue, or
  light grey) regions, separated by tepid regions (white).  The black
  contour lines are streamlines of the flow.  For both small and
  large~$\Pe$ the optimal source converges to an invariant
  eigenfunction.  In all cases there is no source of temperature over
  the elliptic stagnation points, but in the small~$\Pe$ case there
  are sources and sinks over some hyperbolic points.}
\label{fig:srcopt_cell_Pe}
\end{figure}

We consider the cellular flow on the
domain~$\Vol=[0,\Lsc]^2=[0,2\pi]^2$ with streamfunction
\begin{equation}
  \psi(\xc,\yc) = \sqrt{2}\,\sin \xc\, \sin \yc
  \label{eq:sf}
\end{equation}
and velocity field~\hbox{$\uv =
  (\uc_x,\uc_y)=(\pd_y\psi,-\pd_x\psi)$}, normalised to
make~\hbox{$\Ltnorm{\uv} = 1$}.  For this cellular flow there are two
independent optimal source eigenfunctions with degenerate optimal
efficiency; the degeneracy is a consequence of the discrete symmetries
of the flow~\cite{Thiffeault2008}.  In \fref{fig:srcopt_cell_Pe} we
show the source that optimises~$\Eff_0$ for three values of~$\Pe$.  In
the foreground are contour lines of the streamfunction.  The optimal
source distribution appears to become independent of~$\Pe$ both for
small~$\Pe$ and large~$\Pe$, but the distributions are different.  The
transition between the two regimes occurs when~$\Pe$ is of order
unity.  Though the two asymptotic sources are very different, they
respect some general principles: the source is arranged for effective
transport of hot onto cold and vice versa, and regions of high speed
are favoured.  In particular, note that the centre of the rolls has a
nearly zero, flat source distribution in all cases.

Another perhaps surprising aspect of the small~$\Pe$ solution in
\fref{fig:srcopt_cell_Pe} is that it has complicated structure.
In this large diffusivity limit, one would expect diffusion to
dominate and gradients to be smoothed out.  But since our mixing
efficiency~\eref{eq:mixeff} compares the variance to the
unstirred case, which already has very low variance, any amount of
improvement will count.  Hence, the complicated source for
small~$\Pe$ in \fref{fig:srcopt_cell_Pe} only gives a minute
improvement to the efficiency.  The small-$\Pe$ optimal
solution is particular in that it has some hot and cold spots
localised over hyperbolic stagnation points.  This is probably due to
the high speeds along the separatrices being favoured, even at the cost
of straddling hyperbolic stagnation points a little.

\begin{figure}
\begin{indented}
\item[]
\begin{center}
\includegraphics[width=.5\textwidth]{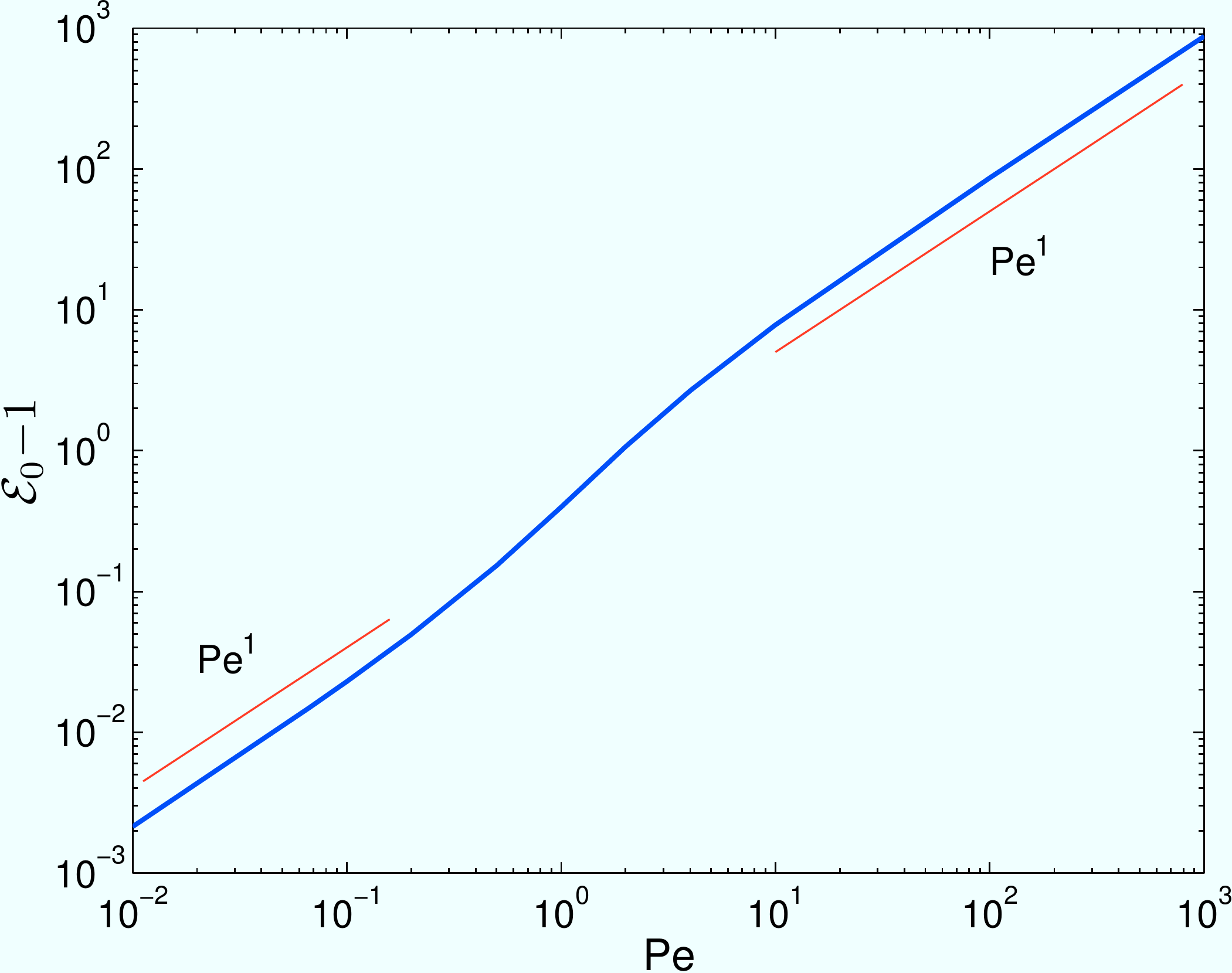}
\end{center}
\end{indented}
\caption{For the flow with streamfunction~\eref{eq:sf}, mixing
  efficiency~$\Eff_0-1$ for the optimal source distribution as a
  function of the P\'eclet number (after Thiffeault \&
  Pavliotis\cite{Thiffeault2008}).}
\label{fig:effplot_cell_Pe}
\end{figure}

In \fref{fig:effplot_cell_Pe} we show the value of the optimal
efficiency~$\Eff_0$ as a function of~$\Pe$.  For large~$\Pe$, the
efficiency typically scales linearly with~$\Pe$: this is the
`classical' scaling discussed in~\cite{Thiffeault2004,
  DoeringThiffeault2006, Shaw2007, Thiffeault2008}.  For small~$\Pe$,
the optimal efficiency also converges towards unity linearly
with~$\Pe$.

In summary, the optimal source distribution becomes independent
of~$\Pe$ for both large and small $\Pe$, but of course for small~$\Pe$
the efficiency gain is minimal (since the~$\Ltwo$ norm of the velocity
is fixed).

\subsubsection{Dependence on norm chosen}
\label{sec:pexpdep}

Our final study will be to examine the behaviour of the optimal
efficiency~$\Eff_\qq$ as~$\pexp$ is varied in~\eref{eq:mixeff}.  In
\sref{sec:diffdep} we used~$\pexp=0$; now we fix~$\Pe=100$, and
allow~$\pexp$ to vary over negative and positive values.
\begin{figure}
\begin{indented}
\item[]
\begin{center}
\subfigure{\includegraphics[width=.25\textwidth]{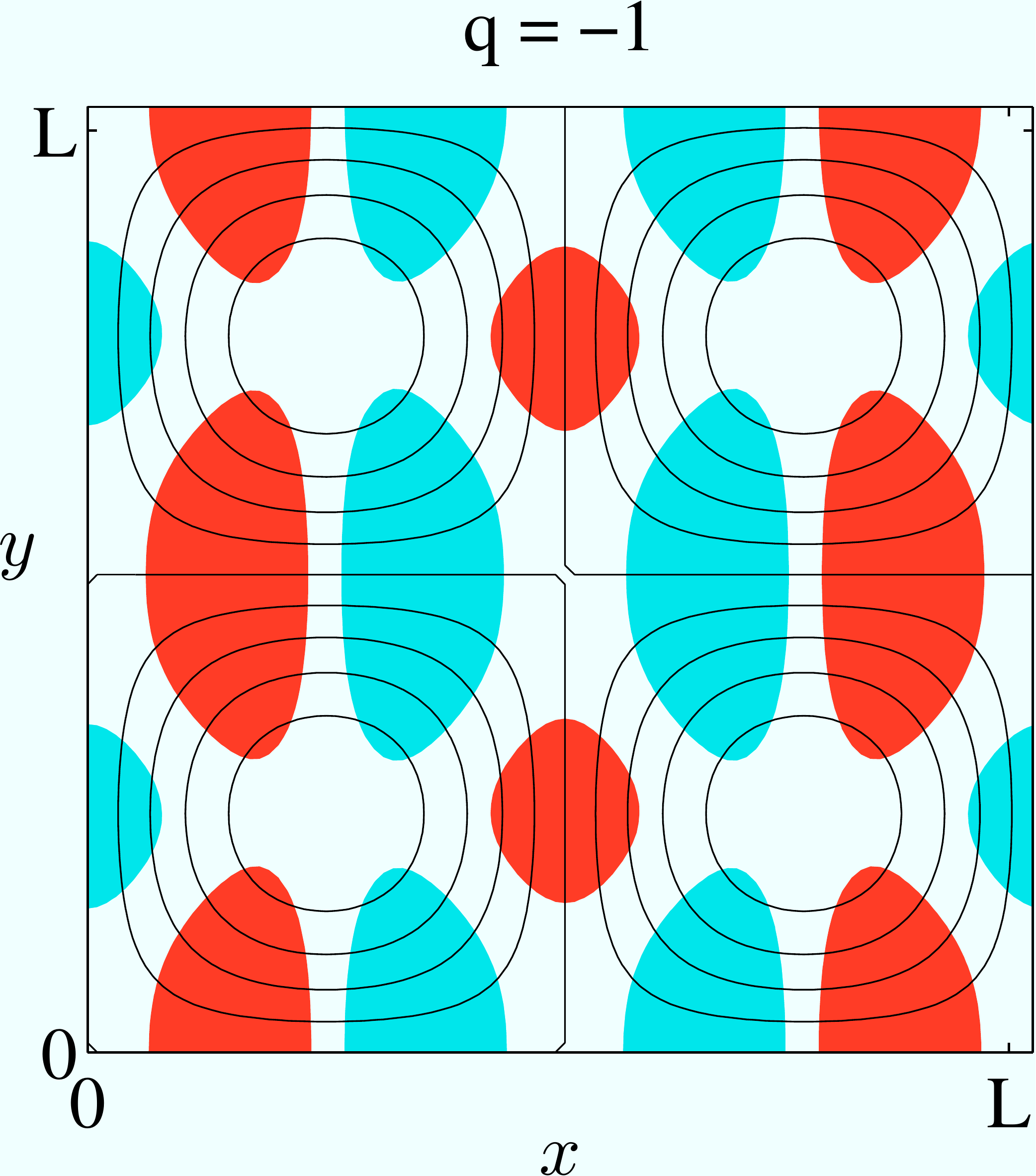}}
\hspace{.5em}
\subfigure{\includegraphics[width=.25\textwidth]{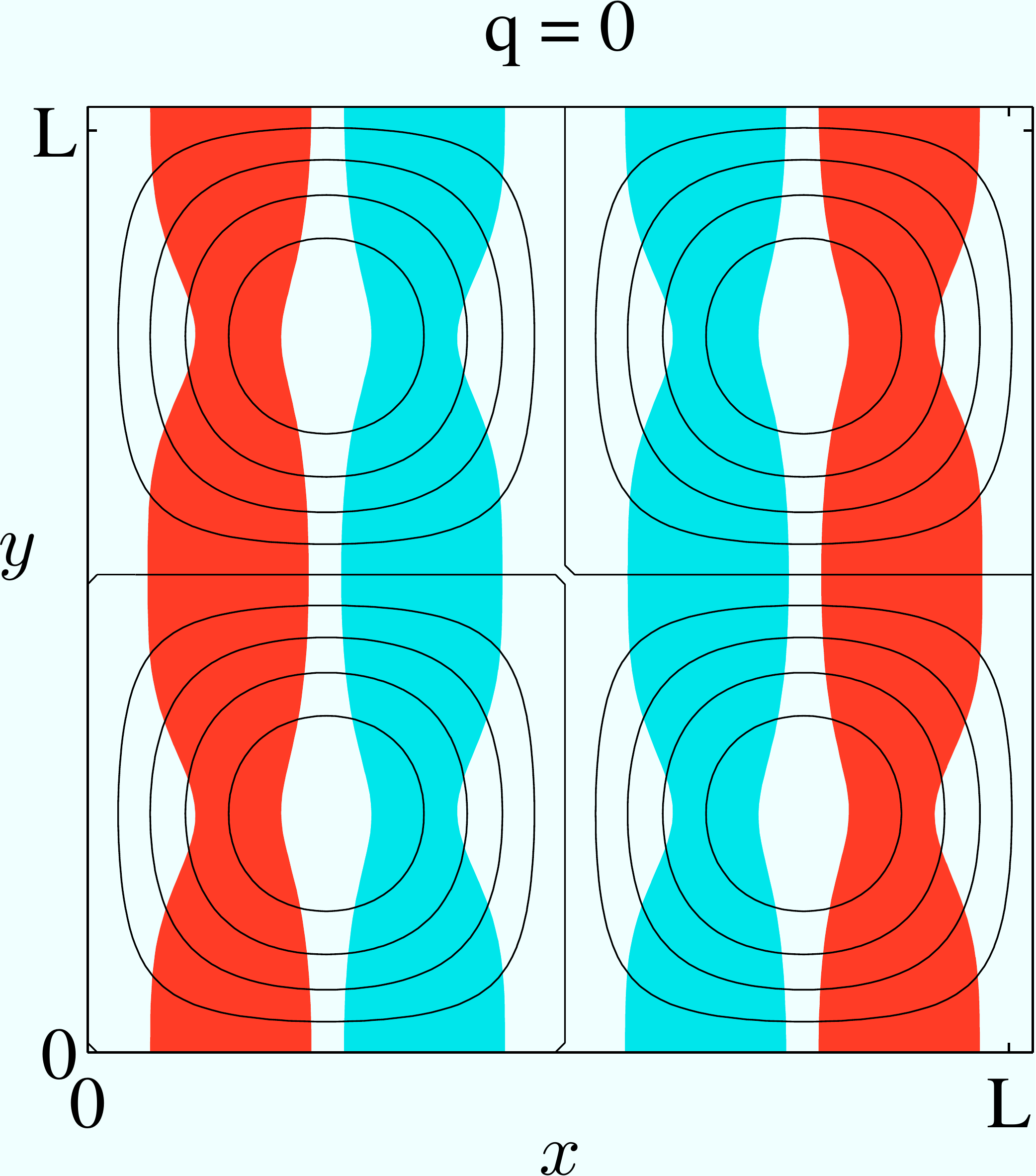}}
\hspace{.5em}
\subfigure{\includegraphics[width=.25\textwidth]{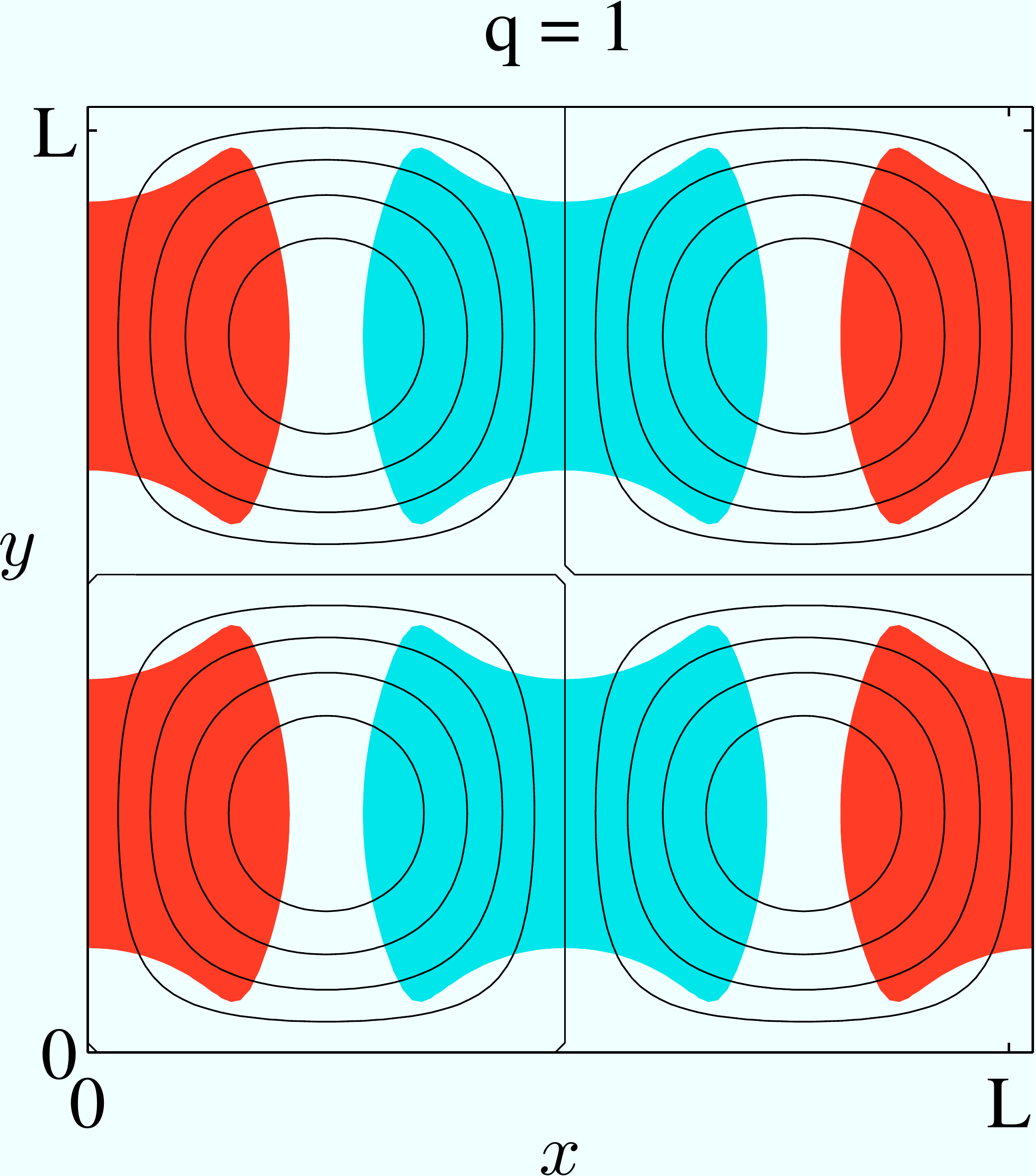}}
\end{center}
\end{indented}
\caption{For the flow with streamfunction~\eref{eq:sf}, optimal source
  distribution~$\src(\xv)$ for~$\Pe=100$ and~$\pexp=-1$, $0$, and~$1$.
  In all cases there are no sources or sinks of temperature over the
  stagnation points.  (See the caption to \fref{fig:srcopt_cell_Pe}
  for a key to the background shading and contours.)}
\label{fig:srcopt_cell_q}
\end{figure}
\Fref{fig:srcopt_cell_q} shows the optimal source distributions
for~$\pexp=-1$, $0$, and~$1$.  For large~$\lvert\pexp\rvert$ (not
shown), the optimal source distribution converges rapidly to invariant
patterns.  The~$\qq=-1$ case in \fref{fig:srcopt_cell_q}
(negative~$\pexp$) shows small, localised sources and sinks.  In
contrast, the~$\qq=1$ case (positive~$\pexp$) shows large, regular
localised sources and sinks.  In fact, what is striking about the
pattern is its simplicity: it is what one might take as a guess at an
efficient source distribution, with no added frills.  Thus, a high
power of~$\pexp$ might be useful in situations where a simple
configuration is preferable due to engineering constraints.  The
reason for the simplicity is that spatial variations in the source
favour the diffusion operator in~$\LL$, and
as~$\pexp\rightarrow\infty$ these are magnified.  Thus, the source
must remain as spatially simple as possible while trying to maximise
alignment with the velocity.  As~$\pexp\rightarrow-\infty$, spatial
variations of the source are downplayed by the norm, allowing more
complexity.

\begin{figure}
\begin{indented}
\item[]
\begin{center}
\includegraphics[width=.45\textwidth]{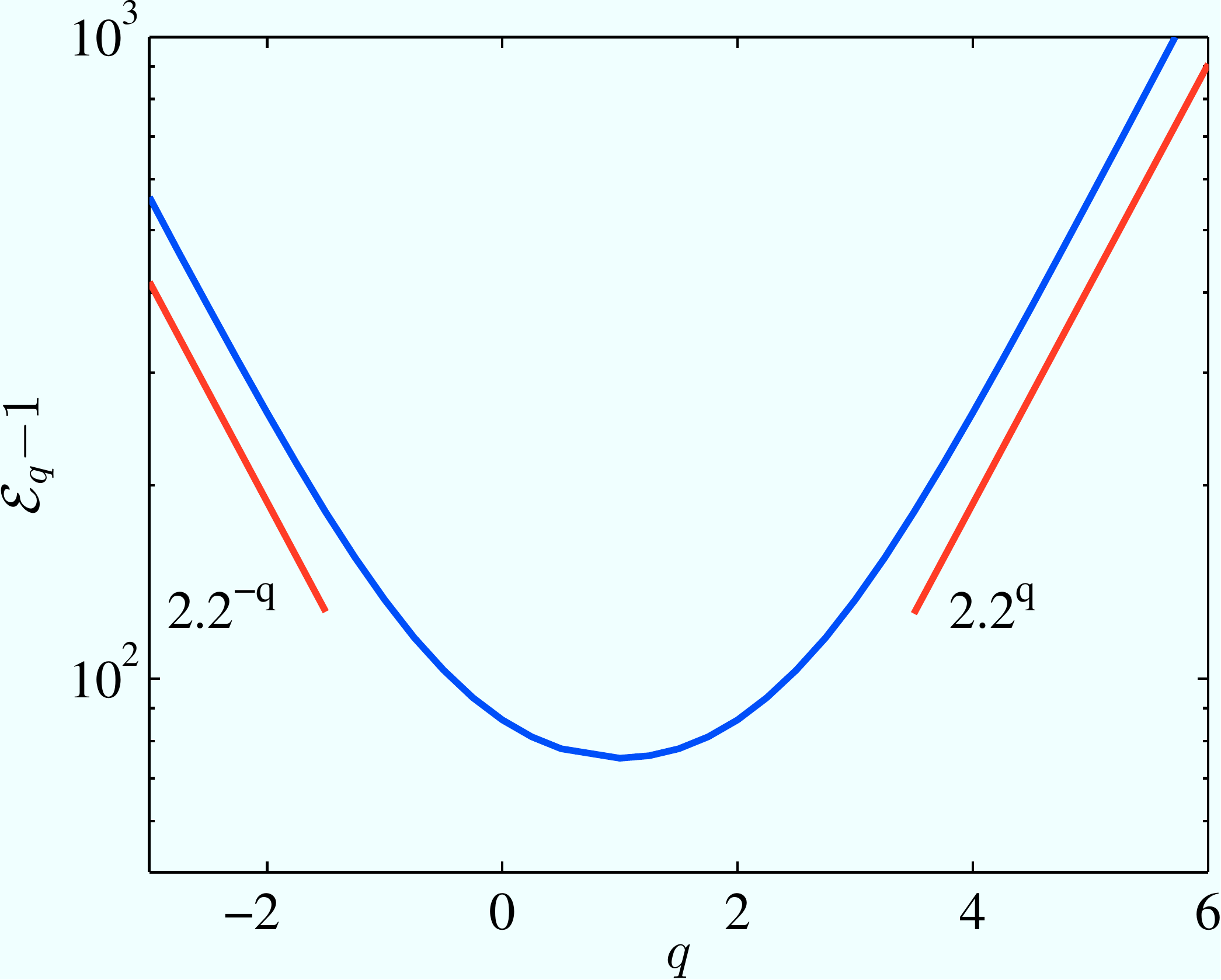}
\end{center}
\end{indented}
\caption{For the flow with streamfunction~\eref{eq:sf}, mixing
  efficiency~$\Eff_\pexp-1$ as a function of the exponent~$\pexp$ for
  optimal source at~$\Pe=100$.  The optimal efficiency is symmetric
  about~\hbox{$\pexp=1$}, and for~$\lvert\pexp\rvert\gg 1$ it grows
  as~$(2.2)^{\lvert\pexp\rvert}$ (after Thiffeault \&
  Pavliotis\cite{Thiffeault2008}).}
\label{fig:effplot_cell_q}
\end{figure}

\Fref{fig:effplot_cell_q} shows how the optimal mixing efficiency
varies as a function of~$\pexp$.
For~$\lvert\pexp\rvert\gg1$, the efficiency scales exponentially
as~$2.2^{\lvert\qq\rvert}$.  Note that the curve is symmetric
about~\hbox{$\pexp=1$}, which leads to a minimum there: whether this
is true in general has not been proved, but no counterexample has been
found.  Thiffeault \& Pavliotis~\cite{Thiffeault2008} provide a
partial proof by explicitly finding the symmetry between the
operators~$\cA_{2-\pexp}$ and~$\cA_\pexp$, but only for small~$\Pe$.

\subsection{Velocity field optimisation}
\label{sec:velopt}

We turn now to a more obviously relevant problem, that of optimising
the stirring velocity field for a given source-sink configuration.  As
in \sref{sec:srcopt}, we restrict to the time-independent problem for
simplicity.  We will need the functional derivative
of~$\normt{\theta}_{\hSobo^\qq}$, which arises from the variation
\begin{equation}
  \delta\normt{\theta}_{\hSobo^\qq}^2
  = \delta\Ltnormt{\mlapl^{\pexp/2} \LL^{-1}\src}^2
  = \delta\savg{(\mlapl^{\pexp/2} \LL^{-1}\src)^2}.
\end{equation}
Since the velocity field only appears in the operator~$\LL$ defined
in~\eref{eq:LLdef}, we have
\begin{equation}
  \delta\normt{\theta}_{\hSobo^\qq}^2
  = 2\savg{\mlapl^{\pexp/2}
    \LL^{-1}\src\, \mlapl^{\pexp/2} \delta\LL^{-1}\src}.
\end{equation}
Using the property~$\delta\LL^{-1}=-\LL^{-1}\,\delta\LL\,\LL^{-1}$
leads to
\begin{align*}
  \delta\normt{\theta}_{\hSobo^\qq}^2
  &= -2\savg{\mlapl^{\pexp/2}
    \LL^{-1}\src\, \mlapl^{\pexp/2} \LL^{-1}\,\delta\LL\,\LL^{-1}\src}\\
  &= -2\savg{\l(\LL^{-1*}\mlapl^{\pexp}
    \LL^{-1}\src\r) \delta\uv\cdot\grad\LL^{-1}\src},
\end{align*}
where we integrated by parts, used the adjoint~$\LL^*$ of~$\LL$, and
substituted~$\delta\LL=\delta\uv\cdot\grad$.  From the
definition~\eref{eq:cA} of the self-adjoint operator~$\cA_\pexp$, we
can then write the functional derivative as
\begin{equation}
  \tfrac12\frac{\delta\normt{\theta}_{\hSobo^\qq}^2}{\delta\uv}
  = -\l(\cA_\pexp^{-1}\src\r) \grad\LL^{-1}\src.
\end{equation}
To formulate an optimisation problem, we also need to add constraints
on~$\uv$: incompressibility and fixed energy.  This is done in the
usual manner by considering Lagrange multipliers in the extended
functional
\begin{equation}
  \func[\uv] = \tfrac12\normt{\theta}_{\hSobo^\qq}^2
  + \tfrac12\mu\,(\Ltnorm{\uv}^2 - \Uc^2) + \savg{\nu\, \div\uv}.
\end{equation}
Here~$\mu$ and~$\nu(\xv)$ are the Lagrange multipliers, with~$\nu$
a function of space since~$\div\uv=0$ is a pointwise constraint.  The
functional derivative of~$\func[\uv]$ then gives the Euler--Lagrange
equation,
\begin{equation}
  \frac{\delta\func[\uv]}{\delta\uv}
  = -\l(\cA_\pexp^{-1}\src\r) \grad\LL^{-1}\src
  + \mu\,\uv - \grad\nu = 0.
  \label{eq:EL}
\end{equation}
to be solved for~$\uv$ for given~$\src$.  Note that~\eref{eq:EL} is
profoundly nonlinear in~$\uv$, since it enters in the nonlocal
operators~$\LL^{-1}$ and~$\cA_\pexp^{-1}$.

Let us restrict to the two-dimensional case, where we can introduce a
streamfunction~$\psi$ with~$\uv = (\pd_y\psi,-\pd_x\psi)$.  Then
taking the curl of~\eref{eq:EL} yields
\begin{equation}
  \mu\,\lapl\psi
  + \brak{\cA_\pexp^{-1}\src}{\LL^{-1}\src}
  = 0,
  \label{eq:EL2}
\end{equation}
where
\begin{equation}
  \brak{f}{g} = \pd_x f\, \pd_y g - \pd_y f\, \pd_x g.
\end{equation}
\Eref{eq:EL2} is a nonlinear eigenvalue problem that can be solved in
several ways.  A direct approach is to start with an initial guess
for~$\psi$ and~$\mu$ and compute the \emph{residual vector}, that
is the amount by which~\eref{eq:EL2} fails to be satisfied.  We also
append the constraint~$(\Ltnorm{\uv}^2-\Uc^2)$ to the residual vector.
Then use a multidimensional nonlinear solver such as Matlab's
\texttt{fsolve} that finds zeroes of the residual vector, by adjusting
the vector~$(\psi,\mu)$.  Here~$\psi$ has been discretised in some
way, either by specifying it on a grid or by expanding it as a Fourier
series.

\begin{figure}
\begin{indented}
\item[]
\begin{center}
\subfigure[]{%
\includegraphics[height=.23\textheight]{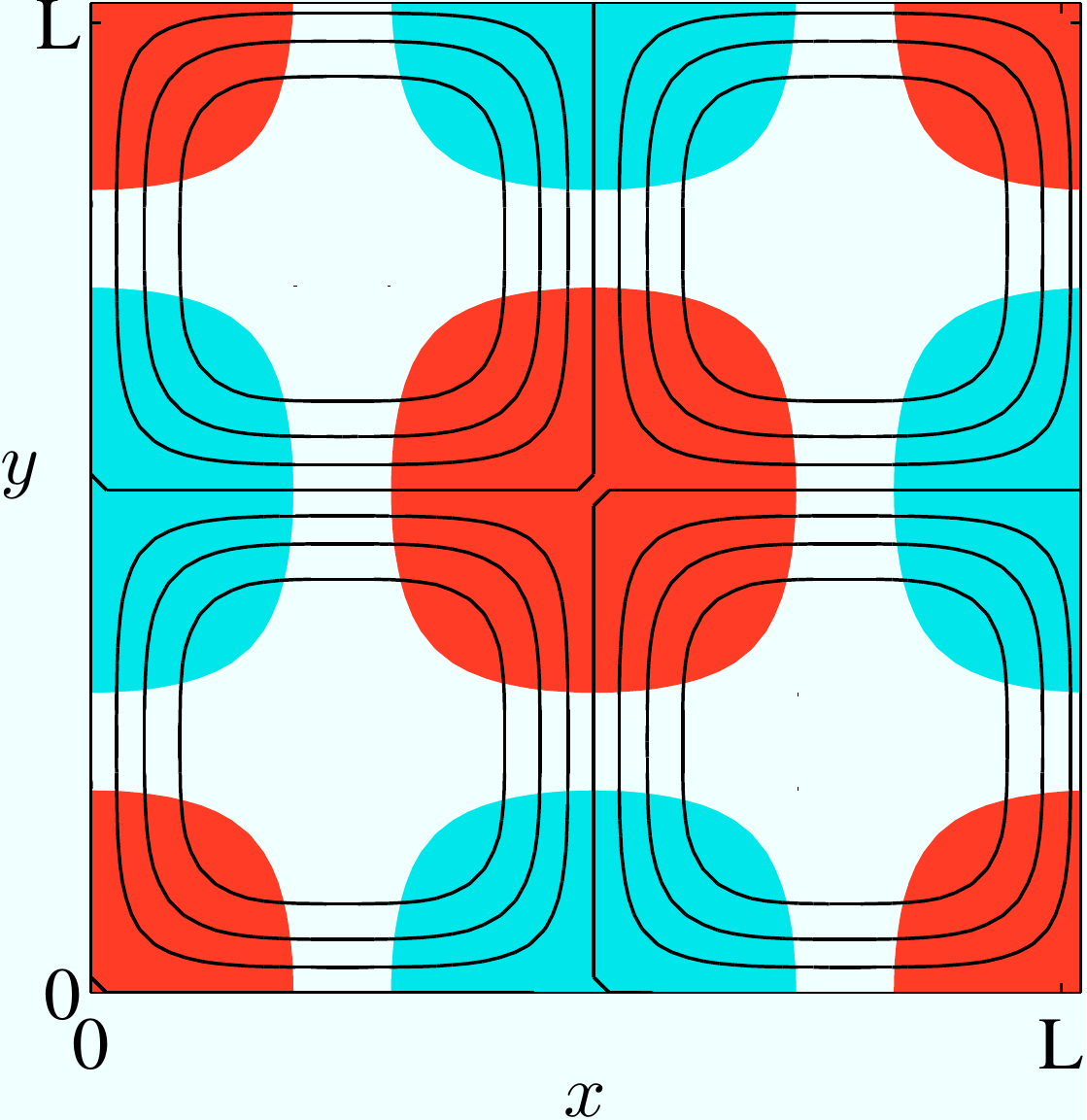}
\label{fig:velopt}
}\hspace{1em}
\subfigure[]{%
\includegraphics[height=.23\textheight]{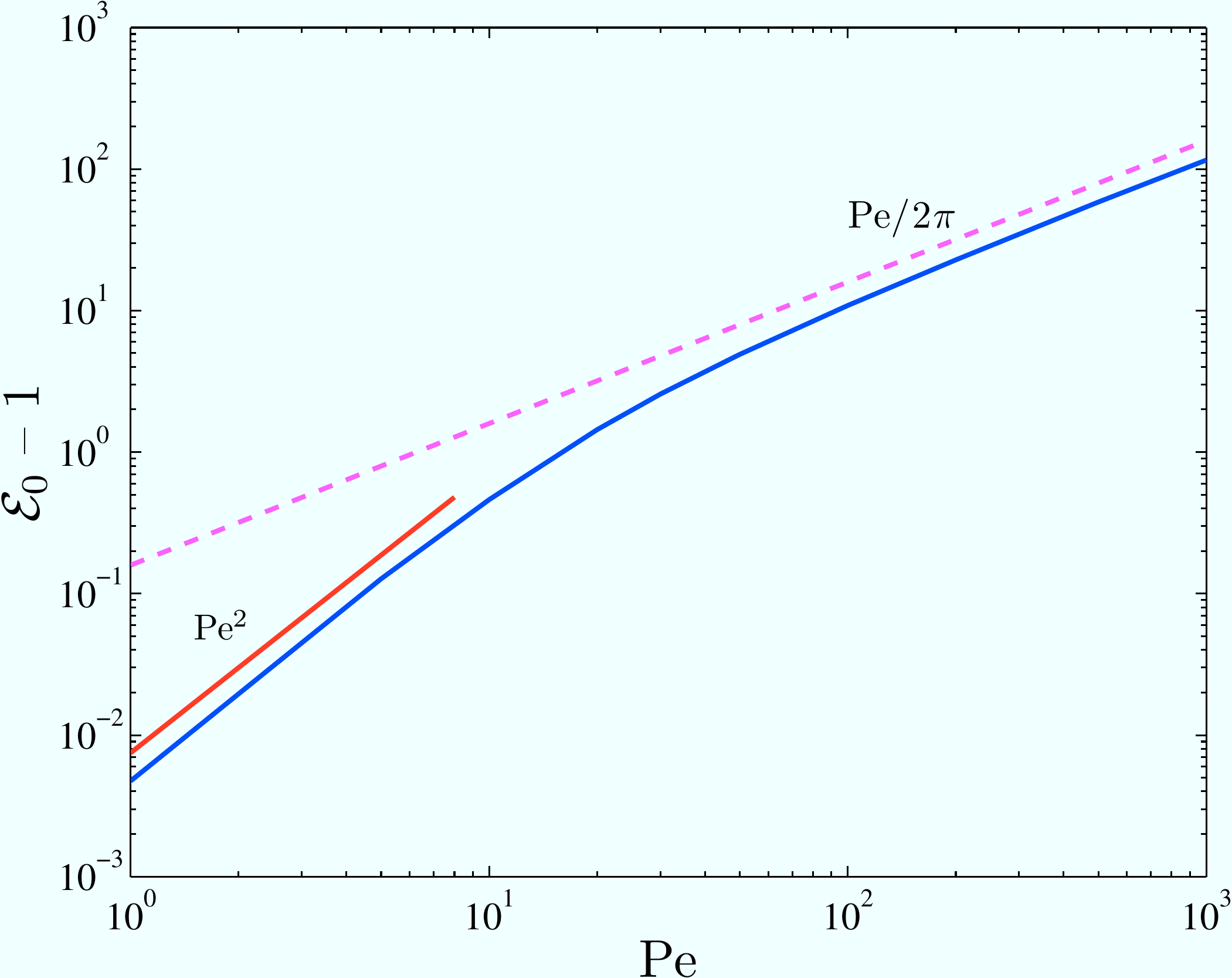}
\label{fig:normHq}
}
\end{center}
\end{indented}
\caption{(a) Optimal stirring velocity field (streamlines) for the
  source~$\cos\xc\cos\yc$, for~$\Pe=10$.  (b) Dependence on P\'eclet
  number of the optimal mixing efficiency~$\Eff_0$.  For small~$\Pe$
  the optimal streamfunction approaches~$\sqrt2\,\sin\xc\sin\yc$.  The
  dashed line is the upper bound~\eref{eq:Varbounduniformsinsin}.}
\end{figure}

\Fref{fig:velopt} shows a solution of~\eref{eq:EL2} with~$\pexp=0$,
$\Uc=1$, and~$\Pe=10$ for the source~$\cos\xc\cos\yc$.
The efficiency corresponding to this solution is~$\Eff_0=1.46$.  The
flow is close to the standard cellular flow~\eref{eq:sf}, but with a
flattened core where velocities are smaller.  Whether this is truly
optimal is a difficult question: \eref{eq:EL2} has many solutions with
different~$\mu$, and unlike the source optimisation case (which leads
to a linear eigenvalue problem) there is no simple way of finding
minimising solutions (but see the upper
bound~\eref{eq:Varbounduniformsinsin} below).  It is an open
challenge to characterise the solutions of~\eref{eq:EL}
and~\eref{eq:EL2} more thoroughly.  The solution in \fref{fig:velopt}
was obtained from the initial guess $\sqrt2\,\sin\xc\sin\yc$, and all
other initial conditions examined gave larger values of the norm.

\Fref{fig:normHq} shows the dependence on~$\Pe$ of the optimal mixing
efficiency~$\Eff_0$, for the fixed source-sink
distribution~$\sin\xc\sin\yc$ of \fref{fig:velopt}.  For smaller
values of~$\Pe$, the optimal~$\Eff_0-1$ is proportional to~$\Pe^2$:
this is the diffusion-dominated regime, where stirring only has a
small effect.  The optimal solution converges
to~$\sqrt2\,\sin\xc\sin\yc$ for small~$\Pe$. For larger values
of~$\Pe$, the optimal solution recovers the `classical' upper bound
scaling, linear in~$\Pe$.

It is instructive to compare the optimal efficiency plotted in
\fref{fig:normHq} with the `global bound'~\eref{eq:Varbounduniform}.
This requires a choice of comparison function~$\varphi$, and the
simplest is to take~$\varphi(\xv)=\src(\xv)=\cos\xc\cos\yc$.  We then
have~$\savg{\varphi\src}=1/4$,~$\norm{\grad\varphi}_{\Linf}=1$
and~$\Ltnorm{\lapl\varphi}=1$ in~\eref{eq:Varbounduniform}.
After normalising by the purely-diffusive solution we obtain
\begin{equation}
  \Eff_0 - 1 \le \frac{\Uc}{\kappa} = \frac{\Pe}{2\pi}\,,
  \qquad \text{(here $\Lsc = 2\pi$)},
  \label{eq:Varbounduniformsinsin}
\end{equation}
which is valid even if the velocity field is allowed to be
time-dependent, in which case~$\Uc$ is defined as in~\eref{eq:Ucdef}.
The bound~\eref{eq:Varbounduniformsinsin} is plotted as a dashed line
in \fref{fig:normHq}, where we see that our optimal solution is
remarkably close to the global upper bound for large~$\Pe$
(about~$37.5\%$ above the optimal solution).  This shows that the
series of inequalities required to
obtain~\eref{eq:Varbounduniformsinsin} do not cause too much loss of
sharpness for large~$\Pe$, but it also implies that even allowing for
arbitrary \emph{time-dependent} stirring cannot improve~$\Eff_0$ by
very much.  The bound~\eref{eq:Varbounduniform} thus helps to
determine if our local optimal solution is anywhere close to being a
global optimum.

\section{Discussion}
\label{sec:discussion}

\begin{figure}
\begin{indented}
\item[]
\begin{center}
\subfigure{%
\includegraphics[width=.32\textwidth]{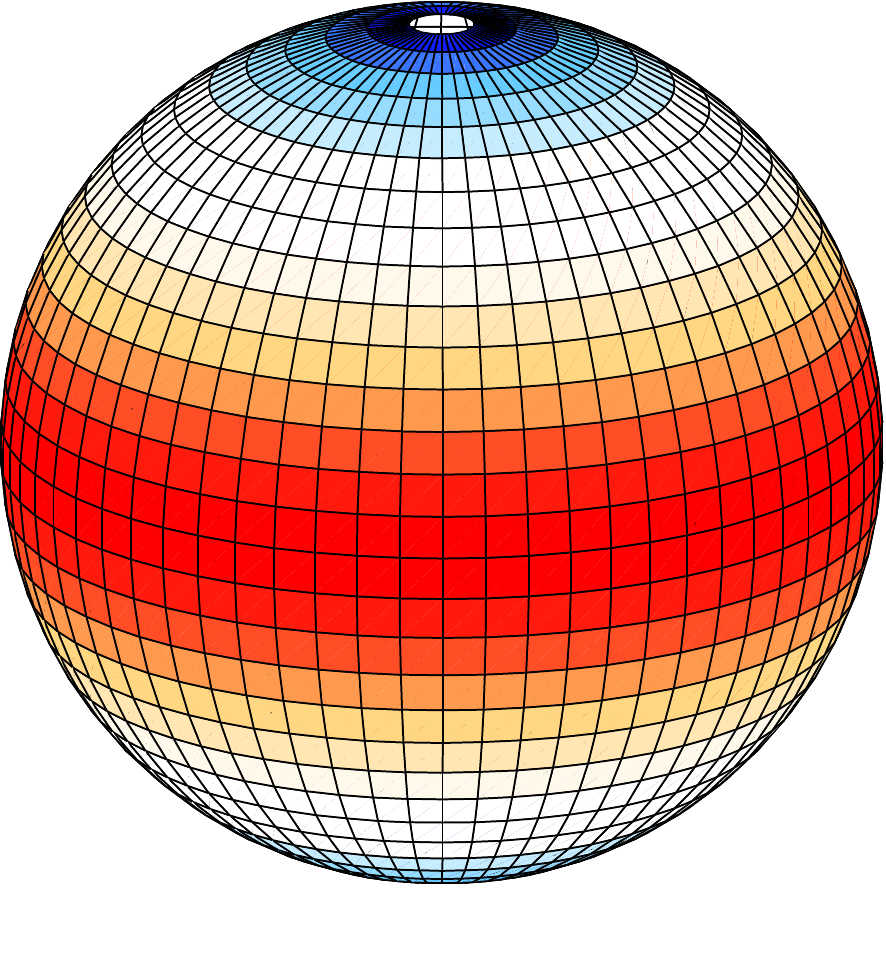}
}\hspace{1em}
\subfigure{%
\includegraphics[width=.45\textwidth]{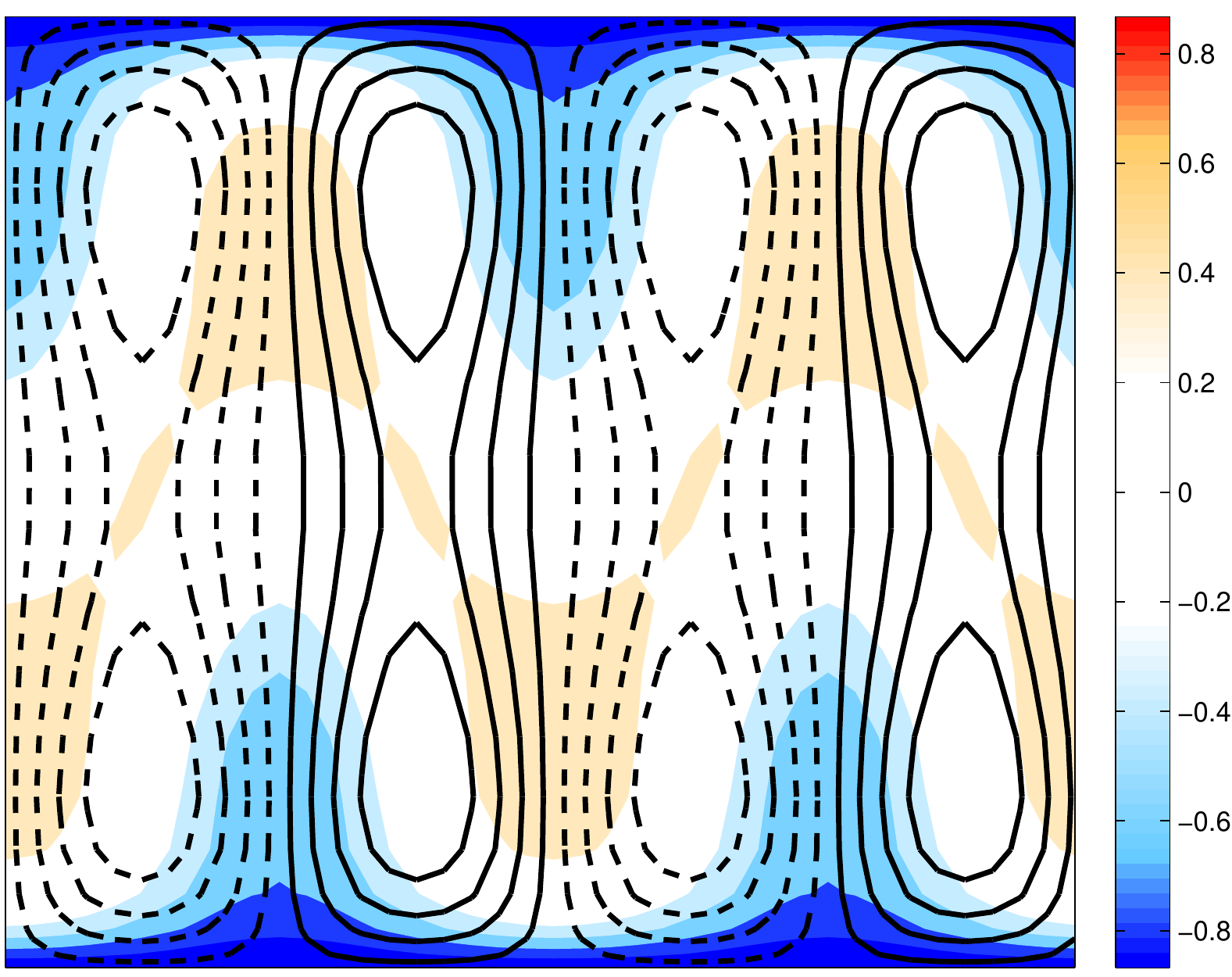}
}
\end{center}
\end{indented}
\caption{Left: a sphere heated at the equator and cooled at the poles.
  Right: longitude--latitude plot of the streamlines of a steady flow
  that maximises~$\Eff_0$, along with the temperature field in the
  background (from O'Rourke~\cite{ORourkeGFD2010}).}
\label{fig:Sphere}
\end{figure}

The literature on stirring, mixing, and transport is enormous, and of
course we only covered a small corner of it, focusing on direct uses
of norms.  The decay of variance itself is an object of study, and
predicting its decay rate in terms of flow characteristics has been a
long-term goal of the theory of mixing.  The dominant approaches are
the \emph{local theory}, based on dynamical systems quantities such as
the statistics of finite-time Lyapunov exponents~\cite{Antonsen1991,
  Shraiman1994, Antonsen1996, Balkovsky1999, Falkovich2001,
  ThiffeaultAosta2004, Tsang2005}, and the \emph{global theory}, which
requires a more thorough analysis of the advection-diffusion
equation~\cite{Pierrehumbert1994, Rothstein1999, Fereday2002,
  Wonhas2002, Sukhatme2002, Thiffeault2003d, Pikovsky2003, Liu2004,
  Haynes2005, Gilbert2006, Popovych2007}.  Neither of these approaches
is particularly well-suited to optimisation, so we do not discuss them
here.  It is also important to note that there are many other measures
of mixing beyond norms --- see for instance~\cite{MattFinn2004,
  Stone2005, Ottino1990, Krasnopolskaya1999, Bigio1990,
  Thiffeault2006, Constantin2008, Stremler2008, Turner2008,
  Turner2009, Gouillart2010b} and references therein.

We have reviewed the reasons why various norms were desirable for
studying mixing, and how they could be used to optimise the rate of
decay for the initial-value, freely-decaying problem.  We then focused
for most of the review on the use of norms in the presence of sources
and sinks in the long-time limit where the system attain an
equilibrium.  Our examples used simple periodic geometries.

Let us briefly discuss the work of Constantin
\etal\cite{Constantin2008}, which contains some of the most important
recent rigorous results (we leave out a few technical conditions).
Their focus is on flows that are \emph{relaxation-enhancing}.  These
are steady flows with the property that for any given~$\time>0$
and~$\varepsilon>0$, it is possible to increase the amplitude of
stirring to make~$\Ltnorm{\theta(\cdot,\time)} < \varepsilon$.  (As
usual we assume mean-zero functions.)  In other words, it is at least
possible to achieve an arbitrary level of mixing, measured according
to the~$\Ltwo$ norm, by stirring hard enough.  Weakly mixing
incompressible flows are always relaxation-enhancing.  Constantin
\etal prove that an incompressible velocity field~$\uv(\xv)$ is
relaxation-enhancing if and only if~$\uv\cdot\grad$ has no
eigenfunctions in~$\Sobo^1$ other than the constant function.  Indeed,
if there exists such eigenfunctions then most initial conditions will
contain some admixture of them, which will not decay.  Current
examples of relaxation-enhancing flows are, however, not very
physical.  It is possible that extending the work of Constantin \etal
to time-dependent velocity fields would greatly increase the range of
flows that are relaxation-enhancing, but this is likely to be
difficult.

There are a number of open problems and areas for further study.  We
mention a few:
\begin{itemize}
\item Find the optimal velocities for more complicated flows, and
  refine the numerical methods needed to do so; for example O'Rourke
  has recently examined optimal transport flows on the
  sphere~\cite{ORourkeGFD2010} (see \fref{fig:Sphere}).
\item Understand better the transition between `transporting' and
  `mixing' flows.  The flows that minimise the norms in the presence
  of sources and sinks are very different from mixing flows.  Is this
  a flaw in the measure?  What would be a better measure?  Are there
  flows and source-sink distributions which are both optimal in the
  sense of minimising norms, but are also optimally mixing in the
  sense of ergodic theory?
\item Since all norms~$\norm{\cdot}_{\hSobo^\qq}$ for~$\qq<0$ act as
  `mix-norms,' that is, they decay if the flow is mixing, which one is
  best?  Do we need to select~$\qq$ according to particular
  applications?  Mathew \etal\cite{Mathew2005, Mathew2007}
  used~$\qq=-1/2$, and in this review we focused more on~$\qq=-1$, but
  there is no clear reason to choose either one at this point.  On the
  plus side, however, both work quite well.
\item In a similar vein as the previous problem, it is not known
  whether there are any advantages in using even more general Sobolev
  norms on the space~$W^{\qq,p}$, $1 \le p \le \infty$, rather than
  on~$W^{\qq,2}=\Sobo^\qq$.  Values of~$p$ other than~$2$ are rarely
  used in the context of mixing, except for~$p=\infty$ which is common
  (see for example~\cite{Turner2009}).  Some rigorous results, such
  as~\cite{Constantin2008}, do not depend on~$p$, which
  suggests the choice of~$p$ matters little.
\item Do the results on `roughness' of the source-sink distribution
  carry over to open flows, where the sources and sinks can be
  regarded as distributed on the boundary rather than in the
  bulk~\cite{Thiffeault2011}?
\item There has been some work on using the norm approach to quantify
  mixing in more complex systems with reactions, for example a
  decaying passive scalar~\cite{ShawGFD2005}, the Fisher
  equation~\cite{Birch2007}, and the Cahn--Hilliard
  equation~\cite{ONaraigh2008}.  In this case how do reaction rates,
  etc., depend on the source-sink structure?  What kinds of flows
  optimise reaction rates?
\end{itemize}

\ack

The author is most indebted to Charlie Doering, who co-wrote many
papers mentioned in this review.  The author also thanks Alexandros
Alexakis, Alex Kiselev, Zhi George Lin, George Mathew, Anna Mazzucato,
Igor Mezi\'{c}, Amanda O'Rourke, Tiffany Shaw, Alexandra Tzella, Jeff
Weiss, and Andrej Zlato\u{s} for many helpful discussions and
comments.  This work began while enjoying the hospitality of the
Institute for Mathematics and its Applications at the University of
Minnesota, and then the Summer Program in Geophysical Fluid Dynamics
at the Woods Hole Oceanographic Institution (both supported by NSF).
The author was funded by the Division of Mathematical Sciences of the
US National Science Foundation, under grant DMS-0806821.

\appendix

\section{Proof of mix-norm theorem}
\label{apx:weakconv}

In this appendix we prove the theorem presented at the end of
\sref{sec:mixing} relating weak convergence to the negative Sobolev
norms.  The proof is from Lin~\etal\cite{Lin2011b}.  Write the norm
for~$\Sobo^\qq(\Vol)$ as
\begin{equation}
  \norm{\fw}_{\Sobo^\qq} =
  \Bigl(\sum_\kv \lambda^{(\qq)}_{\km} \lvert
  {\hat\fw}_\kv\rvert^2\Bigr)^{1/2},
\end{equation}
where~$\lambda^{(\qq)}_{\km}=(1 + \km^2\Lsc^2)^\qq$ for
the norm~\eref{eq:Sobo}.  Suppose that~$\fw(\cdot,\time)$ is uniformly
bounded in~$\Ltwo(\Vol)$, so that~$\Ltnorm{\fw(\cdot,\time)} \le \Cu$,
and~$\lim_{\time\rightarrow\infty}\norm{\fw(\cdot,\time)}_{\Sobo^\qq}
\rightarrow 0$ for some~$\qq<0$.  Then for any~$\gt\in \Ltwo(\Vol)$,
\begin{align*}
  \lvert\pair{\fw}{\gt}\rvert
  &=
  \Biggl\lvert
  \sum_{\km \le \NK}
  \sqrt{\lambda^{(\qq)}_{\km}}\,{\hat\fw}_\kv\,
  \frac{{\hat\gt}_\kv^*}{\sqrt{\lambda^{(\qq)}_{\km}}}
  + \sum_{\km > \NK} {\hat\fw}_\kv\,{\hat\gt}_\kv^*
  \Biggr\rvert \\
  &\le
  \norm{\fw}_{\Sobo^\qq}\,
  \Biggl(\sum_{\km \le \NK}
  \frac{\lvert{\hat\gt}_\kv\rvert^2}{\lambda^{(\qq)}_{\km}}\Biggr)^{1/2}
  + \Ltnorm{\fw}
  \Biggl(\sum_{\km > \NK} \lvert{\hat\gt}_\kv\rvert^2\Biggr)^{1/2}.
\end{align*}
Given~$\epsilon>0$, first choose~$\NK(\epsilon)$ such that
\begin{equation}
  \Biggl(\sum_{\km > \NK(\epsilon)}
  \lvert{\hat\gt}_\kv\rvert^2\Biggr)^{1/2} \le \frac{\epsilon}{2\Cu},
\end{equation}
then choose~$\T(\epsilon)$ such that
\begin{equation}
  \norm{\fw(\cdot,\T(\epsilon))}_{\Sobo^\qq} \le
  \tfrac12\epsilon\,\Biggl(\sum_{\km \le \NK(\epsilon)}
  \frac{\lvert{\hat\gt}_\kv\rvert^2}{\lambda^{(\qq)}_{\km}}\Biggr)^{-1/2}\,,
  \qquad \time > \T(\epsilon).
\end{equation}
We then have
\begin{equation*}
  \lvert\pair{\fw}{\gt}\rvert
  \le
  \tfrac12\l(1 + \Cu^{-1}\Ltnorm{\fw}\r)\epsilon \le \epsilon,
  \qquad \time > \T(\epsilon),
\end{equation*}
which implies that~$\fw$ converges weakly to zero
as~$\time\rightarrow\infty$.  (This is true even for~$\qq=0$.)

Conversely, suppose~$\Ltnorm{\fw(\cdot,\time)} \le \Cu$ for
all~$\time$ and~$\lim_{\time\rightarrow\infty}
\pair{\fw}{\gt}
\rightarrow 0$ for all~$\gt\in \Ltwo(\Vol)$.  By
choosing~$\gt=\exp(-\imi\kv\cdot\xv)$ we see that all the Fourier
coefficients~${\hat\fw}_\kv(\time)\rightarrow 0$
as~$\time\rightarrow\infty$.  Also,
because~$\Ltnorm{\fw(\cdot,\time)}^2 =
\sum_\kv\lvert{\hat\fw}_\kv(\time)\rvert^2\le\Cu^2$ then
each~$\lvert{\hat\fw}_\kv(\time)\rvert \le \Cu$ for all~$\time$.

We have
\begin{align}
  \norm{\fw}_{\Sobo^\qq}^2
  &=
  \sum_{\km \le \NK} \lambda^{(\qq)}_{\km}
  \lvert {\hat\fw}_\kv\rvert^2
  + \sum_{\km > \NK} \lambda^{(\qq)}_{\km}
  \lvert {\hat\fw}_\kv\rvert^2
  \nonumber \\
  &\le
  \sum_{\km \le \NK} \lambda^{(\qq)}_{\km}
  \lvert {\hat\fw}_\kv\rvert^2
  + \lambda^{(\qq)}_{\NK} \Ltnorm{\fw}^2.
  \label{eq:laststep}
\end{align}
For any~$\epsilon>0$, we can choose~$\NK(\epsilon)$ such
that~$\lambda^{(\qq)}_{\km} \Ltnorm{\fw} \le \lambda^{(\qq)}_{\km}\Cu
< \epsilon/2$ for~$\km\ge\NK(\epsilon)$ (this requires~$\qq<0$).  For
any finite~$\NK$, $\sum_{\km\le\NK}
\lambda^{(\qq)}_{\km}\lvert{\hat\fw}_\kv(\time)\rvert^2\rightarrow 0$
as~$\time\rightarrow\infty$, so there exists~$\T(\epsilon)$ such
that~$\sum_{\km\le\NK(\epsilon)}
\lambda^{(\qq)}_{\km}\lvert{\hat\fw}_\kv(\time)\rvert^2 < \epsilon/2$, for
all~$\time>\T(\epsilon)$.  From~\eref{eq:laststep} we
obtain~$\norm{\fw}_{\Sobo^\qq}^2 < \epsilon$ for
all~$\time>\T(\epsilon)$, which proves the result.

\section*{References}

\bibliographystyle{siam}
\bibliography{bib/journals_abbrev,bib/articles}

\end{document}